%% file: survey.tex
\documentclass[acmsmall]{acmart}

\usepackage{enumitem}
\usepackage{multirow}
\usepackage{tabularx}

\usepackage{pgfplots}
\usepackage{xcolor}
\usepackage{tikz}
\usepackage{pgf-pie}
\usepackage{color}
\usetikzlibrary{patterns}

\usepackage{natbib}
\usepackage{lineno}

\usepackage{subcaption}

\AtBeginDocument{%
  \providecommand\BibTeX{{%
    \normalfont B\kern-0.5em{\scshape i\kern-0.25em b}\kern-0.8em\TeX}}}

\setcopyright{acmcopyright}
\copyrightyear{2024}
\acmYear{2024}
\acmDOI{XXXXXXX.XXXXXXX}

\acmJournal{CSUR}
\acmVolume{37}
\acmNumber{4}
\acmArticle{111}
\acmMonth{2}

\AtBeginDocument{%
  \providecommand\BibTeX{{%
    Bib\TeX}}}
\setlist[itemize]{leftmargin=0.3cm,itemsep=0.1cm,topsep=0.1cm}
\begin{document}


\input{command}

\title{A Survey of Distributed Graph Algorithms on Massive Graphs}



\author{Lingkai Meng}
\affiliation{%
  \institution{Antai College of Economics and Management, Shanghai Jiao Tong University}
  \city{Shanghai}
  \country{China}}
\email{mlk123@sjtu.edu.cn}

\author{Yu Shao}
\affiliation{%
  \institution{East China Normal University}
  \city{Shanghai}
  \country{China}}
\email{yushao@stu.ecnu.edu.cn}

\author{Long Yuan}
\authornote{Corresponding author.}
\affiliation{%
  \institution{Nanjing University of Science and Technology}
  \city{Nanjing}
  \country{China}}
\email{longyuan@njust.edu.cn}

\author{Longbin Lai}
\affiliation{%
  \institution{Alibaba Group}
  \city{Hangzhou}
  \country{China}}
\email{longbin.lailb@alibaba-inc.com}

\author{Peng Cheng}
\affiliation{%
  \institution{East China Normal University}
  \city{Shanghai}
  \country{China}}
\email{pcheng@sei.ecnu.edu.cn}

\author{Xue Li}
\affiliation{%
  \institution{Alibaba Group}
  \city{Hangzhou}
  \country{China}}
\email{youli.lx@alibaba-inc.com}

\author{Wenyuan Yu}
\affiliation{%
  \institution{Alibaba Group}
  \city{Hangzhou}
  \country{China}}
\email{wenyuan.ywy@alibaba-inc.com}

\author{Wenjie Zhang}
\affiliation{%
  \institution{University of New South Wales}
  \city{Sydney}
  \country{Australia}}
\email{wenjie.zhang@unsw.edu.au}

\author{Xuemin Lin}
\affiliation{%
  \institution{Antai College of Economics and Management, Shanghai Jiao Tong University}
  \city{Shanghai}
  \country{China}}
\email{xuemin.lin@gmail.com}

\author{Jingren Zhou}
\affiliation{%
  \institution{Alibaba Group}
  \city{Hangzhou}
  \country{China}}
\email{jingren.zhou@alibaba-inc.com}

\renewcommand{\shortauthors}{Lingkai Meng, Yu Shao, and et al.}

\begin{abstract}

Distributed processing of large-scale graph data has many practical applications and has been widely studied. In recent years, a lot of distributed graph processing frameworks and algorithms have been proposed. While many efforts have been devoted to analyzing these, with most analyzing them based on programming models, less research focuses on understanding their challenges in distributed environments.
Applying graph tasks to distributed environments is not easy, often facing numerous challenges through our analysis, including parallelism, load balancing, communication overhead, and bandwidth. In this paper, we provide an extensive overview of the current state-of-the-art in this field by outlining the challenges and solutions of distributed graph algorithms.
We first conduct a systematic analysis of the inherent challenges in distributed graph processing, followed by presenting an overview of existing general solutions.
Subsequently, we survey the challenges highlighted in recent distributed graph processing papers and the strategies adopted to address them. Finally, we discuss the current research trends and identify potential future opportunities.

\end{abstract}

\ccsdesc[500]{Computing methodologies~Distributed algorithms}
\ccsdesc[500]{Theory of computation~Distributed algorithms}
\ccsdesc[500]{Theory of computation~Graph algorithms analysis}

%
\keywords{Distributed Processing, Graph Algorithms, Big Data}

\received{xxx}
\received[revised]{xxx}
\received[accepted]{xxx}

\maketitle

\section{Introduction}
\label{sec:introduction}
\input{section/sec1_Introduction}


\input{section/background}

\section{Distributed Graph Processing: Challenges and Solution Overview}
\label{sec:challenge}
\input{section/challenge}
\section{Distributed Graph Tasks}
\label{sec:algorithms}


In Section~\ref{sec:challenge}, we have explored the inherent challenges in implementing distributed graph algorithms, alongside a discussion of their respective solutions. This section builds upon that foundation, delving deeper into the distinct challenges and solutions pertinent to each algorithmic topic, as introduced in \refsec{background}.
Utilizing the graph (\url{http://gsp.blue/querying?name=graph_algo}) constructed for this survey, we can analyze how challenges are resolved within a given topic, e.g. ``\texttt{Centrality}'', using the query: \texttt{CALL get\_challenge\_papers\_hist\_of\_topic("Centrality")}.

\begin{figure}[htbp]
  
    \centering
    \begin{subfigure}[b]{0.24\textwidth}
      \includegraphics[width=0.88\textwidth]{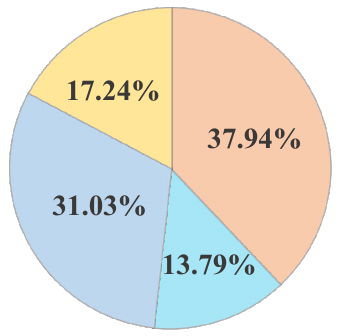}
      \caption{Centrality}
      \label{fig:1a}
    \end{subfigure}
    \begin{subfigure}[b]{0.24\textwidth}
      \includegraphics[width=0.88\textwidth]{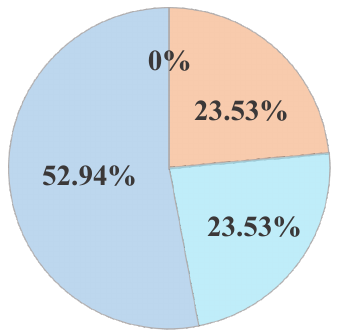}
      \caption{Community Detection}
      \label{fig:1b}
    \end{subfigure}
    \begin{subfigure}[b]{0.24\textwidth}
      \includegraphics[width=0.88\textwidth]{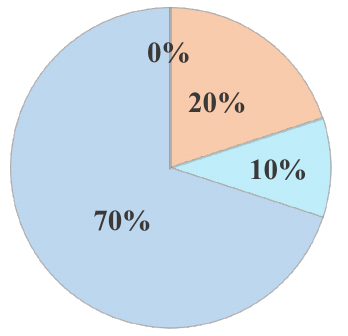}
      \caption{Similarity}
      \label{fig:1c}
    \end{subfigure}
    \begin{subfigure}[b]{0.24\textwidth}
      \includegraphics[width=0.88\textwidth]{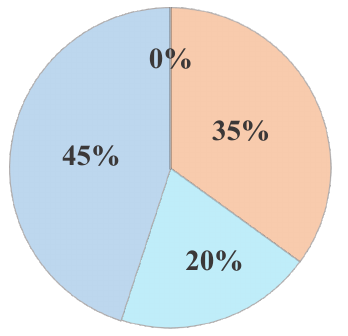}
      \caption{Cohesive Subgraph}
      \label{fig:1d}
    \end{subfigure}


    \begin{subfigure}[b]{0.24\textwidth}
      \includegraphics[width=0.88\textwidth]{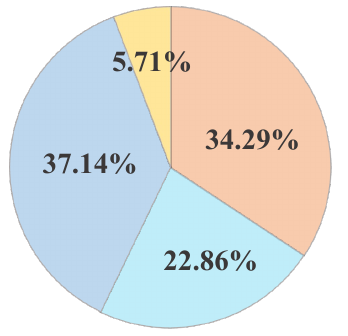}
      \caption{Traversal}
      \label{fig:2a}
    \end{subfigure}
    \begin{subfigure}[b]{0.24\textwidth}
      \includegraphics[width=0.88\textwidth]{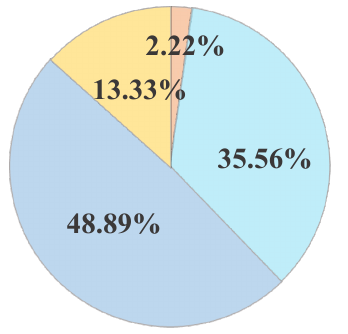}
      \caption{Pattern Matching}
      \label{fig:2b}
    \end{subfigure}
    \begin{subfigure}[b]{0.24\textwidth}
      \includegraphics[width=0.88\textwidth]{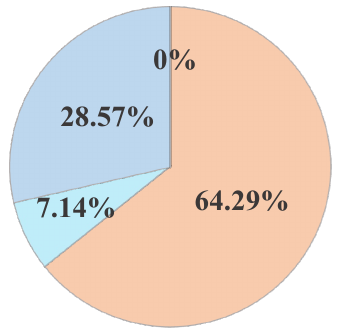}
      \caption{Covering}
      \label{fig:2c}
    \end{subfigure}
    \begin{subfigure}[b]{0.24\textwidth}
      \includegraphics[width=\textwidth]{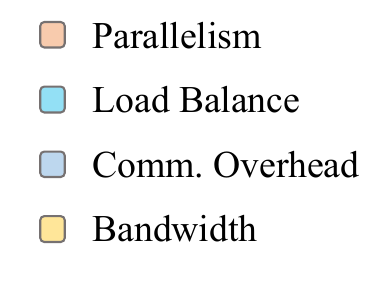}
      \label{fig:2d}
    \end{subfigure}
\caption{The distribution of challenges resolved across various topics}
\label{fig:distribution}


  \end{figure}

The aggregate results of this analysis are depicted in Figure~\ref{fig:distribution}. Our objective is to systematically examine how these challenges differ among the various topics and to critically assess the particular methodologies adopted in existing literature to tackle these challenges.

\subsection{Centrality}
\label{sec:centrality}
\input{section/sec3_Centrality}
\subsection{Community Detection}
\label{sec:community_detection}
\input{section/sec4_Clustering_or_CD}
\subsection{Similarity}
\label{sec:similarity}
\input{section/sec5_Similarity}
\subsection{Cohesive Subgraph}
\label{sec:cohesive_subgraph}

\input{section/sec6_Cohesive_Subgraph}
\subsection{Traversal}
\label{sec:traversal}
\input{section/sec7_Traversal}

\subsection{Pattern Matching}
\label{sec:pattern_matching}
\input{section/sec_Pattern_Matching}

\subsection{Covering}
\label{sec:covering}
\input{section/sec8_Covering}




\section{Application Scenarios of Various Graph Tasks}

\label{sec:application}

\input{section/applications}
\section{Discussion And Opportunities}
\label{sec:discussion_and_opportunity}
\input{section/sec10_Conclusion_and_Future_Directions}
\section{Conclusions}
\label{sec:conclusion}
\input{section/conclusion.tex}

\begin{acks}
Long Yuan is supported by NSFC 62472225. 
Xuemin Lin is supported by NSFC U2241211 and U20B2046.
\end{acks}

\bibliographystyle{ACM-Reference-Format}
\bibliography{survey}

\end{document}

%% file: command.tex
\newcommand{\centrality}{Centrality}
\newcommand{\communitydetection}{Community Detection}
\newcommand{\similarity}{Similarity}
\newcommand{\cohesivesubgraph}{Cohesive Subgraph}
\newcommand{\traversal}{Traversal}
\newcommand{\patternmatching}{Pattern Matching}
\newcommand{\covering}{Covering}
\newcommand{\connectedcomponent}{Connected Component}

\newcommand{\subgraph}{Subgraph-Centric }
\newcommand{\vertex}{Vertex-Centric }
\newcommand{\edge}{Edge-Centric }

\newcommand{\pagerank}{PageRank}
\newcommand{\personalizedpagerank}{Personalized PageRank}
\newcommand{\betweennesscentrality}{Betweenness Centrality}
\newcommand{\closenesscentrality}{Closeness Centrality}

\newcommand{\louvain}{Louvain}
\newcommand{\lpa}{Label Propagation}
\newcommand{\connectedcomponents}{Connected Components}
\newcommand{\jaccard}{Jaccard Similarity}
\newcommand{\cosine}{Cosine Similarity}
\newcommand{\simrank}{SimRank}
\newcommand{\kcore}{$k$-Core}
\newcommand{\kcoredecomposition}{$k$-Core Decomposition}
\newcommand{\ktruss}{$k$-Truss}
\newcommand{\kclique}{$k$-Clique}
\newcommand{\maximalclique}{Maximal Clique}
\newcommand{\trianglecounting}{Triangle Counting}
\newcommand{\subgraphmatching}{Subgraph Matching}
\newcommand{\subgraphmining}{Subgraph Mining}
\newcommand{\simulation}{Simulation}
\newcommand{\cycledetection}{Cycle Detection}
\newcommand{\maximumflow}{Maximum Flow}
\newcommand{\sssp}{Single Source Shortest Path}
\newcommand{\mst}{Minimum Spanning Tree}
\newcommand{\mvc}{Minimum Vertex Covering}
\newcommand{\maximummatch}{Maximum Matching}
\newcommand{\coloring}{Graph Coloring}

\newcommand{\update}[1]{\textcolor{blue}{$\Rightarrow$ UPDATE: #1}}
\newcommand{\todo}[1]{\textcolor{red}{$\Rightarrow$ TODO: #1}}
\long\def\comment#1{}
\newcommand{\stitle}[1]{\noindent\underline{\textbf{#1}}}
\newcommand{\stitlered}[1]{\noindent\textcolor{blue}{\underline{\textbf{#1}}}}

\newcommand{\push}{\textit{push} }
\newcommand{\pull}{\textit{pull} }


\newcommand{\vg}{$V$}
\newcommand{\eg}{$E$}
\newcommand{\weight}{$w$}
\newcommand{\inneighbor}{$N_{in}$}
\newcommand{\outneighbor}{$N_{out}$}
\newcommand{\neighbor}{$N$}

\newcommand{\reffig}[1]{Figure~\ref{fig:#1}}
\newcommand{\refsec}[1]{Section~\ref{sec:#1}}
\newcommand{\reftable}[1]{Table~\ref{tab:#1}}
\newcommand{\refalg}[1]{Algorithm~\ref{alg:#1}}
\newcommand{\refeq}[1]{Equation~\ref{eq:#1}}
\newcommand{\refdef}[1]{Definition~\ref{def:#1}}





\newcounter{remark}[section]
\renewcommand{\theremark}{\nthesection.\arabic{remark}}
\newenvironment{remark}{\begin{em}
        \refstepcounter{remark}
        {\vspace{1ex}\noindent\bf Remark \theremark:}}{
        \end{em}\eop\vspace{0.5ex}} 

\newcommand{\proofsketch}{\noindent{\bf Proof Sketch: }}
\newcommand{\myproof}{\noindent{\emph {Proof:} }}

\newcommand{\nthesection}{\arabic{section}}

\newcommand{\eop}{\hspace*{\fill}\mbox{$\Box$}}


\newcommand{\sstitle}[1]{\vspace{1ex} \noindent{\textit{ #1}}}
\newcommand{\ssstitle}[1]{\vspace{1ex} \noindent{\textbf{ #1}}}

\newcommand{\kw}[1]{{\ensuremath {\mathsf{#1}}}\xspace}
\newcommand{\bkw}[1]{{\ensuremath {\mathsf{\textbf{#1}}}}\xspace}

\newcommand{\kwnospace}[1]{{\ensuremath {\mathsf{#1}}}}
\newcommand{\ltt}{\kw{LTT}}
\newcommand{\arr}{\kw{arrive}}
\newcommand{\vp}{\kw{p}}

\newcommand{\bellf}{{\sc Bellman-Ford}\xspace}
\newcommand{\bfalgo}{{\sc OR}\xspace}

\newcommand{\aalgo}{{\sc KDXZ}\xspace}

\newcommand{\dijk}{{\sc Dijkstra}\xspace}
\newcommand{\dalgo}{{\sc Two-Step-LTT}\xspace}

\newcommand{\dtalgo}{{\sc DOT}\xspace}

\newcommand{\genfunc}{{\sl timeRefinement}\xspace}
\newcommand{\pathc}{{\sl pathSelection}\xspace}
\newcommand{\fifo}{{\sl FIFO}\xspace}

\newcommand{\st}{starting time\xspace}
\newcommand{\sti}{starting-time interval\xspace}
\newcommand{\stsi}{starting-time subinterval\xspace}
\newcommand{\stsis}{starting-time subintervals\xspace}
\newcommand{\stis}{starting-time intervals\xspace}
\newcommand{\ti}{time interval\xspace}
\newcommand{\tis}{time intervals\xspace}
\newcommand{\ttime}{travel time\xspace}
\newcommand{\ttimea}{travel time\xspace}
\newcommand{\at}{arrival time\xspace}
\newcommand{\ata}{arrival-time\xspace}
\newcommand{\atf}{arrival-time function\xspace}
\newcommand{\ati}{arrival-time interval\xspace}
\newcommand{\ats}{arrival times\xspace}
\newcommand{\ed}{edge delay\xspace}
\newcommand{\eda}{edge-delay\xspace}
\newcommand{\edf}{edge-delay function\xspace}
\newcommand{\eds}{edge delays\xspace}
\newcommand{\wt}{waiting time\xspace}

\newcommand{\g}{\overline{g}}
\newcommand{\iend}{\tau}

\newcommand{\argmin}{\operatornamewithlimits{argmin}}

\newcommand{\myhead}[1]{\vspace{.05in} \noindent {\bf #1.}~~}
\newcommand{\cond}[1]{(\emph{#1})~}
\newcommand{\op}[1]{(\emph{#1})~}

\newcommand{\qc}{\ensuremath{Q^c}}

\newcommand{\rewrite}{\kw{XPathToReg}}

\newcommand{\upparen}[1]{\ensuremath{\mathrm{(}}{#1}\ensuremath{\mathrm{)}}}
\newcommand{\func}[2]{\funcname{#1}\upparen{\ensuremath{#2}}}
\newcommand{\funcname}[1]{\ensuremath{\mathit{#1}}}

\newcommand\AS{\textbf{as}\ }
\newcommand{\xsltsize}{\small}

\newcommand{\X}{{\cal X}}
\newcommand{\sem}[1]{[\![#1]\!]}
\newcommand{\NN}[2]{#1\sem{#2}}
\newcommand{\pcdata}{{\tt str}\xspace}

\newcommand{\exa}[2]{{\tt\begin{tabbing}\hspace{#1}\=\hspace{#1}\=\+\kill #2\end{tabbing}}}
\newcommand{\ra}{\rightarrow}
\newcommand{\la}{\leftarrow}
\newcommand{\rsa}{\_} 
\newcommand{\Ed}[2]{E_{{\scriptsize \mbox{#1} \rsa \mbox{#2}}}}
\newenvironment{bi}{\begin{itemize}
        \setlength{\topsep}{0.5ex}\setlength{\itemsep}{0ex}\vspace{-0.6ex}}
        {\end{itemize}\vspace{-1ex}}
\newenvironment{be}{\begin{enumerate}
        \setlength{\topsep}{0.5ex}\setlength{\itemsep}{0ex}\vspace{-0.6ex}}
        {\end{itemize}\vspace{-1ex}}
\newcommand{\ei}{\end{itemize}}
\newcommand{\ee}{\end{enumerate}}

\newcommand{\mat}[2]{{\begin{tabbing}\hspace{#1}\=\+\kill #2\end{tabbing}}}
\newcommand{\m}{\hspace{0.05in}}
\newcommand{\ls}{\hspace{0.1in}}
\newcommand{\beqn}{\begin{eqnarray*}}
\newcommand{\eeqn}{\end{eqnarray*}}

\newcounter{ccc}
\newcommand{\bcc}{\setcounter{ccc}{1}\theccc.}
\newcommand{\icc}{\addtocounter{ccc}{1}\theccc.}

\newcommand{\oneurl}[1]{\texttt{#1}}
\newcommand{\tabstrut}{\rule{0pt}{4pt}\vspace{-0.1in}}
\newcommand{\tabstruct}{\rule{0pt}{8pt}\\[-2ex]}
\newcommand{\stab}{\rule{0pt}{8pt}\\[-2.2ex]}
\newcommand{\sstab}{\rule{0pt}{8pt}\\[-2.2ex]}

\newcommand{\eat}[1]{}

\newfloat{tcm}{thp}{loa}
\floatname{tcm}{Recursive \sql}

\def\subfigcapskip{2pt}


\newcommand{\rdms}{{\sc rdbms}\xspace}
\newcommand{\sql}{{\sc sql}\xspace}
\newcommand{\dbms}{{\sc dbms}\xspace}

\newcommand{\cfig}{Fig.~}
\newcommand{\ctab}{Table~}
\newcommand{\csec}{Section~}
\newcommand{\cdef}{Definition~}
\newcommand{\cthm}{Theorem~}
\newcommand{\clem}{Lemma~}
\newcommand{\cequ}[1]{Equation~(#1)}
\newcommand{\SG}{\mathbf{SG}}
\newcommand{\SA}{\mathbf{SA}}
\renewcommand{\AA}{\mathbf{AA}}

\newcommand{\revise}[1]{\textcolor{blue}{#1}}

\newcommand{\xml}{{\sl XML}\xspace}
\newcommand{\xlink}{{\sl XLink}\xspace}
\newcommand{\xpath}{{\sl XPath}\xspace}
\newcommand{\xpointer}{{\sl XPointer}\xspace}
\newcommand{\rdf}{{\sl RDF}\xspace}
\newcommand{\tc}{{\sl TC}\xspace}
\newcommand{\dfs}{{\sl DFS}\xspace}
\newcommand{\DAG}{{\sl DAG}\xspace}
\newcommand{\DAGs}{{\sl DAG}s\xspace}
\newcommand{\grail}{{\sl GRAIL}\xspace}
\newcommand{\yesgrail}{{\sl Yes-GRAIL}\xspace}
\newcommand{\code}{\kw{code}}
\newcommand{\sit}{\kw{sit}}
\newcommand{\psit}{{\cal P}_{sit}}
\newcommand{\yescode}{{\sl Yes-Label}\xspace}
\newcommand{\nocode}{{\sl No-Label}\xspace}
\newcommand{\entry}{\kw{entry}\xspace}
\newcommand{\yngindex}{{\sl YNG-Index}\xspace}
\newcommand{\rqrun}{{\sl RQ-Run}\xspace}
\newcommand{\citeseerx}{{\sl citeseerx}\xspace}
\newcommand{\gouniprot}{{\sl go-uniprot}\xspace}
\newcommand{\uniprot}{{\sl uniprot150}\xspace}

\long\def\comment#1{}

\newcommand{\scc}{strongly connected component\xspace}
\newcommand{\sccs}{strongly connected components\xspace}
\newcommand{\sscc}{\kw{SCC}}
\newcommand{\ssccs}{\kwnospace{SCC}s\xspace}
\newcommand{\sccg}{\kwnospace{SCC}\textrm{-}\kw{Graph}}
\newcommand{\strongc}{\leftrightarrow}
\newcommand{\nstrongc}{\nleftrightarrow}
\newcommand{\emscc}{\kwnospace{EM}\textrm{-}\kw{SCC}}
\newcommand{\dfsscc}{\kwnospace{DFS}\textrm{-}\kw{SCC}}
\newcommand{\dfstree}{\kwnospace{DFS}\textrm{-}\kw{Tree}}
\newcommand{\len}{\kw{len}}
\newcommand{\dep}{\kw{depth}}
\newcommand{\tdep}{\kw{drank}}
\newcommand{\tlink}{\kw{dlink}}
\newcommand{\vedges}{up-edges\xspace}
\newcommand{\vedge}{up-edge\xspace}
\newcommand{\cvedge}{Up-Edge\xspace}

\newcommand{\drsscc}{\kwnospace{1P}\textrm{-}\kw{SCC}}
\newcommand{\drssccb}{\kwnospace{1PB}\textrm{-}\kw{SCC}}

\newcommand{\Bdrsscc}{\kwnospace{B}\textrm{-}\kwnospace{BR'}\textrm{-}\kw{SCC}}

\newcommand{\deprtree}{depth-ranked tree\xspace}
\newcommand{\cdeprtree}{Depth-Ranked Tree\xspace}
\newcommand{\drtree}{\kwnospace{BR}\textrm{-}\kw{Tree}}
\newcommand{\drplustree}{\kwnospace{BR}$^+$\textrm{-}\kw{Tree}}
\newcommand{\drscc}{\kwnospace{2P}\textrm{-}\kw{SCC}}
\newcommand{\updatedrank}{\kwnospace{update}\textrm{-}\kw{drank}}

\newcommand{\drtreeconstruct}{\kwnospace{Tree}\textrm{-}\kw{Construction}}
\newcommand{\drtreesearch}{\kwnospace{Tree}\textrm{-}\kw{Search}}
\newcommand{\depthrerank}{\kw{pushdown}}
\newcommand{\itrerank}{\kwnospace{iterative}\textrm{-}\kw{rerank}}
\newcommand{\drr}{\Downarrow}
\newcommand{\reach}{\kw{Rset}}
\newcommand{\earlyrejection}{\kwnospace{early}\textrm{-}\kw{rejection}}
\newcommand{\earlyacceptance}{\kwnospace{early}\textrm{-}\kw{acceptance}}
\newcommand{\greduce}{\earlyacceptance}
\newcommand{\drea}{\kwnospace{1P}\textrm{/}\kw{ER}}
\newcommand{\myinf}{\kw{INF}}

\newcommand{\gcloud}{\kw{GCloud}\xspace}
\newcommand{\degree}{\kw{Degree}\xspace}
\newcommand{\bfs}{\kw{BFS}\xspace}
\newcommand{\keysearch}{\kw{KWS}\xspace}
\newcommand{\cc}{\kw{CC}\xspace}
\newcommand{\msf}{\kw{MSF}\xspace}
\newcommand{\dmax}{\kw{rmax}}
\newcommand{\mystar}{\kw{star}\xspace}
\newcommand{\twitter}{{\sl{Twitter-2010}}\xspace}
\newcommand{\friendster}{{\sl{Friendster}}\xspace}
\definecolor{lgray}{gray}{0.85}
\definecolor{llgray}{gray}{0.9}
\newcommand{\mycc}{CC\xspace}
\newcommand{\myccs}{CCs\xspace}
\newcommand{\mymsf}{MSF\xspace}
\newcommand{\oneroundmsf}{\kw{OneRoundMSF}}
\newcommand{\multiroundmsf}{\kw{MultiRoundMSF}}

\newcommand{\hashtomin}{\kw{HashToMin}\xspace}
\newcommand{\hashgreatertomin}{\kw{HashGToMin}\xspace}
\newcommand{\pramsimulation}{\kwnospace{PRAM}\textrm{-}\kw{Simulation}\xspace}

\newcommand{\pagerankpig}{\kwnospace{PageRank}\textrm{-}\kw{Pig}\xspace}
\newcommand{\bfspig}{\kwnospace{BFS}\textrm{-}\kw{Pig}\xspace}
\newcommand{\keysearchpig}{\kwnospace{KWS}\textrm{-}\kw{Pig}\xspace}

\newcommand{\ttwig}{\kwnospace{TwinTwig}\xspace}
\newcommand{\ttwigs}{\kwnospace{TwinTwig}s\xspace}
\newcommand{\ttjoin}{\kwnospace{TwinTwig}\kw{Join}}
\newcommand{\sdec}{\kwnospace{SDEC}\xspace}
\newcommand{\subgenum}{\kwnospace{SubgraphEnum}\xspace}
\newcommand{\mymap}{\kwnospace{map}\xspace}
\newcommand{\myreduce}{\kwnospace{reduce}\xspace}
\newcommand{\cascadejoin}{\kwnospace{Edge}\kw{Join}}
\newcommand{\starjoin}{\kwnospace{Star}\kw{Join}}
\newcommand{\multiwayjoin}{\kwnospace{Multiway}\kw{Join}}
\newcommand{\cost}{\kwnospace{cost}\xspace}
\newcommand{\mysize}{\kwnospace{card}\xspace}
\newcommand{\er}{\kwnospace{ER}\xspace}
\newcommand{\optdec}{\kwnospace{Optimal}\textrm{-}\kwnospace{Decomp}\xspace}

\newcommand{\ttone}{\kwnospace{TT1}\xspace}
\newcommand{\tttwo}{\kwnospace{TT2}\xspace}
\newcommand{\ttthree}{\kwnospace{TT3}\xspace}

\newcommand{\alEdge}{\kwnospace{Edge}\xspace}
\newcommand{\alMul}{\kwnospace{Mul}\xspace}
\newcommand{\alStar}{\kwnospace{Star}\xspace}

\newcommand{\alTTBO}{\kwnospace{TTBS}\xspace}
\newcommand{\alTTNLB}{\kwnospace{TTOA}\xspace}
\newcommand{\alTTLB}{\kwnospace{TTLB}\xspace}
\newcommand{\alTTFil}{\kwnospace{TT}\xspace}

\newcommand{\refthm}[1]{Theorem~\ref{thm:#1}}
\newcommand{\reflem}[1]{Lemma~\ref{lem:#1}}
\newcommand{\refex}[1]{Example~\ref{ex:#1}}

\makeatletter
\newcommand{\rmnum}[1]{\romannumeral #1}
\newcommand{\Rmnum}[1]{\expandafter\@slowromancap\romannumeral #1@}
\makeatother

\newcommand{\enumall}{\kwnospace{Enum}\kw{All}}
\newcommand{\enumsub}{\kwnospace{Enum}\kw{Sub}}
\newcommand{\cliqueall}{\kwnospace{Clique}\kw{All}}
\newcommand{\cliquesub}{\kwnospace{Clique}\kw{Sub}}
\newcommand{\maxcover}{\kwnospace{Max}\kw{Cover}}
\newcommand{\enumk}{\kwnospace{Enum}\kw{K}}
\newcommand{\priv}{\kw{priv}}
\newcommand{\cov}{\kw{cov}}
\newcommand{\enumkbasic}{\kwnospace{Enum}\kwnospace{K}\kw{Basic}}
\newcommand{\enumkopt}{\kwnospace{Enum}\kwnospace{K}\kw{Opt}}
\newcommand{\candmaintainbasic}{\kwnospace{Cand}\kwnospace{Maintain}\kw{Basic}}
\newcommand{\candmaintain}{\kwnospace{Cand}\kw{Maintain}}
\newcommand{\pnpindex}{\kwnospace{PNP}\textrm{-}\kw{Index}}
\newcommand{\rcov}{\kw{rcov}}
\newcommand{\rpriv}{\kw{rpriv}}
\newcommand{\insertc}{\kw{Insert}}
\newcommand{\deletec}{\kw{Delete}}
\newcommand{\calp}{\kwnospace{Cal}\kw{P}}

\newcommand{\myscore}{\kw{score}}
\newcommand{\initk}{\kw{InitK}}
\newcommand{\globalpruning}{\kw{GlobalPruning}}
\newcommand{\localpruning}{\kw{LocalPruning}}
\newcommand{\cliquek}{\kw{CliqueK}}
\newcommand{\mycolor}{\kw{color}}
\newcommand{\mycore}{\kw{core}}
\newcommand{\cliquegreedy}{\kw{CliqueGreedy}}

\newcommand{\sops}{\kw{SOPS}}
\newcommand{\gops}{\kw{GOPS}}
\newcommand{\sieve}{\kw{SIEVE}}
\newcommand{\enumkglobal}{\kw{Global}}
\newcommand{\enumklocal}{\kw{Local}}

\newcommand{\dsgoogle}{\textit{Google}\xspace}
\newcommand{\dsskitter}{\textit{Skitter}\xspace}
\newcommand{\dsyoutube}{\textit{Youtube}\xspace}
\newcommand{\dsundirecteda}{Different $\epsilon$ ($\mu$ = 2)\xspace}
\newcommand{\dsundirectedb}{Different $\epsilon$ ($\mu$ = 3)\xspace}
\newcommand{\dsundirectedc}{Different $\epsilon$ ($\mu$ = 4)\xspace}
\newcommand{\dsundirectedd}{Different $\mu$ ($\epsilon$ = 0.6)\xspace}

\newcommand{\dsdirecteda}{Different $\epsilon_f$ ($\mu$=2, $\epsilon_c$=0.45)\xspace}
\newcommand{\dsdirectedb}{Different $\epsilon_c$ ($\mu$=2, $\epsilon_f$=0.45)\xspace}
\newcommand{\dsdirectedc}{Different $\epsilon_f$ ($\mu$=3, $\epsilon_c$=0.45)\xspace}
\newcommand{\dsdirectedd}{Different $\epsilon_c$ ($\mu$=3, $\epsilon_f$=0.45)\xspace}
\newcommand{\dsdirectede}{Different $\epsilon_c$ and $\epsilon_f$($\mu$ = 2)\xspace}
\newcommand{\dsdirectedf}{Different $\mu$ ($\epsilon_c$=$\epsilon_f$=0.45)\xspace}

\newcommand{\dswebgoogle}{Google\xspace}
\newcommand{\dsLiveJournal}{LiveJournal\xspace}
\newcommand{\dswebbase}{Webbase\xspace}
\newcommand{\dsuk}{UK-2002\xspace}
\newcommand{\dsukt}{UK-2005\xspace}
\newcommand{\dsemail}{email\xspace}
\newcommand{\dswiki}{Wiki\xspace}
\newcommand{\dspokec}{Pokec\xspace}
\newcommand{\dsenwiki}{EnWiki-2020\xspace}
\newcommand{\dsfriendster}{Friendster\xspace}

\newcommand{\dscaseaa}{$\epsilon_c$ = 0.2, $\epsilon_f$ = 0.3, $\mu$ = 2\xspace}
\newcommand{\dscaseab}{$\epsilon$ = 0.55, $\mu$ = 2\xspace}

\newcommand{\dscaseba}{$\epsilon_c$ = 0.4, $\epsilon_f$ = 0.2, $\mu$ = 2\xspace}
\newcommand{\dscasebb}{$\epsilon$ = 0.52, $\mu$ = 2\xspace}

\newcommand{\dscaseca}{$\epsilon_c$ = 0.16, $\epsilon_f$ = 0.18, $\mu$ = 2\xspace}
\newcommand{\dscasecb}{$\epsilon$ = 0.51, $\mu$ = 2\xspace}

\newcommand{\dsuntri}{Undirected\xspace}
\newcommand{\dscycletri}{Cycle\xspace}
\newcommand{\dsflowtri}{Flow\xspace}

\newcommand{\nbr}{\kw{nbr}\xspace}
\newcommand{\nbrin}{\kw{nbr^-}}
\newcommand{\nbrout}{\kw{nbr^+}}
\newcommand{\mydeg}{\kw{deg}}
\newcommand{\mydegin}{\kw{deg^-}}
\newcommand{\mydegout}{\kw{deg^+}}
\newcommand{\scan}{\kw{SCAN}}
\newcommand{\dscan}{\kw{DSCAN}}
\newcommand{\clusteringq}{$Q_{\epsilon_c, \epsilon_f, \mu}$\xspace}
\newcommand{\dscanindex}{\kwnospace{DSCAN}\textrm{-}\kw{Index}}
\newcommand{\dscanquerynaive}{\kwnospace{DSCAN}\textrm{-}\kwnospace{NaiveIndex}\textrm{-}\kw{Query}}
\newcommand{\dscanquery}{\kwnospace{DSCAN}\textrm{-}\kwnospace{Index}\textrm{-}\kw{Query}}
\newcommand{\dscanquerynospace}{\kwnospace{DSCAN}\textrm{-}\kwnospace{Query}}
\newcommand{\dscanindexcons}{\kwnospace{DSCAN}\textrm{-}\kwnospace{Index}\textrm{-}\kw{Cons^+}}
\newcommand{\dscanindexconsnospace}{\kwnospace{DSCAN}\textrm{-}\kwnospace{Index}\textrm{-}\kwnospace{Cons^+}}
\newcommand{\dscanindexconsp}{\kwnospace{DSCAN}\textrm{-}\kwnospace{Index}\textrm{-}\kw{Cons}}
\newcommand{\dscanindexconsnaive}{\kwnospace{DSCAN}\textrm{-}\kwnospace{Index}\textrm{-}\kw{ConsNaive}}

\newcommand{\trianglec}{\kw{CountTriangles}}
\newcommand{\trianglecnospace}{\kwnospace{CountTriangles}}
\newcommand{\trianglecp}{\kw{CountTriangles^+}}
\newcommand{\trianglecpnospace}{\kwnospace{CountTriangles^+}}

\newcommand{\dscanindexins}{\kwnospace{DSCAN}\textrm{-}\kwnospace{Index}\textrm{-}\kw{Ins^+}}
\newcommand{\dscanindexinsnospace}{\kwnospace{DSCAN}\textrm{-}\kwnospace{Index}\textrm{-}\kwnospace{Ins^+}}
\newcommand{\dscanindexinsp}{\kwnospace{DSCAN}\textrm{-}\kwnospace{Index}\textrm{-}\kw{Ins}}
\newcommand{\dscanindexdel}{\kwnospace{DSCAN}\textrm{-}\kwnospace{Index}\textrm{-}\kw{Del^+}}
\newcommand{\dscanindexdelnospace}{\kwnospace{DSCAN}\textrm{-}\kwnospace{Index}\textrm{-}\kwnospace{Del^+}}
\newcommand{\dscanindexdelp}{\kwnospace{DSCAN}\textrm{-}\kwnospace{Index}\textrm{-}\kw{Del}}
\newcommand{\pdscan}{\kw{pDSCAN}}
\newcommand{\dscannaive}{\kwnospace{DSCAN}\textrm{-}\kw{Naive}}

%% file: section/sec1_Introduction.tex


Graph is a high-dimensional structure that models point-to-point relationships among entities. Due to its strong representation ability, the graph is widely applied in social network analysis~\cite{DBLP:journals/cn/BroderKMRRSTW00}, road network routing~\cite{DBLP:conf/soda/GoldbergH05}, and biological structure forecasting~\cite{DBLP:conf/ismb/BorgwardtOSVSK05}. With the development of information science and big data applications~\cite{DBLP:journals/tvcg/AbelloHK06, DBLP:journals/tcad/DaiHCZSL00Y19} in recent years, the scales of graph data sets become too large for single machines due to their limited storage and computing power. To support the queries and analyses on massive graphs, researchers proposed many distributed graph algorithms and systems to store a large-scale graph in multiple machines separately and compute collaboratively, such as Pregel~\cite{DBLP:conf/sigmod/MalewiczABDHLC10}, Giraph~\cite{avery2011giraph}, GraphX~\cite{DBLP:conf/osdi/GonzalezXDCFS14}, and GraphScope~\cite{DBLP:journals/pvldb/FanHLLLLQ0WXYYY21}.

Recent years have seen a surge in research on distributed graph algorithms, with a focus on developing specific algorithms like PageRank, label propagation, and triangle counting, or addressing challenges such as workload scheduling and machine-to-machine communication. However, comprehensive surveys providing a comprehensive view of the field remain limited.
This paper aims to bridge that gap by consolidating research on distributed graph algorithms for massive graphs over the past decade, as featured in prestigious conferences and journals like SIGMOD, VLDB, PPoPP, SC, TPDS, and TC. We distill four primary and recurrently addressed challenges among these papers:

\begin{itemize}

    \item  \textbf{Parallelism} is a major objective that requires processing multiple operations at the same time and reducing iteration rounds. Achieving efficient parallelism is challenging due to the inherent sequential dependencies in graph analysis tasks. These dependencies require certain computations to be completed before others can begin, limiting the ability to execute multiple operations simultaneously.
 
    \item \textbf{Load balance} aims to distribute work evenly across vertices and improve the utilization of computing resources. This helps prevent some machines from becoming overloaded while others remain idle. Graphs often have skewed degree distributions, where a small number of vertices have many connections while the majority have few. This imbalance can lead to uneven distribution of workload across machines.

    \item \textbf{Communication} refers to the exchange of messages between vertices, which is an expensive operation compared to random memory access. Distributed graph processing involves frequent exchanges of data between machines. These communications can become a significant overhead, especially in large-scale graphs, due to the complex and dense interconnections between vertices. Optimizing communication overhead can improve execution efficiency in practice.

    \item \textbf{Bandwidth} limits the size of messages transmitted between vertices. Some algorithms require a large amount of bandwidth, which may not be feasible in certain frameworks.
    The primary reasons for the bandwidth limitations challenge in distributed graph algorithms are twofold: first, high-degree vertices necessitate substantial bandwidth for communication; second, the frequent transmission of small messages can result in inefficient bandwidth utilization.
\end{itemize}

To address the challenges, a lot of open-source distributed graph processing frameworks~(e.g., Pregel~\cite{DBLP:conf/sigmod/MalewiczABDHLC10} and GPS~\cite{DBLP:conf/ssdbm/SalihogluW13}) have been proposed. Generic solutions (e.g., parallel looping, message receiving and sending, and broadcasting) are abstracted in these frameworks. Users have the flexibility to develop graph algorithms following high-level functionalities, effectively abstracting away the complexities of underlying implementation details. However, these solutions are highly diverse and tailored for specific algorithms due to the irregularity of graph algorithms, and there is no one unified pattern that can fit all graph algorithms.

In addition, various graph tasks are addressed by distributed graph algorithms in the existing studies. To clearly introduce them, we classify the widely studied graph tasks, including PageRank, Louvain, SimRank, max flow, subgraph matching, and vertex cover, into seven topics: \textit{Centrality}, \textit{Community Detection}, \textit{Similarity}, \textit{Cohesive Subgraph}, \textit{Traversal}, \textit{Pattern Matching}, and \textit{Covering}.

\begin{figure}[h]
    \centering
    \includegraphics[width=0.92\textwidth]{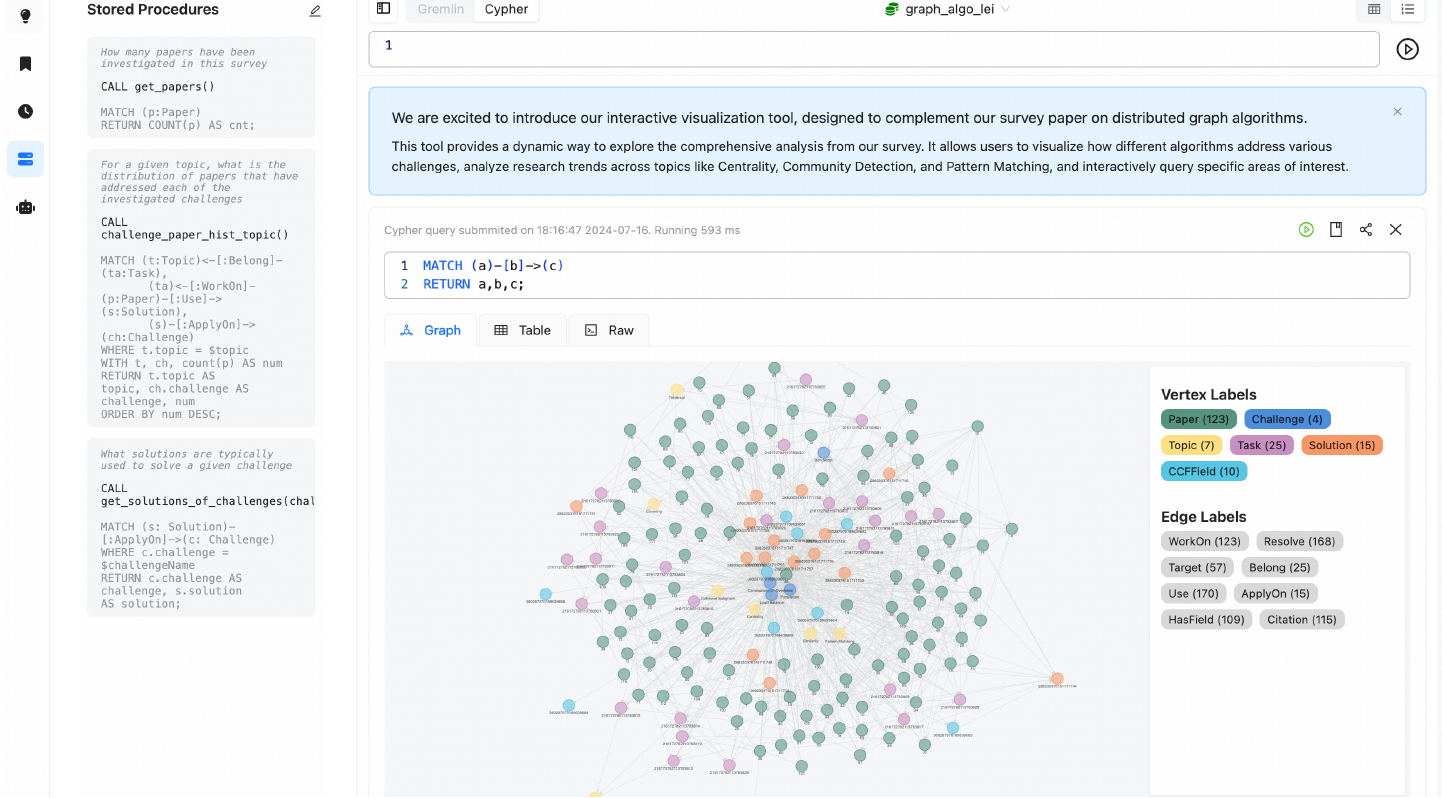}
    \caption{A comprehensive graph that encapsulates the surveyed material.}
    \label{fig:survey_graph}
\end{figure}

In this paper, we first introduce the generic solutions for the four challenges and then
dissect the proportion of research papers that address the challenges across different algorithmic topics. Moreover, we delve into the reasons behind the varying degrees of attention that certain challenges receive within specific topics. For example, 70\% of the papers related to \textit{Similarity} topic (\reffig{1c}) have focused on reducing communication overhead. Through the analysis, we show some deep views of the research in distributed graph algorithms and also give insights into the potentially promising directions for future studies. A unique contribution of this work is the construction of a comprehensive graph that encapsulates the surveyed material, as shown in \reffig{survey_graph}. This graph maps out the intricate connections among papers, topics, algorithms, solutions, and challenges, etc., providing a visual narrative of the landscape. Readers can play with the graph using an online interactive tool (\url{http://gsp.blue/querying?name=graph_algo}), which includes several examples to guide readers on how to use the tool.

\noindent{\underline{\textbf{Contribution}}.} Existing surveys focus on either specific distributed challenges~(e.g., load balance~\cite{DBLP:journals/tpds/Jiang16}) or particular distributed algorithms~(e.g., pattern matching~\cite{DBLP:journals/csur/BouhenniYNK21}). However, our survey targets challenges faced by different distributed graph algorithms in processing, with irregular computation taken into consideration.
Specifically, we make the main contributions as follows:

\begin{itemize}
    \item We provide an overview of the main challenges in distributed graph algorithms and the different solutions proposed to tackle them. This provides a comprehensive understanding of distributed graph processing.

    \item We survey various distributed graph algorithms and categorize them into seven topics based on the challenges they address.

    \item For each topic of distributed graph algorithms, we conduct a thorough analysis of existing efforts. We also summarize the main challenges they address and provide unique insights into the underlying reasons.

\end{itemize}

\begin{figure}[h]
    \centering
    \includegraphics[width=0.99\textwidth]{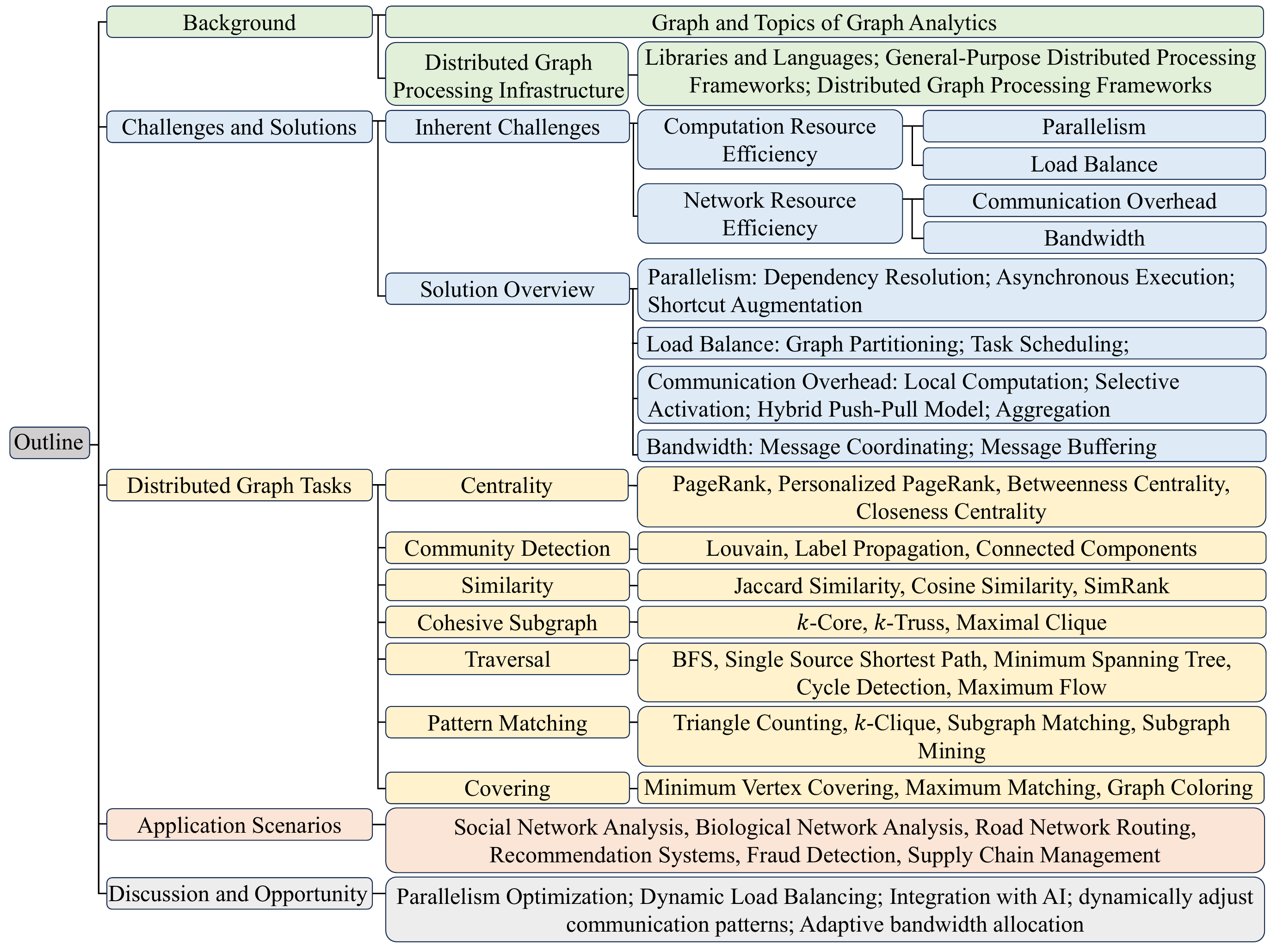}
    \caption{Outline of the Survey}
    \label{fig:outline}
\end{figure}

The outline of our survey is shown in Figure~\ref{fig:outline} and the rest of this paper is organized as follows. Section~\ref{sec:background} provides a review of existing distributed graph systems and computing processing. Section~\ref{sec:challenge} summarizes some challenges and solutions, which are not common for single-machine algorithms. Section~\ref{sec:algorithms} describes popular distributed graph algorithms in detail, highlighting differences from their single-machine versions.
Section~\ref{sec:application} presents various application scenarios for different graph tasks.
Section~\ref{sec:discussion_and_opportunity} discusses prevalent research trends and potential research opportunities. Section~\ref{sec:conclusion} concludes this survey.


%% file: section/background.tex
\section{Background}
\label{sec:background}


\subsection{Graph and Topics of Graph Analytics}

We first define the graph and other relevant concepts that are used throughout this paper.

\begin{definition}[Graph]
    A graph $G$ comprises a set $V(G)$ of $n$ vertices and a set $E(G)$ of $m$ edges. Vertices represent entities, and edges signify the relationships between entities. This paper primarily focuses on simple graphs, namely having at most one edge between any pair of vertices. Graphs are categorized into directed and undirected graphs.
    In an undirected graph, an edge~$(u, v)$ denotes an unordered pair of vertices, implying that $(u, v) = (v, u)$. Conversely, in a directed graph, an edge $(u, v)$ represents an ordered pair of vertices, where $(u, v) \neq (v, u)$. Additionally, when edges are assigned a float number based on edge attributes, the graph is termed a weighted graph.
\end{definition}
\begin{definition}[Density]
    The density of a graph is defined as the ratio of the edge number $m$ and the maximum possible edge number~($n^2$ in directed graphs and $n(n-1)/2$ in undirected graphs). Generally, a graph is sparse if $m \ll O(n^2)$ and dense otherwise.
\end{definition}
\begin{definition}[Neighbors and Degree]
    In a directed graph $G$, for any given vertex $u \in V(G)$, we define the in-neighbor and out-neighbor set of $u$ as $N_{in}(u) = \{v | (v, u) \in E(G)\}$ and $N_{out}(u) = \{v | (u, v) \in E(G)\}$. The in-degree and out-degree of vertex $u$ are the sizes of $N_{in}(u)$ and $N_{out}(u)$, respectively. In the context of an undirected graph $G$, the neighbor set for a vertex $u \in V(G)$ is defined as $N(u) = \{v | (u, v) \in E(G)\}$, and its degree is the size of the neighbor set. 
\end{definition}
\begin{definition}[Path]
    A path in a graph is a sequence of vertices where each pair of consecutive vertices is connected by an edge. Specifically, a path $P$ can be defined as $P = (v_1, v_2, \ldots, v_k)$, where each pair $(v_i, v_{i+1}) \in E(G)$ for all $1 \leq i < k$. The length of the path is determined by the number of vertices it contains, which in this case is $k$.
    In a weighted graph, where edges carry weights, the distance is defined as the minimum sum of the weights of all edges $(v_i, v_{i+1})$ along all paths. Consequently, the shortest path length and distance from vertex $u$ to vertex $v$ are the minimum length and the minimum sum of weights, respectively, among all possible paths connecting $u$ and $v$.
\end{definition}
\begin{definition}[Diameter]
The diameter of a graph is defined as the longest shortest path \emph{length} between any pair of vertices. In other words, it represents the largest length that must be traversed to connect any two vertices in the graph using the shortest path between them.
\end{definition}
\begin{definition}[Cycle]
    A cycle is a special case of a path that starts and ends at the same vertex, i.e., $C = (v_1, v_2, \ldots, v_k)$ with $v_1 = v_k$. The length of a cycle is the number of vertices it contains.
\end{definition}
\begin{definition}[Tree]
    A tree, commonly defined within the scope of an undirected graph, is a connected graph that is free of cycles.
    A forest is a \emph{disjoint} set of trees.
\end{definition}
\begin{definition}[Subgraph]
    A subgraph $H$ of a graph $G$ is another graph where $V(H) \subset V(G)$, $E(H) \subset E(G)$, and for each $(u, v) \in E(H)$, $u, v \in V(H)$.
\end{definition}

Due to the diverse application scenarios of graph analytics, a broad range of topics within this field are explored in the literature. In this paper, we concentrate on specific categories of graph analytics that are frequently examined in the context of distributed graph algorithms. A similar taxonomy can be found in other works \cite{DBLP:conf/tpctc/Dominguez-SalMMBL10, DBLP:journals/csur/HeidariSCB18,DBLP:conf/icde/LiM0LYQ0Z23}. These categories are: (1) \textbf{Centrality}: Focus on determining the relative significance of each vertex within a complex graph. (2) \textbf{Community Detection}: Aim at segregating the vertices of a graph into distinct communities, characterized by denser connections within the same community compared to those with vertices outside the community. (3) \textbf{Similarity}: Concern with assessing the likeness between two vertices in a graph, based on defined criteria.
 (4) \textbf{Cohesive Subgraph}: Involve identifying subgraphs that meet specific cohesion criteria. (5) \textbf{Traversal}: Entail visiting vertices in a graph following a certain procedure, to review or modify the vertices' attributes. (6) \textbf{Pattern Matching}: Search for specific input patterns within a graph.
 (7) \textbf{Covering}: Address combinatorial optimization problems in graphs, often linked to minimum vertex-covering challenges. 
These categories represent key areas in graph analytics, each with distinct methodologies and applications in distributed graph algorithms, as will be discussed in detail in~\refsec{algorithms}.

\subsection{Distributed Graph Processing Infrastructure}

Processing a large volume of data on a cluster of machines has been studied for decades in computer science. Many tools and systems have been developed to facilitate distributed computing, which can be utilized for distributed graph processing as well. We divide them into three categories based on their closeness to distributed graph processing as follows:

\subsubsection{Distributed Computing Libraries and Languages} To support efficient distributed algorithm and system development, distributed computing libraries based on traditional sequential languages are designed, such as MPI \cite{forum1994mpi}, Java Remote Method Invocation \cite{RN16}, PyCOMPSs \cite{DBLP:journals/cse/BadiaCELL22}, and OpenSHMEM \cite{DBLP:conf/pgas/ChapmanCPPKKS10}. Among them, MPI is one of the earliest libraries designed for distributed computing. Strictly, MPI is a library standard specification for message passing. It defines a model of distributed computing where each process has its own local memory, and data is explicitly shared by passing messages between processes. MPI  includes point-to-point and collective communication routines, as well as support for process groups, communication contexts, and application topologies \cite{the1993mpi}. MPI has now emerged as the de facto standard for message passing on computer clusters \cite{DBLP:conf/hipc/Gropp01}.

Besides the distributed computing libraries, programming languages specifically for implementing distributed applications are also created  \cite{DBLP:journals/csur/BalST89}. Unlike  sequential programming languages, distributed programming languages aim to address problems such as concurrency, communication, and synchronization in distributed computing from the language level to simplify distributed programming. They generally provide  built-in  primitives for task distribution, communication, and fault-tolerance. \mbox{Representative languages included  X10~\cite{DBLP:conf/oopsla/CharlesGSDKEPS05}, Haskell \cite{DBLP:journals/jfp/TrinderLP02},  JoCaml \cite{fournet2002jocaml}, and Erlang \cite{armstrong2010erlang}.}

\subsubsection{General-Purpose Distributed Processing Frameworks} Although distributed computing libraries and languages enable efficient distributed computing, the applications still put lots of effort into the details of running a distributed program, such as the issues of parallelizing the computation, distributing the data, and handling failures. To address this complexity, general-purpose distributed processing frameworks are designed.

MapReduce \cite{DBLP:conf/osdi/DeanG04} is a simple and powerful framework introduced by Google for scalable distributed computing, simplifying the development of applications that process vast amounts of data. It takes away the complexity of distributed programming by exposing two processing steps that programmers need to implement: Map and  Reduce. In the Map step, data is broken down into key-value pairs and then distributed to different computers for processing, and in the Reduce step, all computed results are combined and summarized based on key. MapReduce and its variants~\cite{white2012hadoop} reduce the complexity of developing distributed applications. However, MapReduce reads and writes intermediate data to disks, which incurs a significant performance penalty. To address the deficiency of MapReduce,  Spark  \cite{DBLP:conf/hotcloud/ZahariaCFSS10} shares a similar programming model to MapReduce but extends it with a data-sharing abstraction  called resilient distributed dataset (RDD) \cite{DBLP:conf/nsdi/ZahariaCDDMMFSS12}, which can be explicitly cached in memory  and reused  in multiple MapReduce-like parallel operations. Due to the introduction of RDD, Spark achieves significant performance improvement compared to MapReduce \cite{DBLP:journals/cacm/ZahariaXWDADMRV16}. In addition to MapReduce and Spark, other general-purpose distributed processing frameworks are also designed, such as Storm \cite{toshniwal2014storm},  Flink \cite{DBLP:journals/debu/CarboneKEMHT15}, and Ray \cite{DBLP:conf/osdi/MoritzNWTLLEYPJ18}. \cite{DBLP:journals/csur/SakrLF13,DBLP:journals/sp/CzarnulPD20,DBLP:journals/comsur/ShirazGKB13} provide  comprehensive surveys on general-purpose distributed processing frameworks.

\subsubsection{Distributed Graph Processing Frameworks} General-purpose distributed processing frameworks facilitate efficient distributed computing by isolating the applications from the lower-level details. However, the general-purpose distributed processing frameworks are inefficient in processing big graph data due to the skewed degree distributions of graph data and poor locality during the computation  \cite{DBLP:journals/csur/HeidariSCB18}. As a consequence, distributed graph processing systems started to emerge with Google's Pregel \cite{DBLP:conf/sigmod/MalewiczABDHLC10} for better performance and scalability.

Pregel \cite{DBLP:conf/sigmod/MalewiczABDHLC10}  is based on the \vertex programming model in which the graph computation is divided into a series of synchronized iterations, each of which is called a superstep. During a single superstep, each vertex executes a user-defined function in parallel. This function typically processes data based on the vertex's current state and the messages it received from other vertices~\cite{DBLP:journals/tkde/KalavriVH18}. 
Besides Pregel, representative \vertex graph processing framework includes  GraphLab~\cite{DBLP:journals/pvldb/LowGKBGH12}, PowerGraph \cite{DBLP:conf/osdi/GonzalezLGBG12}, GPS \cite{DBLP:conf/ssdbm/SalihogluW13}, Giraph \cite{DBLP:journals/pvldb/HanD15}, and Pregel+ \cite{DBLP:conf/www/YanCLN15}. 

Beyond the \vertex based framework, the \edge and \subgraph based distributed graph processing frameworks have also been developed. The \edge model, pioneered by X-Stream  \cite{DBLP:conf/sosp/RoyMZ13} and WolfGraph \cite{DBLP:journals/fgcs/ZhuHLM20}, focuses on edges rather than vertices, offering better performance for non-uniform graphs by streamlining edge interactions and avoiding random memory access \cite{DBLP:journals/csur/HeidariSCB18}.
Despite its programming complexity, the \edge model boosts performance for algorithms that require pre-sorted edge lists \cite{DBLP:journals/fgcs/ZhuHLM20, DBLP:conf/sosp/RoyMZ13}.
The \subgraph model concentrates on entire subgraphs. This model minimizes communication overhead, making it ideal for tasks where frequent vertex interaction is crucial, such as in social network analysis. Frameworks like Giraph++ \cite{DBLP:journals/pvldb/TianBCTM13}, GoFFish \cite{DBLP:conf/europar/SimmhanKWNRRP14}, and Blogel \cite{DBLP:journals/pvldb/YanCLN14} adapt this model.
Despite these frameworks' advancements, distributed graph algorithms still face challenges like parallelism, load balance, communication, and bandwidth due to the complexity and irregularity of graph data, as discussed in Section \ref{sec:introduction}.

%% file: section/challenge.tex

Distributed graph processing is able to handle graphs of very large scale via interconnected computers. However, the shift from single-machine computing to distributed computing introduces challenges arising from the inherent characteristics of distributed systems and graphs, making them essential considerations in the design of distributed graph algorithms. In this section, we conduct a systematic analysis of the inherent challenges in distributed graph processing (\refsec{inherentc}) and provide an overview of existing solutions (\refsec{solution}).

\subsection{Inherent Challenges in Distributed Graph Processing}
\label{sec:inherentc}

\begin{figure}[h]
\centering

\begin{subfigure}{0.98\linewidth}
    \centering
    \includegraphics[width=1\linewidth]{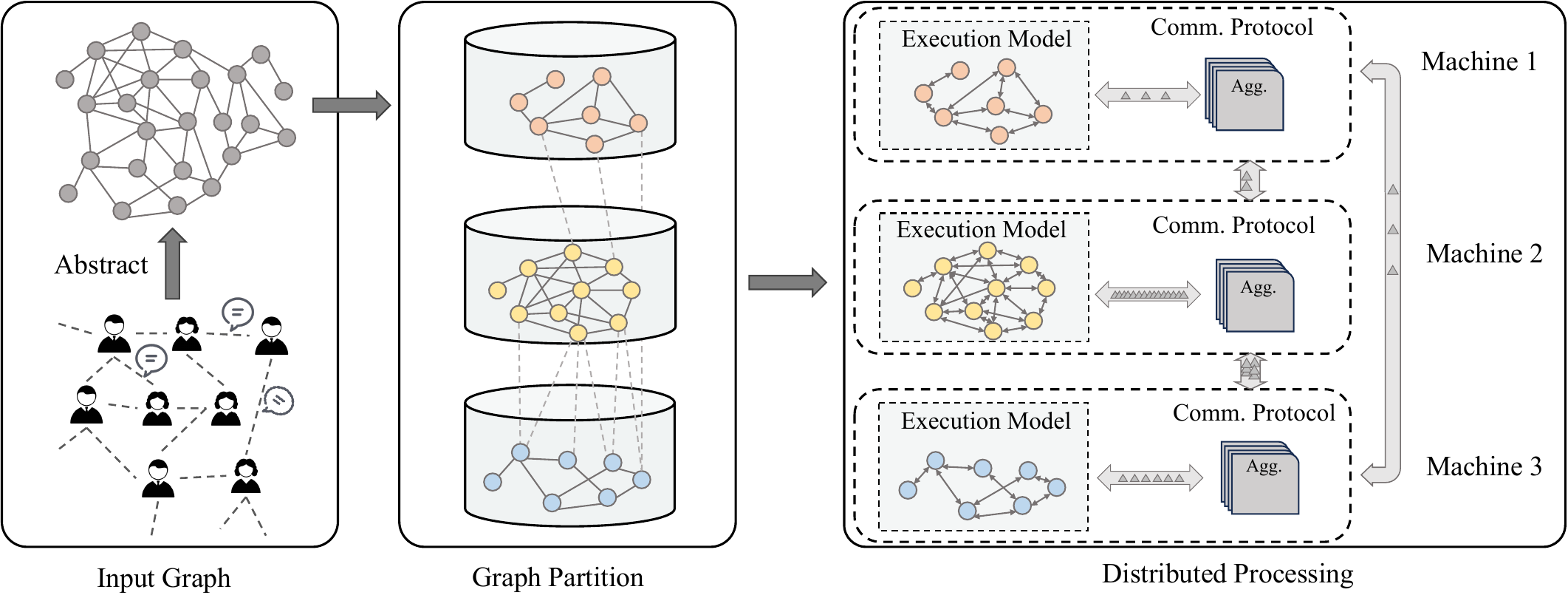}
\end{subfigure}
\caption{The framework of the distributed graph processing. The graph is divided into several partitions that are assigned to different machines. Vertices in different partitions require a global communication to send messages while vertices in the same partition can exchange messages by local communication. The volume of one communication has a limitation called bandwidth, indicating the largest size of a single message. }
\label{fig:distributedsystem}
\end{figure}

In a distributed system, comprising multiple interconnected machines, each serves as an independent computational unit, often dispersed across different locations. This setup, as depicted in \reffig{distributedsystem}, harnesses collective computing power for efficient data processing. However, it also introduces significant challenges in computation and network resource utilization, aspects that are particularly critical in the context of distributed graph processing.

\noindent\underline{\textbf{Computation Resource Efficiency:}} Distributed systems are characterized by their vast and scalable computation resources, which enable the system to process massive volumes of graph data and execute complex graph computations. Therefore, it is of great importance to fully exploit the computation resource in the system when designing distributed graph algorithms. Different from the centralized graph algorithms in which all the instructions are executed in a single machine, distributed graph algorithms need the collaboration and cooperation of multiple machines to finish a task, which brings the challenges of \textit{parallelism} and \textit{load balance}.

\begin{itemize}
   \item \emph{Parallelism}: Parallelism in distributed graph processing involves executing multiple computations concurrently across various machines. This approach entails segmenting a larger graph analysis task into smaller, manageable subtasks. These subtasks are then distributed among different machines, enabling them to be executed simultaneously. Such a strategy not only facilitates efficient resource utilization but also significantly reduces the overall computation time, thereby enhancing the performance of graph processing tasks. However, graph analysis tasks often exhibit inherent sequential dependencies \cite{DBLP:conf/icdcs/HuaFAQLSJ16, DBLP:conf/ppopp/AlabandiPB20, DBLP:journals/vldb/ZhangYWTCS17}, making the realization of parallelism in distributed graph algorithms a complex endeavor. A profound understanding of the fundamental nature of these tasks is essential to discern independent subtasks that can be parallelized effectively. This necessitates a careful analysis to strike a balance between preserving the integrity of the sequential dependencies and optimizing for parallel execution.

    \item \emph{Load Balance}: Load balance in distributed graph processing ensures that the computational workload is evenly distributed across all machines. An imbalance in load can lead to inefficiencies: some machines may complete their tasks quickly and remain underutilized, while others (often known as stragglers) struggle with ongoing computations, ultimately delaying the overall process. This imbalance is particularly problematic in distributed graph processing, where the irregular nature of computations arises from non-uniform   \cite{DBLP:journals/siamrev/ClausetSN09} degree distribution and topological asymmetry. Albeit crucial, effectively resolving load imbalance is complex. It demands not only precise initial workload quantification but also ongoing adjustments during runtime to address any imbalances.
\end{itemize}

\noindent\underline{\textbf{Network Resource Efficiency:}} In distributed systems, where machines communicate via networks, efficiently using network resources becomes vital, particularly in graph processing. The inherent complexity of graph data, marked by intricate structures and irregular vertex connections, often requires operations on a single vertex to interact with multiple others. This scenario leads to frequent and extensive network data exchanges especially when interconnected vertices are spread in different machines. Consequently, two main challenges arise in terms of network resource efficiency.




\begin{itemize}

    \item \emph{Communication Overhead}: Communication overhead in distributed systems, defined by the network resource usage from message exchanges, is largely dependent on data transmission volumes. In distributed graph processing, the necessity to communicate across machines for accessing vertices or edges on different machines increases network communication. Inefficient management of these data exchanges can lead to significant network congestion, turning network communication into a critical bottleneck for overall computing performance. Thus, managing communication overhead is crucial for optimizing the efficiency and effectiveness of distributed graph processing.

\item \emph{Bandwidth}: Bandwidth in distributed systems represents the maximum data transfer capacity between machines in each round of message passing. Limited by hardware and network infrastructure, bandwidth is not infinitely scalable. In distributed graph processing, vertices with high degrees demand substantial bandwidth due to intensive communication with neighbors \cite{DBLP:conf/podc/Censor-HillelGL20} and frequent visits from multiple vertices, common in random walk algorithms \cite{DBLP:conf/aaai/Luo19}. Additionally, low bandwidth utilization is also a challenge. For tasks, like triangle counting, BFS, and connected components, numerous small messages  are exchanged between low-degree vertices, containing only neighbor information. On the other hand, each message exchange using the message-passing interface, like MPI, introduces additional overhead in the form of header information and handshake protocol messages resulting in a diminished ratio of actual payload data and thereby leading to an inefficient utilization of bandwidth resources \cite{DBLP:conf/sc/SteilRIPSP21}. For these reasons, effectively and efficiently optimizing bandwidth utilization is challenging in  distributed graph processing. 
    \end{itemize}

\noindent\textbf{Remark.} Besides the above challenges, there are still other challenges like fault tolerance \cite{DBLP:journals/tpds/ZhouGG21, DBLP:conf/aaai/GuptaS23}   that need to  be considered in distributed graph processing.  However, distributed graph processing often addresses them following the established solutions for general distributed algorithms in the literature.  Therefore, we focus on the above four challenges in this survey and  omit other challenges (Interested readers can  refer to \cite{DBLP:conf/sosp/TeixeiraFSSZA15,DBLP:conf/asplos/ChenQ23a,DBLP:conf/sc/WangCMYC22,DBLP:conf/bigdataconf/MandalH17} for the solutions to these challenges).



\subsection{Solution Overview}
\label{sec:solution}
Following our analysis of the inherent challenges in distributed graph processing in \refsec{inherentc}, this section summarizes various solutions developed to address these challenges, particularly in the field of distributed graph processing, and provides an overview of the common techniques in the detailed algorithms presented in \refsec{algorithms}.

\subsubsection{Computation Resource Efficiency Optimization}
This section focuses on the solutions to optimize the computation resource efficiency including  parallelism and load balance.

\noindent\underline{\textbf{Parallelism:}} Breaking down the graph analysis task into smaller subtasks that can be processed independently reduces the overall execution time by allowing multiple computations to proceed concurrently in the distributed system. To achieve an effective parallelism, various techniques, including \textit{dependency resolution}, \textit{asynchronous execution}, and  \textit{shortcut augmentation}, have been proposed.

\begin{itemize}
\item \emph{Dependency {Resolution}}: Graph analysis tasks generally have an intrinsic dependency order on the computation. For example, the order to visit the vertices in the graph traversal issues is  specific, and any algorithms on the graph traversal have to follow this order to ensure correctness. The implied dependency order significantly limits the potential parallelism of the distributed graph algorithms.  Dependency resolution focuses on identifying and exploiting the inherent dependencies among subtasks in the original graph analysis task to ensure correct and efficient parallel execution. Given the dependency order is unique to each specific graph analysis task, the dependency resolution methods also vary accordingly. The specific details on these dependency resolution methods are presented in \refsec{algorithms} \cite{DBLP:journals/pvldb/LiaoLJHXC22,DBLP:journals/jpdc/ChanSS21,DBLP:conf/bigdataconf/MandalH17,DBLP:journals/tpds/MontresorPM13,DBLP:conf/icde/LiuLHXG23,DBLP:conf/bigdataconf/ChenCC14,DBLP:conf/infocom/CrescenziFP20,DBLP:conf/ppopp/HoangPDGYPR19,DBLP:conf/icdcs/HuaFAQLSJ16}.


\item \emph{Asynchronous Execution}: Synchronization (Figure \ref{fig:Synchronous_Asynchronous} (a)) is a coordination mechanism used to ensure that multiple concurrent processes in distributed graph processing do not cause data inconsistency and are able to successfully complete the task without conflict. In distributed graph processing, synchronous execution  is the easiest synchronization mechanism in which  all computations are divided into discrete global iterations or super-steps. During each super-step, all machines perform their computations simultaneously locally, and at the end of each super-step, there is a barrier synchronization point where all machines must wait until every other machine has finished its computations and communications \cite{DBLP:journals/ai/ModiSTY05, DBLP:conf/nips/AsuncionSW08}. This can ensures consistency but often lead to poor computation performance due to \mbox{waiting times caused by slower machines or stragglers.}

Asynchronous execution (Figure \ref{fig:Synchronous_Asynchronous} (b)), instead, allows machines to operate independently without waiting for each other at barrier synchronization points. Once a machine finishes its computation, it immediately sends its updated information to its neighbors and proceeds with the next round of computation. This approach can significantly speed up the computation process as there’s no idle waiting time, and it can also converge faster in certain scenarios. Therefore, many distributed graph algorithms  \cite{DBLP:conf/icassp/HeW21,DBLP:journals/dc/MashreghiK21,DBLP:journals/jpdc/RemisGAN18,DBLP:journals/vldb/ZhangYWTCS17} exploit the asynchronous execution to enhance parallelism.

\begin{figure}[h]
\centering
    \begin{subfigure}{0.55\linewidth}
		\centering
		\includegraphics[width=1.0\linewidth]{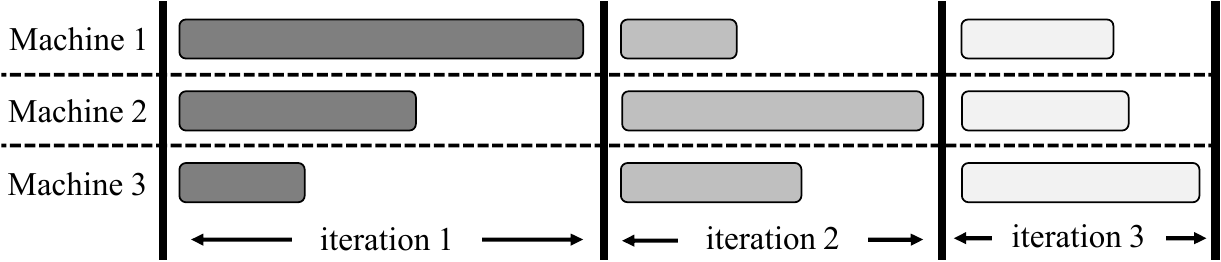}
		\caption{The Synchronous Execution}
        \label{fig:challenge_asynchronous_a}
	\end{subfigure}
    \hspace{1em}
    \begin{subfigure}{0.41\linewidth}
		\centering
		\includegraphics[width=1.0\linewidth]{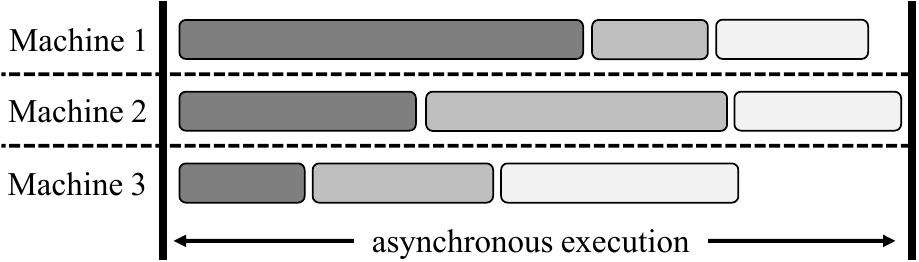}
		\caption{The Asynchronous Execution}
        \label{fig:challenge_asynchronous_b}
    \end{subfigure}
\caption{
    An example of synchronous and asynchronous execution: In (a), vertical lines represent synchronous signals, where each machine must wait until all other machines have completed their workloads. In (b), asynchronous execution involves no waiting time, allowing each machine to compute independently, thereby increasing overall parallelism.
}
\label{fig:Synchronous_Asynchronous}
\end{figure}

\item  \emph{Shortcut Augmentation}:  Shortcut augmentation aims to reduce the number of communication rounds required by adding additional edges, known as shortcuts, to the original graph. These shortcuts serve as direct links between distant vertices and enable more efficient information exchange among them. By introducing shortcuts between vertices that are far apart, the algorithm can effectively reduce the number of rounds required to disseminate information or perform computations on the graph. This reduction in communication rounds leads to significant improvements in both time and resource utilization. Shortcuts augmentation optimization is widely used in  path-related algorithms,  in which shortcuts can significantly reduce the length of a path {\cite{DBLP:journals/tcs/SarmaMPU15, DBLP:journals/isci/LuoWK22, DBLP:conf/podc/CaoFR21}}, and thus achieve faster convergence of the algorithm.

           
            
    
\end{itemize}
\noindent\underline{\textbf{Load balance:}} Uneven workload distribution is a critical challenge encountered in distributed graph computing, as it often leads to load imbalance and hampers optimal performance. In the literature,  \textit{graph partitioning} and  \textit{task scheduling} have emerged as the primary approaches to address the load balance challenge.  

\begin{itemize}
    \item \emph{Graph Partitioning}: Partitioning-based methods aim to distribute the graph data evenly by estimating the computational requirements of individual vertices and edges and partitioning them in advance. By carefully assigning vertices and edges to different machines, these methods strive to achieve load balance and improve overall system efficiency. Partition-based methods (Figure \ref{fig:challenge_partition}) can be further divided into three categories: (1) Vertex partitioning which divides the graph into  subgraphs by assigning vertices to the different partition sets while minimizing edge cuts concerning load balance constraint. (2) Edge partitioning divides the graph into subgraphs by assigning edges to the different partition sets while considering a maximum load balance and minimum vertex cut. (3) Hybrid partitioning exploits the interior structure, such as  vertex in-degree, the vertex out-degree, the degeneracy number, and the core number, of the graph to perform partitioning for a better-balanced workload~\cite{DBLP:conf/sigmod/GuoCCLL17, DBLP:conf/www/Lin19, DBLP:journals/tpds/ChakaravarthyCM17,DBLP:journals/tsc/XuCF16, DBLP:journals/tpds/PandeyWZTZLLHDL21, DBLP:journals/vldb/ZhangYWTCS17,DBLP:conf/ppopp/MalekiNLGPP16, DBLP:conf/sigmod/Yu0KLLCY20,DBLP:conf/podc/Censor-HillelLV22,DBLP:journals/tjs/SattarA22,DBLP:conf/ipps/StrauszVGBH22,DBLP:conf/hpec/GhoshH20,DBLP:conf/hpec/Ghosh22,DBLP:conf/sc/BogleBDRS20}. Moreover, recent algorithms also adopt the dynamic re-partitioning strategy in which the graph is dynamically redistributed across machines to adapt to changes in the graph analysis workload or resource availability to further improve the load balance \cite{DBLP:journals/tpds/JainFPGA17,DBLP:conf/asplos/ChenQ23a,DBLP:journals/is/AlemiH19,DBLP:journals/pvldb/0012XFC00M20}.
    
    \begin{figure}[h]
        \centering
            \begin{subfigure}{0.3\linewidth}
                \centering
                \includegraphics[width=0.72\linewidth]{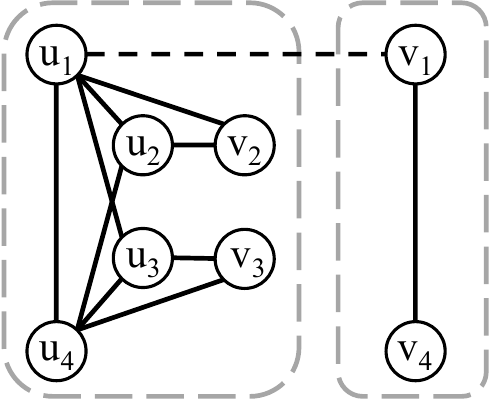}
                \caption{The Vertex Partitioning}
                \label{fig:challenge_partition_vertex}
            \end{subfigure}
        \hspace{-1em}
            \begin{subfigure}{0.39\linewidth}
                \centering
                \includegraphics[width=0.95\linewidth]{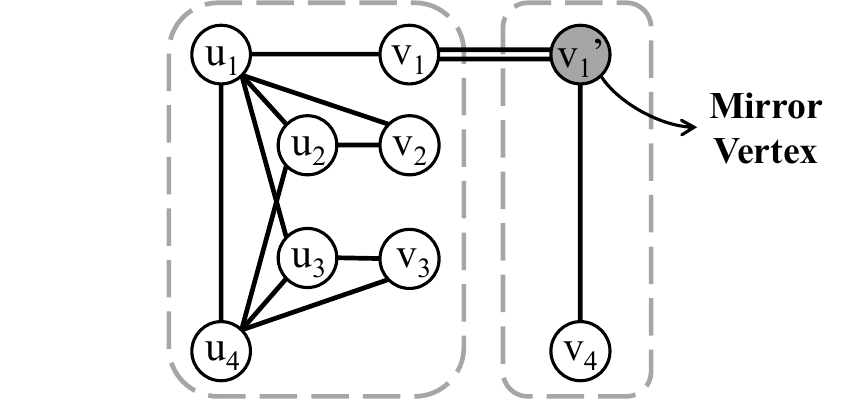}
                \caption{The Edge Partitioning}
                \label{fig:challenge_partition_edge}
            \end{subfigure}
        \hspace{0em}
            \begin{subfigure}{0.28\linewidth}
                \centering
                \includegraphics[width=0.72\linewidth]{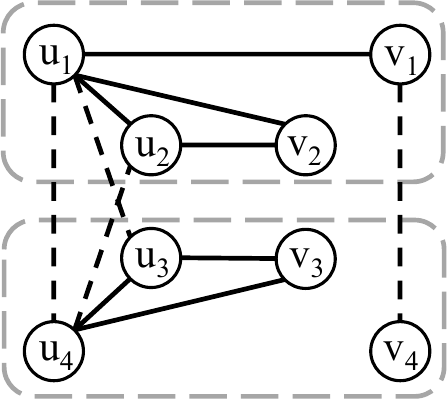}
                \caption{The Hybrid Partitioning}
                \label{fig:challenge_partition_hybrid}
            \end{subfigure}
         \caption{An example of partitioning methods: (a) vertex partitioning with 1 edge-cut, (b) edge partitioning with 1 vertex-cut (requiring mirror vertices for synchronization). Minimum-cut can cause unbalanced partitions. The hybrid method (c) with 3 edge-cuts considers both cut numbers and vertex distribution for better balance.}
        \label{fig:challenge_partition}
    \end{figure}

    \item \emph{Task Scheduling}: Scheduling-based methods focus on dynamically scheduling the computation resources (e.g., processors). By monitoring the workload and resource availability in real-time, these methods enable the system to dynamically distribute workload and adjust resource allocation across machines as the computation progresses, aiming to maintain load balance and optimize the utilization of computation resources as a whole. A typical task scheduling implementation generally encompasses a task queue, responsible for holding pending tasks available for execution, and a task scheduler that selects the subsequent task for execution using diverse policies \cite{DBLP:conf/eurosys/ChenLZYYC18,DBLP:conf/sigmod/YangL0H021}.

\end{itemize}

\subsubsection{Optimizing Network Resource Efficiency}

This section  focuses on the solutions to address  communication overhead and bandwidth challenges regarding network resource efficiency.

\noindent\underline{\textbf{Communication Overhead:}} In distributed graph processing, vertices in different machines frequently exchange messages, resulting in substantial communication overhead. To address this challenge, researchers have devoted considerable efforts to developing effective solutions to optimize communication overhead, which are categorized as follows:

\begin{figure}[h]
    \centering
    
        \begin{subfigure}{0.5\linewidth}
            \centering
            \includegraphics[width=0.7\linewidth]{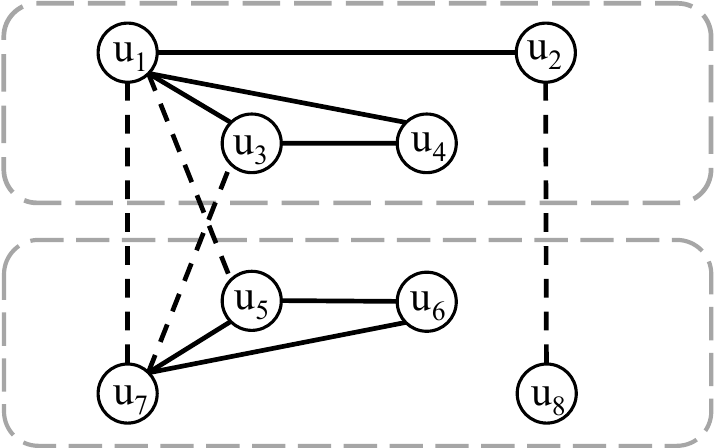}
            \caption{Vertex-Centric Model}
        \end{subfigure}
        \hspace{-1em}
        \begin{subfigure}{0.5\linewidth}
            \centering
            \includegraphics[width=0.75\linewidth]{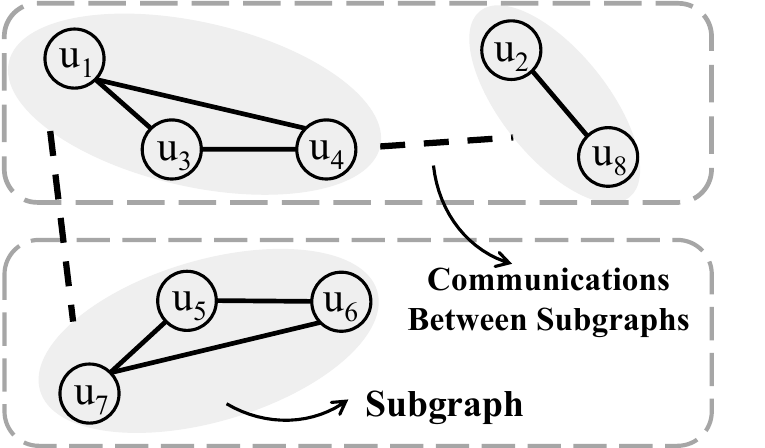}
            \caption{Subgraph-Centric Model}
            \label{fig:challenge_subgraph}
        \end{subfigure}
    \caption{An example of the \subgraph~ model is shown in (b). Compared to the \vertex model in (a), where communication occurs between vertices, in (b), local computations are carried out within subgraphs until convergence, and then communication only occurs between subgraphs.}
    \label{fig:Subgraph-Centric Model}
    \end{figure}

\begin{itemize}
   \item  \emph{Local Computation}: By exploiting the inherent locality of the graph analysis task, local computation aims to perform the computation primarily based on locally available information at each machine to avoid message exchange to reduce communication overhead \cite{DBLP:conf/dasc/SattarA18,DBLP:journals/tjs/SattarA22,DBLP:journals/ijcsa/CosulschiGSS15}. To achieve this goal, a widely adopted solution is to aggregate information from neighbors in advance, and thus accessing the information about these neighbors only requires local communication or local shared memory \cite{DBLP:conf/ppopp/CaoWWLMYC22}. Moreover, \subgraph computation (Figure \ref{fig:Subgraph-Centric Model}) \cite{DBLP:conf/usenix/LakhotiaKP18, DBLP:journals/tpds/MontresorPM13, DBLP:conf/sigmod/ShaoCC14,DBLP:conf/comad/BadamS14,DBLP:journals/pvldb/LiaoLJHXC22,DBLP:conf/ppopp/MalekiNLGPP16} is another popular local computation variant to reduce communication overhead.  Subgraph-centric computation decomposes the graph into smaller subgraphs. Each machine in a distributed system is tasked with processing a specific subgraph independently, leveraging local information and neighboring subgraphs' data as necessary. This localized approach minimizes the need for frequent expensive inter-machine communication.

\item \emph{Selective Activation}: Selective activation solution selects a subset of vertices to be activated in each iteration, thereby mitigating the need for extensive communication across all edges during the distributed graph processing. Two prevalent methods for selective activation have been proposed in the literature. The randomized method activates vertices according to certain probabilistic criteria following the intuition that  deleting one edge from the graph rarely changes the required proprieties such as shortest path distances ~\cite{DBLP:conf/cluster/SariyuceSKC13} and cycle-freeness~\cite{DBLP:conf/spaa/FraigniaudO17}, which effectively prevents unnecessary communication on inactive vertices \cite{DBLP:journals/dc/DaniGHP23}. The deterministic method often adopts an activating threshold, whereby a vertex only sends messages when it satisfies a specific condition based on the threshold \cite{DBLP:journals/jpdc/ChanSS21,DBLP:journals/tpds/MontresorPM13,DBLP:journals/tpds/WengZLPL22,DBLP:journals/tpds/LulliCDLR17}. By leveraging these selective activation solutions, distributed graph algorithms can effectively reduce the communication overhead.

\item \emph{Hybrid Push-Pull Model}: The \push and \pull are two fundamental communication paradigms in distributed graph processing.
In the \push model, each vertex sends its updating messages or partial computation results to its neighboring vertices, while in the \pull model, vertices retrieve information from their neighbors as needed to complete their computations. Generally, these two paradigms each have their own merits regarding communication efficiency in distributed graph processing. For instance, the \push model might be a better choice for the initial stage with fewer active vertices, while the \pull model might perform better on few inactive vertices~\cite{DBLP:journals/jpdc/RemisGAN18, DBLP:conf/ppopp/CaoWWLMYC22}. 
To combine the strengths of both models to reduce the communication overhead, hybrid solutions that automatically switch between the \push model and the \pull model are proposed \cite{DBLP:journals/tpds/ChakaravarthyCM17, DBLP:conf/hpec/0001FZHBLG19}.

\comment{For instance, the \push model might be a better choice for vertices with low degree while the \pull model might perform better on the vertices with high degree \todo{ \cite{} add more reference}. To combine the strengths of both models to reduce the communication overhead, hybrid solutions that automatically switch between the \push model and the \pull model are proposed \cite{DBLP:journals/jpdc/RemisGAN18, DBLP:conf/ppopp/CaoWWLMYC22, DBLP:journals/tpds/ChakaravarthyCM17}.}

\item \emph{Aggregation}: Aggregation aims to reduce the amount of data transmitted between machines by merging multiple data elements and eliminating duplicates or redundant elements. A prominent method is to aggregate messages locally before sending them to other machines, wherein only essential information is retained. This method not only reduces the overall communication volume but also effectively diminishes the rounds of communications required \cite{DBLP:journals/dpd/FengCLQZY18,DBLP:conf/ipps/SandersU23}. In addition, some algorithms \cite{DBLP:conf/podc/Censor-HillelLV22, DBLP:conf/ipps/Lamm022} focus on simplifying the graph structure by aggregating the information of vertices/edges within a certain area where communication primarily occur between these vertices/edges. This simplification facilitates the communication  between these aggregated vertices/edges, thereby significantly alleviating the overall communication overhead.
    


\end{itemize}

\noindent\underline{\textbf{Bandwidth:}} Frequent data exchange between machines also poses significant strain on the available network bandwidth. To mitigate the requirement of bandwidth in distributed graph processing, various solutions are designed in the literature, \mbox{including  \textit{message coordination} and  \textit{dense representation}.}

\begin{itemize}


    \item\emph{Message Coordinating}:  In distributed graph processing, a vertex may receive a substantial number of messages within a single iteration and the available network bandwidth cannot accommodate the simultaneous transmission of these messages to the neighbors, message coordinating methods prioritize the messages based on certain criteria such that a message with potentially better result or a farther destination is assigned higher priority while less promising messages may be pruned by later updates, and transmit the messages based on their priorities accordingly \cite{DBLP:conf/aaai/Luo19, DBLP:journals/isci/LuoWK22, DBLP:conf/ppopp/HoangPDGYPR19, DBLP:conf/icdcs/HuaFAQLSJ16}. Message coordinating allows for more efficient utilization of available network resources and helps mitigate the impact of bandwidth limitations on distributed graph processing systems.



\item \emph{{Message Buffering}}: Batch processing is commonly used to tackle the issue of low bandwidth utilization in distributed graph processing. It utilizes a message buffer queue within each machine, wherein messages intended for transmission are temporarily stored. Once the buffer reaches its capacity, all accumulated messages are combined into a single batch and subsequently transmitted to other machines. Although this approach may introduce a delay in communication, it yields significant benefits in terms of optimizing bandwidth resource utilization and reducing the processing time associated with sending and receiving messages \cite{DBLP:conf/ipps/SandersU23,DBLP:conf/hpec/Ghosh22,DBLP:conf/hpec/GhoshH20,DBLP:conf/sc/SteilRIPSP21,DBLP:conf/sc/WangCMYC22}.


\end{itemize}

%% file: section/sec3_Centrality.tex
\centrality~measures in graph theory are instrumental for quantifying the relative importance of vertices within a complex network. 
PageRank and personalized PageRank are relatively easy to parallelize as they only require neighbor information for computation. In contrast, Betweenness centrality and closeness centrality assess vertex influence based on shortest paths, requiring global graph information. This complexity poses challenges for designing distributed algorithms. Due to these characteristics, the latter is more suitable for identifying critical nodes that control the flow of communication or resources in networks compared to the former.

As \reffig{distribution}(a) illustrates, parallelism in centrality tasks is a popular research focus ($37.93\%$). This is primarily due to the inherent independence of centrality computations for each vertex, which lends itself well to parallel processing approaches. Surprisingly, load balancing—a fundamental challenge in distributed environments—has garnered limited attention in this domain, accounting for only $13.79\%$ of the research. This could be attributed to two potential factors. Firstly, the computational requirements for centrality calculations on each vertex are relatively straightforward, such as the summation process in PageRank algorithms. Secondly, in cases where the centrality task is complex, such as in calculating betweenness centrality, the research community may prioritize developing efficient parallelization techniques over addressing load balancing issues.
A significant portion of the centrality research, precisely $31.03\%$, is devoted to minimizing communication overhead, while $17.24\%$ of the studies aim to address bandwidth constraints.


\comment{
\begin{figure}[h]
\centering
\begin{tikzpicture}[scale=0.8]
\pie[text=legend, color={cyan!30!gray!30, red!30!blue!30, cyan!30, yellow!40!brown!30}]{36.37/Parallelism, 4.54/Load Balance, 40.9/Communication Overhead, 18.18/Bandwidth}
\end{tikzpicture}
\label{challenges_of_Community_Detection}
\caption{Community Detection}
\end{figure}
}

\stitle{\pagerank},
initially conceptualized by in~\cite{pagerank}, is designed to assess the importance of webpages, represented as vertices in a web graph, interconnected by hyperlinks symbolizing the edges. The \pagerank~score, symbolized by $\pi$, is derived from the stationary probabilities of an $\alpha$-decay random walk. This process starts from a vertex selected uniformly from all $n$ vertices at random. Subsequently, at each step, there exists a probability $\alpha$ of initiating a new random walk from a random vertex, and a complementary probability $1-\alpha$ of transitioning to an adjacent vertex. The \pagerank~score of a vertex $u$, denoted as $\pi(u)$, quantifies the relative importance of vertex $u$ within the web graph, which can be formulated as
$
\pi(u) = (1 - \alpha) \cdot \frac{1}{n} + \alpha \cdot \sum_{v \in N_{in}(u)} \frac{\pi(v)}{N_{out}(v)}.
$
Typically, algorithms for computing \pagerank~can be categorized into three distinct types:

(1) Iterative algorithms: Examples include PCPM~\cite{DBLP:conf/usenix/LakhotiaKP18} and SubgraphRank~\cite{DBLP:conf/comad/BadamS14}. These algorithms iteratively update \pagerank~values until converge.

(2) Random-walk-based algorithms: This category comprises algorithms including IPRA~\cite{DBLP:journals/tcs/SarmaMPU15}, RP~\cite{DBLP:journals/isci/LuoWK22}, MRP~\cite{DBLP:conf/aaai/Luo19, DBLP:journals/isci/LuoWK22}, and FP~\cite{DBLP:conf/icml/Luo20}, which generate and simulate multiple random walks and estimate PageRank~values through the destination vertices of these walks.

(3) Eigenvector-based algorithms: The \pagerank~values as fixed points $\pi$ for all vertices represent an eigenvector. Rather than directly using the original equation, the Stochastic Approximation (SA) algorithm~\cite{DBLP:conf/icassp/HeW21} iteratively approximates the stable state $\pi$ by constructing a differential equation, enabling gradual convergence towards the solution.

Optimizing parallelism is crucial, especially in random-walk-based algorithms. These algorithms generate multiple independent random walks to approximate \pagerank values, benefiting significantly from parallel processing.
Specifically, IPRA introduces a method using short walks as efficient indices in the graph, reducing the number of processing rounds needed. Similarly, MRP starts with short walks that are then combined to form longer walks, further enhancing efficiency.
In contrast, the eigenvector-based SA algorithm employs asynchronous computation. It updates \pagerank values by selecting individual vertices in each iteration, allowing subsequent iterations to proceed asynchronously as long as they do not update the same vertices simultaneously.

As a prevalent challenge in distributed computing, load balancing can impact the efficiency of \pagerank~algorithms. The computation of a vertex's \pagerank~is influenced by its connectivity, resulting in vertices with higher degrees engaging in more intensive operations of \pagerank~values. To mitigate this imbalance, iterative algorithms like PCPM and SubgraphRank employ the \subgraph~ computation model, which divides the graph into multiple partitions, each managed by a dedicated thread. However, balancing the workload, which includes both internal computation and global communication, remains complex. They use tools like OpenMP for dynamic scheduling of vertex computation, rather than adhering to an ideal static partitioning.


\pagerank~algorithms face significant challenges in minimizing communication overhead. 
A major advantage of this model is its ability to replace inter-partition vertex communication with local memory access. Additionally, it allows multiple communications from one vertex to all destination vertices within the same partition to be consolidated into a single data transfer, further reducing communication overhead. The SA algorithm optimizes communication costs by using the \pull mode. In each iteration, a selected vertex updates its \pagerank value by ``pulling'' information from its neighbors. This eliminates the need for the vertex to ``push'' its updated \pagerank value to others, thereby halving communication overhead.

Bandwidth constraints can pose challenges, particularly in random walk-based algorithms, where a proliferation of random walks may converge on a single vertex, creating a bottleneck. To mitigate this, both RP and MRP algorithms employ a coordinator mechanism to evenly distribute walks among all neighboring vertices, thus preventing concentration on a single vertex. 
In contrast, the FP algorithm adopts a more bandwidth-efficient approach by transmitting only the expected number of paths, rather than sending individual path messages. 

\stitle{\personalizedpagerank}
modifies the standard \pagerank~algorithm by enabling users to specify a start vertex, denoted as $s$. Unlike traditional \pagerank~, where the random walk commences from and restarts at a randomly selected vertex with probability $\alpha$,
personalized \pagerank~consistently initializes and restarts the walk at the specified source vertex $s$.
\eat{
This alteration leads to a modification in the algorithm's formulation, which can be represented as
    $\pi(u) = (1 - \alpha) \cdot e_s(u) + \alpha \cdot \sum_{v \in N_{in}(u)} \frac{\pi(v)}{N_{out}(v)}$,
where $e_s$ is an indicator such that $e_s(u)=1$ if $u=s$, and $e_s(u)=0$ otherwise.
}

Given the similarities between personalized \pagerank~and the standard \pagerank~algorithm, most techniques for addressing distributed computing challenges are comparable. For instance, the \subgraph~ computation model ~\cite{DBLP:conf/sigmod/GuoCCLL17} and shared-memory approaches ~\cite{DBLP:journals/kais/SoudaniFN19} effectively reduce internal communication within partitions. Additionally, employing a coordinator between two partitions can significantly minimize redundant global communications, as demonstrated in ~\cite{DBLP:conf/sigmod/GuoCCLL17}. 

We highlight techniques employed in personalized \pagerank~to achieve optimal load distribution. The study~\cite{DBLP:conf/sigmod/GuoCCLL17} proposes a theoretically perfect partition utilizing hierarchical methods. Moreover, load imbalance can also arise during sampling in weighted graphs, where the next hop in random walks is uneven, necessitating weighted sampling. The Alias method, capable of constant-time sampling for weighted elements, and its extension, the Alias Tree, are effective solutions for handling extremely large numbers of neighbors in memory-limited scenarios~\cite{DBLP:conf/www/Lin19}.




\stitle{\betweennesscentrality}
 measures the extent to which a specific vertex lies on the shortest paths between other vertices within a graph. The underlying premise is that important vertices are more likely to be on many of these paths, influencing communication flow across the graph. Given $C_b(u)$ as the normalized betweenness centrality value for a vertex $u$,
 most algorithms~\cite{DBLP:conf/infocom/CrescenziFP20, DBLP:conf/ppopp/HoangPDGYPR19, DBLP:journals/tpds/JamourSK18, DBLP:conf/icdcs/HuaFAQLSJ16} follow the Brandes' algorithm~\cite{Brandes2001Faster} to compute  $C_b(u)$ as:
    $C_b(u) =
    \frac{1}{(n-1)(n-2)} \sum_{s \in V \backslash \{u\}} \delta_{s\bullet}(u), \text{  and }
    \delta_{s \bullet}(u) = \sum_{t \in V \backslash\{s, u\}} \frac{\sigma _{su} \cdot \sigma_{ut}}{\sigma_{st}}
    $
, where $\sigma_{st}$ is the number of shortest paths from a vertex $s$ to another vertex $t$. The value
of $\delta_{s \bullet}(u)$ can be recursively computed as:
$
    \delta_{s\bullet}(u) = \sum_{u \in P_s(w)}\frac{\sigma_{su}}{\sigma_{sw}} \cdot (1 + \delta_{s\bullet}(w))
$
, where $P_s(w)$ is the set of predecessors of $w$ in the SSSP Directed Acyclic Graph~(SSSP DAG). 

To compute betweenness centrality values of all vertices, Brandes' algorithm comprises two primary steps: (1) SSSP Step involves constructing a Directed Acyclic Graph (DAG) from each vertex in the graph; (2) Brandes' Step recursively computes the dependency score $\delta_{s\bullet}(u)$ and systematically accumulate these scores to determine the value of $u$.

\eat{
\begin{itemize}
    \item \textbf{SSSP Step:} This step involves constructing a Directed Acyclic Graph (DAG) from each vertex in the graph. The DAG represents all the shortest paths originating from that vertex. During this process, the algorithm also counts the number of shortest paths ($\delta_{su}$) from the source vertex $s$ to every other vertex $u$. Most algorithms focus on unweighted graphs so this
    \item  \textbf{Brandes' Step:} Recursively compute the dependency score $\delta_{s\bullet}(u)$ using \refeq{bc_delta} and systematically accumulate these scores to determine the betweenness centrality of $u$.
\end{itemize}
}

In adapting Brandes' algorithm for distributed computing, a key consideration is parallelizing the computation across $n$ SSSP DAGs to enhance parallelism. This involves executing the SSSP and Brandes' steps concurrently from all vertices. In unweighted graphs, BFS is used for this parallel execution, computing $\sigma_{su}$ and $\delta_{s\bullet}(u)$ in the BFS order~\cite{DBLP:journals/tpds/JamourSK18, DBLP:conf/icdcs/HuaFAQLSJ16}. For weighted graphs, alternatives like the Bellman-Ford~\cite{DBLP:journals/tpds/BusatoB16} and Lenzen-Peleg~\cite{DBLP:journals/dc/LenzenPP19} algorithms are employed~\cite{DBLP:conf/infocom/CrescenziFP20, DBLP:conf/ppopp/HoangPDGYPR19}.

The communication demands of the SSSP step, especially on weighted graphs, are high due to frequent message passing. To mitigate this, the algorithm developed by Crescenzi et al.~\cite{DBLP:conf/infocom/CrescenziFP20}, which is based on Bellman-Ford, updates $\sigma_{su}$ and $\delta_{s\bullet}(u)$ with differential values instead of full recomputation. Alternatively, the Lenzen-Peleg-based algorithm in~\cite{DBLP:conf/ppopp/HoangPDGYPR19} adopts an early termination strategy called Finalizer to stop the SSSP step as long as all vertices have received correct $\sigma_{su}$, which is shown to be effective when graph diameters are small.

Addressing bandwidth limitations involves coordinating the order of message sending. In the SSSP step, messages concerning shorter distances are prioritized~\cite{DBLP:conf/ppopp/HoangPDGYPR19}, as longer distances are more likely to change. In the Brandes' step, priority is reverted to longer distances due to their greater number of hops to the source, as discussed in ~\cite{DBLP:conf/icdcs/HuaFAQLSJ16}.  Given the inherent difference between the two steps, the algorithm in~\cite{DBLP:conf/ppopp/HoangPDGYPR19} attempts to track the round numbers at which updates occur in the SSSP step and then apply these in reverse order during the Brandes' step. 
Additionally, Pebble BFS~\cite{DBLP:conf/podc/HolzerW12}, particularly effective in parallel SSSP for unweighted graphs, adopts a staggered BFS approach. The algorithm~\cite{DBLP:conf/icdcs/HuaFAQLSJ16} utilizes the idea of visiting one vertex (instead of multiple vertices simultaneously) per round, thereby avoiding message congestion to some large-degree vertices. 

\stitle{\closenesscentrality}
quantifies the average shortest path length from a given vertex to all other vertices in the network, suggesting that a vertex of higher importance typically exhibits shorter average distances to other vertices. The normalized closeness centrality value $C_c(u)$ for a vertex $u$ is calculated as
    $C_c(u) = \frac{n-1}{\sum_{v \in V} d(u, v)}$,
where $d(u,v)$ is the shortest  distance from $u$ to $v$.

The calculation of closeness centrality values, much like the betweenness centrality, confronts considerable communication overhead. This is particularly severe in incremental closeness centrality, where the introduction of new edges necessitates the reevaluation of closeness centrality values. The STREAMER algorithm~\cite{DBLP:conf/cluster/SariyuceSKC13} effectively addresses this by utilizing a level structure to filter out edges that do not impact shortest path distances, which is a common scenario in real-world graphs. To further refine the process, the DACCER algorithm~\cite{DBLP:journals/cn/WehmuthZ13} employs an approximation of the ``relative rank''. This is based on the observation that vertices with higher closeness centrality tend to have larger neighborhoods with higher-degree vertices. By aggregating degree information from neighbors, each vertex can estimate its closeness centrality rank with reduced communication.

%% file: section/sec4_Clustering_or_CD.tex
Community detection aims to partition vertices in a graph into different communities where vertices within a community are more closely linked to each other than to external vertices.
Louvain optimizes modularity through hierarchical clustering, offering effectiveness at the cost of computational efficiency. In contrast, label propagation iteratively assigns labels based on neighbor majority voting, providing efficiency but potentially leading to unstable community detection, suitable for real-time scenarios.
The connected components algorithm identifies subgraphs where all vertices are connected directly or indirectly, helping to understand the graph's basic structure.

As Figure \ref{fig:distribution} (b) illustrates, a notable proportion of research (23.53\%) focuses on enhancing parallelism in community detection algorithms. The computation of a vertex's community state primarily relies on information that is local to the vertex, making the process inherently amenable to parallel execution across multiple processors. Load balancing is another critical aspect, with many studies (23.53\%) aiming to resolve the skewed workload associated with high-degree vertices.
The primary challenge in community detection, tackled by over half of the research (52.9\%), is minimizing communication overhead.
Despite the frequent exchanges between vertices in community detection algorithms, where each piece of communicated data is generally small, our survey reveals no studies have specifically tackled bandwidth limitations in community detection.  
\stitle{\louvain} is widely regarded as one of the most popular methods in this topic~\cite{blondel2008fast}. Central to this algorithm is the concept of modularity, denoted by $Q$, which serves as a measure to assess the quality of the identified communities within a graph.
The modularity is given by the equation:
$
    Q = \frac{1}{2m}\sum_{i,j}\left[A_{ij} - \frac{k_ik_j}{2m}\right]\delta(c_i,c_j)
    $
, where $A_{ij}$ represents the weight of the edge connecting vertices $i$ and $j$, $k_i  = \sum_{j}A_{ij}$ is the sum of the weights of all edges attached to vertex $i$, $c_i$ denotes the community to which vertex $i$ belongs, $\delta(u, v)$ outputs $1$ if $u = v$ and $0$ otherwise, and $m = \frac{1}{2}\sum_{ij}A_{ij}$ is the sum of weights of all edges.

At the beginning of the algorithm, each vertex is treated as a separate community.
It then proceeds iteratively, examining each vertex $i$ and evaluating the gain in modularity that would result from moving $i$ to a neighboring community $j$.
The vertex $i$ is placed in the community that yields the highest increase in modularity.
The process of moving vertices between communities is repeated until the modularity no longer increases.


Load balancing plays a critical role in the efficiency of \louvain~ algorithms in distributed environments, since the workload of a single machine is typically proportional to the degree of the vertices it processes, thus a few high-degree vertices can lead to significant skew in the workload distribution.
Jianping Zeng, et al. \cite{DBLP:conf/cluster/ZengY18} extend the graph partitioning method based on vertex delegates,  initially proposed by Pearce et al. \cite{DBLP:conf/sc/PearceGA14}.
The method described involves duplicating high-degree vertices across all machines in a distributed system. The edges connected to these high-degree vertices are then reassigned to ensure a balanced distribution of edges, and consequently, computational work, across the different machines in the system.
Another distributed algorithm, called DPLAL \cite{DBLP:journals/tjs/SattarA22}, employs the well-known graph partitioning tool  METIS \cite{karypis1997metis}, which is designed to minimize the number of cut edges or the volume of communication needed between different parts of the graph.

Reducing communication overhead is also a prevalent challenge in \louvain~ algorithms.
Some recent studies \cite{DBLP:conf/ipps/GhoshHTKLCKG18,DBLP:conf/cluster/ZengY18} reduce this overhead by utilizing ``ghost'' vertices, which are a set of replicas of the vertices that are connected to local vertices but are processed by other machines.
These ghost vertices help to maintain the local integrity of the graph, thereby reducing inter-machine communication.
The paper \cite{DBLP:conf/ipps/GhoshHTKLCKG18} also presents two heuristics for optimizing the distributed \louvain~ algorithm, one is setting varying thresholds during different stages of iterations to decide termination, effectively reducing the total number of iterations during the algorithm; another is setting vertices with minor changes in modularity during iterations as ``inactive'', to save computational and communication resources.
Moreover, some papers \cite{DBLP:conf/dasc/SattarA18,DBLP:journals/tjs/SattarA22} adopt the Subgraph-Centric model. Each machine collects information about adjacent vertices, then updates local community states independently, without inter-machine communication. 
\stitle{\lpa} algorithm (LPA) \cite{raghavan2007near} is an effective approach to detecting communities by propagating labels throughout the network. The algorithm starts by assigning a unique label to each vertex, which represents the community to which the vertex currently belongs.
In each subsequent iteration, every vertex considers the labels of its immediate neighbors and adopts the label that is most frequent among them.
The algorithm continues to iterate through the vertices and update their labels until convergence, i.e., when each vertex has the majority label of its neighbors or the maximum number of iterations defined by the user is reached.

Implementing distributed label propagation algorithms involves balancing parallelism with the quality of identified communities. Misordered computations can result in overly large, impractical communities.
The EILPA algorithm~\cite{azaouzi2017evidential} identifies influential vertices, termed ``leaders'', using a personalized PageRank model. Label propagation starts from these leader vertices, ensuring a structured, influence-based community detection.
The PSPLAP algorithm~\cite{ma2018psplpa} also considers vertex influence by assigning weights via $k$-shell decomposition and calculating propagation probabilities and vertex similarities based on these weights. Vertices update their labels during each iteration using these probabilities and similarities, aiming for more coherent community \mbox{detection by reducing label updating randomness typical in label propagation algorithms.}

Communication overhead is another prevalent challenge when implementing the label propagation algorithm in distributed environments.
To tackle this challenge, Xu Liu, et al. \cite{DBLP:conf/hpec/0001FZHBLG19} propose a dynamic mechanism that alternates between \push~ and \pull~ strategies to reduce communication costs.
Initially, \push~ is used for rapid community formation, and then the algorithm switches to \pull~ in later iterations to consider neighbors' label quality.

\stitle{\connectedcomponents} refer to subgraphs where every vertex can be reached from any other vertex. Computing connected components typically involves traversing all vertices and edges in a graph based on Breadth-First-Search (BFS) or Depth-First-Search (DFS), resulting in a linear time complexity. However, in distributed environments, the irregularity of the graph leads to centralized workload and communication in high-degree vertices, which are the primary focus for optimizing connected component computation.

For improving parallelism, the paper \cite{DBLP:journals/tac/ReedRBP23} leverages the max-consensus-like protocol where each vertex only considers the maximum value among its in-neighbors to update its own value. The algorithm can terminate in $D + 2$ iterations where $D$ is the diameter of the graph.
Benefiting from this protocol, each vertex only needs to receive information from its in-neighbors, which also helps to improve computational and communication efficiency.

Balancing workload is also important for improving the efficiency of connected component computation.
In the paper of Sebastian Lamm, et al. \cite{DBLP:conf/ipps/Lamm022}, high-degree vertices are replicated to different machines to alleviate the load imbalance issues caused by processing such vertices on a single machine.
Chirag Jain, et al. \cite{DBLP:journals/tpds/JainFPGA17} adopt the strategy of dynamically redistributing vertices. Due to the vertex pruning strategy used, vertices may be removed from machines after each iteration, leading to load imbalance issues. In their work, after each iteration, some vertices will be redistributed to achieve load balancing.

To minimize communication overhead, selective activation strategies are adopted by a number of papers \cite{DBLP:journals/tpds/LulliCDLR17,DBLP:journals/tpds/JainFPGA17}.
CRACKER \cite{DBLP:journals/tpds/LulliCDLR17} identifies seed vertices to construct trees for each connected component. Vertices are iteratively removed from the graph and added to the tree, reducing communication volume by trimming vertices during this process.
Another work \cite{DBLP:journals/tpds/JainFPGA17} removes completed connected components in each iteration to decrease the number of active vertices in subsequent iterations.
The paper \cite{DBLP:conf/ipps/Lamm022} proposes a graph contraction pre-processing strategy. It conducts local BFS on each machine and merges vertices within the same connected component into a single vertex, avoiding multiple messages to common neighbors.
In addition, Xing Feng, et al. \cite{DBLP:journals/dpd/FengCLQZY18} adopts the local computation strategy, where they first run BFS locally to label vertices in the same connected component with the same color. Then, inter-machine communication is carried out to merge subgraphs by merging different colors received by the same vertex. To further reduce the amount of communication, messages are locally aggregated before being sent to other machines.

%% file: section/sec5_Similarity.tex

This topic refers to evaluating the similarity of the two vertices in a graph. Vertices with a high similarity score indicate that they have something in common in structure or attribute.
Jaccard similarity and cosine similarity are basic measurement methods using set operations, requiring only neighboring information for computation, thus enabling high parallelization. In contrast, SimRank calculates similarity by considering the entire graph's information. Despite its lower efficiency, it can uncover latent relationships and similarities hidden within the graph structure.

According to \reffig{1c}, efforts are divided, with $20\%$ focusing on enhancing parallelism and $10\%$ on improving load balancing. This distribution underscores the challenge in similarity computation across all vertex pairs, necessitating extensive inter-vertex communication. Consequently, the predominant emphasis of research, at $70\%$, is on reducing communication overhead. It's noteworthy, however, that bandwidth constraints have not received significant attention in this topic. 


\noindent\underline{\textbf{\jaccard~ and \cosine}}
are commonly used to measure the similarity of vertices.
Jaccard similarity measures the similarity between two sets based on their shared elements. Specifically, it is defined as the size of the intersection of two sets divided by the size of their union, which is:
    $Jaccard(A, B) = \frac{|A \cap B|}{|A \cup B|}$.
The value of Jaccard similarity ranges from 0 to 1, and if either $A$ or $B$ is an empty set, the value is 0.
In the context of graph data, sets $A$ and $B$ are typically defined as the sets containing all the neighbors of a given vertex.
cosine similarity is another measure of similarity between two sets, which is defined as:
    $Cosine(A, B) = |A \cap B|/\sqrt{|A| \times |B|}$.



Communication overhead poses a significant challenge, especially when computing similarity values between all vertex pairs, as each vertex is required to exchange information with nearly all others.
WHIMP \cite{DBLP:conf/www/SharmaSG17} calculates the incidence vector of vertices in the graph to identify all pairs whose cosine similarity values exceed a specific threshold.
SimilarityAtScale \cite{DBLP:conf/ipps/BestaKMKRHS20} also adopts the method of compressing vectors to reduce the volume of communication between machines during the computation of Jaccard similarity values.
It first divides the vector into smaller batches, filters out zero rows using distributed sparse vectors, and then converts segments of these rows into more efficient bit vectors for processing.
Finally, a study \cite{DBLP:journals/ijcsa/CosulschiGSS15} adopts a local computation strategy, decomposing the computational tasks and allocating them to different machines. In this process, there will be communication of the intermediate results for the local solutions.
\noindent\underline{\textbf{\simrank}} is an algorithm based on the observation that ``two objects are considered similar if they are referenced by similar objects'' \cite{DBLP:conf/kdd/JehW02}, i.e., the similarity of two vertices is determined by its neighbors' similarities. The formal definition is:
$
    s(a, b) = \frac{C}{|I(a)|\cdot|I(b)|}\sum_{i=1}^{|I(a)|}\sum_{j=1}^{|I(b)|}s(I_i(a), I_j(b))
    $
, where $C$ is a constant between 0 and 1, and the set of in-neighbors of $a/b$ is denoted as $I(a)/I(b)$, where $I_i(a)/I_j(b)$ represent the $i$-th$/$$j$-th in-neighbor of $a/b$. Note that $s(a,b)=0$ when $I(a)=\emptyset$ or $I(b)=\emptyset$ and $s(a,b)=1$ if $a=b$.
Computing \simrank~ typically involves two primary algorithms:

(1) Iterative algorithm: The algorithm iteratively computes the \simrank~ scores according to its definition until convergence or a fixed number of iterations.

(2) Monte Carlo algorithm: It is a Random-Walk-based algorithm, where the \simrank~ score between two vertices is based on the expected ﬁrst meeting time of random walkers starting from each vertex and traversing the graph in reverse. 


Optimizing parallelism is significant, particularly for those algorithms that involve random walks. Each walk simulation is independent and suitable for parallel processing. However, the sequence-dependent nature of individual random walks, where the next vertex selected depends on the previous one, presents parallel processing challenges.
DISK \cite{DBLP:journals/pvldb/0012XFC00M20} adopts the approach of constructing forests instead of random walks, thus each vertex can independently build its next layer of child vertices, enhancing parallelism.
In CloudWalker \cite{DBLP:journals/pvldb/LiFLCCL15}, the recursive computation of \simrank~ is transformed into solving a linear system by estimating a diagonal matrix, which represents attributes of each vertex. Subsequently, this linear system is solved using the Jacobi method, which eliminates the recursive dependencies when computing \simrank~ values.

To balance workload, DISK \cite{DBLP:journals/pvldb/0012XFC00M20} declares two load thresholds $W_{max}$ and $W_{min}$, and after each iteration, some vertices from high-load machines (larger than $W_{max}$) will be moved to low-load machines (smaller than $W_{min}$) to achieve dynamic load balancing.

Many papers \cite{DBLP:journals/pvldb/LiFLCCL15,DBLP:conf/icde/LuoGZY17} adopt the method of data sharing
for reducing communication overhead.
CloudWalker \cite{DBLP:journals/pvldb/LiFLCCL15} caches the intermediate results during the computation process, specifically computed by counting the landing positions of all walkers at some vertices, for saving the network communication cost.
Besides, UniWalk \cite{DBLP:conf/icde/LuoGZY17} adopts a path-sharing method. Specifically, when computing the \simrank~ scores for multiple vertices, it allows the reuse of certain paths, thereby reducing the amount of data that needs to be communicated or processed.
Another solution, pruning, is also used to reduce communication overhead.
Siqiang Luo, et al. \cite{DBLP:journals/corr/abs-2304-04015} adopt the method of truncating the length of each random walk and limiting the number of random walks to be sampled for accelerating computation and reducing the volume of communication.
Another tree-based algorithm, DISK \cite{DBLP:journals/pvldb/0012XFC00M20}, builds the forest from leaves to roots, and the \simrank~ values are computed based on the overlapping leaf vertices.
To optimize both computational and communication overhead, each vertex samples its in-neighbors with a certain probability during the forest construction.

%% file: section/sec6_Cohesive_Subgraph.tex
In graph theory, cohesive subgraphs are defined as parts of the original graph where vertices are densely connected with each other. In this survey, we focus on three popular cohesive subgraphs, $k$-core, $k$-truss, and maximal clique. The level of density they identify differs due to their different definitions, with $k$-Core < $k$-Truss < maximal clique. Correspondingly, their computational efficiency decreases in the same order, that is, $k$-Core > $k$-Truss > maximal clique. Therefore, each has its own applications depending on the required density and efficiency.

Computing these subgraphs, varying in density requirements, necessitates traversing all vertices or edges, posing significant challenges. As Figure \ref{fig:distribution} (d) shows, parallelism ($35\%$) presents a challenge, especially in tasks like $k$-core and $k$-truss that require iterative updates. Load balancing, crucial in computationally demanding tasks like maximal clique, is the focus of $20\%$ of studies, as imbalanced workloads can severely impact overall performance. Moreover, $45\%$ of research focuses on reducing communication overhead due to frequent vertex-data communications. The simplicity of exchanged data results in \mbox{bandwidth limitations being a minor issue.}



\stitle{$k$-Core} of a graph, as defined in \cite{DBLP:journals/vldb/MalliarosGPV20}, is the maximal subgraph in which each vertex has at least a degree of $k$. The core value (i.e. coreness) of a vertex is the highest $k$ for which it is part of the $k$-core. The $k$-core decomposition algorithm is usually used to compute $k$-core, which involves iteratively computing these core values for all vertices, starting with a specified $k$ (typically the largest degree of the graph) and then decrementing $k$ in subsequent iterations.
The algorithm functions by ``deleting'' vertices (and the associated edges) with degrees less than the current $k$, while simultaneously updating the degrees of the remaining vertices.
This process continues until all vertices are assigned a core value, marking the completion of the computation.



Parallelism poses a significant challenge in $k$-core decomposition~ algorithms, particularly within the ``vertex-deletion'' approach. In this method, the deletion of a vertex necessitates the sequential updating of its neighbors' degrees, which complicates the efforts of parallelization. To tackle this issue, a series of studies~\cite{DBLP:journals/pvldb/LiaoLJHXC22,DBLP:journals/tpds/MontresorPM13,DBLP:journals/tpds/WengZLPL22,DBLP:journals/jpdc/ChanSS21,DBLP:conf/bigdataconf/MandalH17} have proposed a distributed $k$-core decomposition~ algorithm that decouples the ``deleting'' and ``updating'' phases. 
Initially, the deletion phase is parallelized, ignoring potential update conflicts. Then, an updating phase adjusts the degrees of vertices affected in the deletion phase.
It is important to note, however, that this distributed version of the \kcore~ algorithm may require more iterations (than the \mbox{sequential counterpart) to achieve convergence.}

To minimize communication overhead, The \subgraph~ approach is widely adopted, as highlighted in \cite{DBLP:journals/pvldb/LiaoLJHXC22,DBLP:journals/tpds/MontresorPM13}, as a natural fit for the distributed $k$-core decomposition~ algorithm as previously mentioned. Alongside the \subgraph~ model, the works in \cite{DBLP:journals/tpds/MontresorPM13, DBLP:journals/tpds/WengZLPL22} focus on reducing communication overhead using the \vertex model. In this approach, each vertex does not broadcast messages to all neighbors but selectively to those whose coreness values are anticipated to change. Moreover, T.-H. Hubert Chan, et al. \cite{DBLP:journals/jpdc/ChanSS21} devised an approximate $k$-core decomposition algorithm that imposes an iteration limit. They demonstrated that, for any $\gamma > 2$, $\left\lceil {\log n}/{\log(\frac{\gamma}{2})} \right\rceil$ iterations are sufficient to achieve an approximate core value for all vertices, at most $\gamma\times$ the actual value. Differing from the traditional ``vertex-deleting'' approach, this algorithm considers multiple thresholds in each iteration to expedite the process and removes vertices whose degree falls below the highest threshold. Furthermore, each vertex communicates only with a specific subset $\Lambda$ of vertices, which effectively reduces the communication volume to $\log_2 |\Lambda|$ bits each iteration. Upon receipt of a message, a subroutine is invoked to adjust $\Lambda$ if necessary.

\stitle{\ktruss}
is defined as a subgraph where each edge in the subgraph is part of at least $(k-2)$ triangles. Similar to the computation of a $k$-core, the process of determining a $k$-truss involves iteratively deleting edges that are part of fewer than $(k-1)$ triangles until convergence is reached. The time complexity of calculating a \ktruss~ is primarily constrained by the process of listing triangles, which, in the worst-case scenario, is $O(m^{1.5})$.

Similar to the computation of $k$-core, parallelizing the $k$-truss algorithm is challenging due to its inherently sequential nature. To address this, Pei-Ling Chen et al. \cite{DBLP:conf/bigdataconf/ChenCC14} proposed a distributed $k$-truss algorithm that splits the computation into two phases: ``deleting'' and ``updating'', aiding in parallel execution management.
Additionally, Yingxia Shao et al.~\cite{DBLP:conf/sigmod/ShaoCC14} introduced PETA, which constructs a Triangle-Complete (TC) subgraph for each vertex to enhance the $k$-truss algorithm's efficiency within its partition, minimizing communication during triangle counting.

Research also dives into load balancing. Following the initial placement of all vertices and their neighbor lists, KTMiner \cite{DBLP:journals/is/AlemiH19} involves strategically reorganizing and redistributing the vertices. This approach ensures even workload distribution \mbox{among all machines involved in the computation.}

Communication overhead remains a significant challenge in $k$-truss computations. 
Pei-Ling Chen et al. \cite{DBLP:conf/bigdataconf/ChenCC14} proposed an algorithm centered on the \edge~ model, which proves more apt for $k$-truss computations than the \vertex~ model. Their approach starts by transforming the original graph into a line graph, where each vertex represents an edge from the original graph. In this line graph, vertices are connected if their corresponding edges in the original graph share a common vertex.
To further minimize communication overhead, they introduced a pruning strategy: in the line graph, two vertices are adjacent only if their corresponding edges in the original graph form a triangle with another edge.
Moreover, KTMiner \cite{DBLP:journals/is/AlemiH19} enhances efficiency by pre-computing neighbor lists for each vertex and its neighbors, allowing triangles to be calculated locally. This method effectively eliminates the need for cross-machine communication between vertices.

\stitle{\maximalclique}
in a graph refers to a clique—a subgraph where each vertex is connected to all others—that cannot be expanded by adding another vertex.
Enumerating maximal cliques is a local task, allowing each vertex to independently participate without requiring intermediate results from others. This feature makes it suitable for parallel processing. However, its computational complexity requires $O(3^{n/3})$ in the worst case \cite{DBLP:conf/cocoon/TomitaTT04}, where $n$ represents the number of vertices.
Due to non-uniform subproblem sizes, and the unbalanced lengths of search paths in different subproblems, balancing the workload is a significant challenge many researchers focus on.

For workload balancing, Xu et al. \cite{DBLP:journals/tsc/XuCF16,DBLP:conf/bigdata/XuCFB14} initiate the process by sorting vertices using three distinct methods: degeneracy ordering, degree ordering, and core number ordering. Following this, the algorithm distributes the neighbor lists for each vertex. Specifically, a vertex $v$ and its corresponding neighbor list are allocated to the machine responsible for processing $v$. For each neighbor $u$ of $v$, if $u$ appears after $v$ in the ordering, the information about $v$ is also transmitted to the machine that processes $u$. This method ensures that the data is shared efficiently between machines, enhancing the overall execution of the task.
Another algorithm, PECO, proposed in \cite{DBLP:journals/jpdc/SvendsenMT15}, also adopts the strategy of sorting the vertices before distributing them to different machines. For sorting methods, degree ordering, triangle ordering, lexicographic ordering and random ordering are used and the experiment shows that the distribution of work is better with the degree ordering.

%% file: section/sec7_Traversal.tex
\traversal, a fundamental method in graph analysis, involves visiting each vertex or edge in a graph to identify paths, trees, or cycles. This process is crucial for uncovering the structural information of a graph. Each traversal algorithm, such as single-source shortest path (SSSP) for finding the shortest path and maximum flow for capacity planning in flow networks, serves specific applications. However, these algorithms all require global updates and synchronization during computation, posing significant challenges in distributed environments.

As indicated in \reffig{2a}, the studies on parallelism focus on identifying sub-structures that have strong dependencies among vertices during traversal, which account for $34.29\%$ of the research efforts in traversal.
Additionally, about $22.86\%$ of works consider load balancing, especially when the \subgraph~ model is adopted.
A significant number of papers ($37.14\%$) have contributed to the reduction of communication overhead in traversal: 
the inherent redundancy in visiting vertices and the need for transferring messages make it a communication-intensive task. 
Finally, There is little ($5.71\%$) research specifically targeting bandwidth optimization.

\stitle{Breadth-First-Search~(BFS)} is utilized to systematically visit all vertices starting from a specified source vertex, facing challenges due to irregular vertex access patterns despite its simplicity.


The nature of the workload in BFS processes, where a vertex becomes active only upon visitation, poses a challenge in maintaining load balance. To address this, BFS-4K~\cite{DBLP:journals/tpds/BusatoB15} employs a dynamic thread scheduling strategy for child kernels to effectively balance the workload. Additionally, a delegate-based approach~\cite{DBLP:conf/ppopp/CaoWWLMYC22} involves assigning delegates to high-degree vertices in different machines. This method ensures that access to these vertices primarily involves local communication, thereby reducing the reliance on more costly global communication.

To address the challenge of communication overhead in BFS, particularly due to redundant communications with already visited vertices, a direct solution is to ``mask'' these vertices, omitting them from future iterations as detailed in~\cite{DBLP:journals/jpdc/BernaschiCMV15}. Additionally, optimizing BFS based on its stages further reduces this overhead. Initially, when few vertices have been visited, a top-down (\push model) is more effective. As the number of visited vertices grows, a bottom-up (\pull model) approach becomes preferable to limit redundant communications. Identifying the ideal point to switch between these models is key, as explored in ~\cite{DBLP:journals/jpdc/RemisGAN18, DBLP:conf/ppopp/CaoWWLMYC22}.

\stitle{\sssp~(SSSP)} aims to find the shortest path from a source vertex to all others. 
The discussions of SSSP are predominantly on weighted graphs. While the sequential Bellman-Ford and Dijkstra algorithms are hard to parallelize, 
there are four major methodologies for implementing SSSP in distributed environments:

(1) Subgraph SSSP: Applying Dijkstra's algorithm on subgraphs within a distributed framework and exchanging distance messages between partitions~\cite{DBLP:conf/ppopp/MalekiNLGPP16}.

(2) $\Delta$-stepping: The $\Delta$-stepping method operates by selecting vertices within a specific distance range, denoted by $[k\Delta, (k{+}1)\Delta)$. This approach prioritizes the relaxation of shorter edges that fall below a threshold $\Delta$ until reaching convergence~\cite{DBLP:journals/tpds/ChakaravarthyCM17, DBLP:conf/sc/WangCMYC22}.

    \comment{
    In each step, Bellman-Ford and Dijkstra relax on all and only one vertex(vertices), respectively. Therefore, Bellman-Ford wastes too many redundant communications while Dijkstra does not take parallelism perfectly. $\Delta$-stepping, instead, combining these two algorithms that relax on a certain number of vertices. Specifically, $\Delta$-stepping will choose vertices whose current distance is in the range $[k\Delta, (k{+}1)\Delta)$~(called $\Delta$-bucket). Edges shorter than $\Delta$ within the bucket will relax first until convergence, and then relax remaining long edges at most once~\cite{DBLP:journals/tpds/ChakaravarthyCM17, DBLP:conf/sc/WangCMYC22}. Specifically, Bellman-Ford is $n$-stepping and Dijkstra is $1$-stepping.
    }

(3) Skeleton: These algorithms~\cite{DBLP:journals/dc/ChechikM22, DBLP:journals/siamcomp/HenzingerKN21, DBLP:conf/podc/CaoFR21, DBLP:journals/jacm/Elkin20a, DBLP:conf/focs/ForsterN18}, which offer faster processing with approximation, construct a ``skeleton'' graph (a subgraph) and compute ``hopsets'' to expedite finding the shortest paths. Skeleton-based algorithms compute distances from skeleton vertices to other graph vertices, adding as shortcuts into the original graph. These shortcuts can significantly decrease the number of hops (vertices) of a shortest path.

    \comment{
    Skeleton-based algorithms have a faster processing speed with approximation, which builds a skeleton graph on the original graph first with few vertices, and computes hopsets from skeleton vertices. With a high probability, a shortest path will pass through these skeleton vertices, so these hopsets can help to reduce iteration rounds~\cite{DBLP:journals/dc/ChechikM22, DBLP:journals/siamcomp/HenzingerKN21, DBLP:conf/podc/CaoFR21, DBLP:journals/jacm/Elkin20a, DBLP:conf/focs/ForsterN18}.
    }

(4) Transshipment: A broader approach that generalizes the SSSP task, often modeled as a linear programming problem~\cite{DBLP:journals/siamcomp/BeckerFKL21} or linear oblivious routing~\cite{DBLP:conf/soda/ZuzicGYHS22, DBLP:conf/stoc/RozhonGHZL22} for approximating distances.

    \comment{
    Transshipment is a generalization of the SSSP task also known as uncapacitated min-cost flow. A common approach is to further model this transshipment problem as linear programming~\cite{DBLP:journals/siamcomp/BeckerFKL21}. Some recent works focus on linear oblivious routing, i.e., a cost approximating matrix that approximates the shortest path distance for a given query~\cite{DBLP:conf/soda/ZuzicGYHS22, DBLP:conf/stoc/RozhonGHZL22}.
    }

Skeleton-based algorithms focus on achieving higher parallelism. In these algorithms, parallelism largely depends on the quality of the ``hopset''. Utilizing an approximate ``hopset'' with a shorter hop distance (the number of hops required between vertices) has proven more practical~\cite{DBLP:journals/siamcomp/HenzingerKN21}. CFR algorithms introduce a tradeoff parameter to balance hop distance, ``hopset'' size, and computation time~\cite{DBLP:conf/podc/CaoFR21}. Moreover, a two-level skeleton approach, which constructs an additional skeleton atop the original, achieves even greater efficiency, closely approaching theoretical limits~\cite{DBLP:journals/dc/ChechikM22}.

In the context of optimizing load imbalance, Delta-stepping algorithms skillfully integrate aspects of Bellman-Ford and Dijkstra for distributed environments. Nevertheless, they are susceptible to load unbalancing, particularly when high-degree vertices are processed simultaneously, leading to excessive communications. A practical solution~\cite{DBLP:journals/tpds/ChakaravarthyCM17} involves distributing these vertices across different machines. For updating long edges, instead of transmitting sparse pairs of delayed distance messages, a more efficient approach~\cite{DBLP:conf/sc/WangCMYC22} is to aggregate these as a dense distance vector. This uniform communication strategy not only streamlines the process but also aids in balancing the communication load across the system.

High communication overhead, a prevalent challenge in SSSP algorithms, results from extensive distance relaxing and updating processes. One approach to address this is dynamically switching between \push and \pull models at different stages, as proposed in \cite{DBLP:journals/tpds/ChakaravarthyCM17}. Another optimization involves using a hierarchical sliding window instead of a fixed $\Delta$-bucket to enhance communication efficiency~\cite{DBLP:conf/sc/WangCMYC22}. The Subgraph SSSP algorithm~\cite{DBLP:conf/ppopp/MalekiNLGPP16} reduces communication costs by computing partial shortest path distances within individual partitions. In transshipment algorithms, gradient descent techniques are employed to minimize communication by optimizing update directions~\cite{DBLP:journals/siamcomp/BeckerFKL21}.

\stitle{\mst~(MST)} is a connected subgraph that encompasses all $n$ vertices and $n-1$ edges, forming a tree. It is characterized by the minimum aggregated weight of its edges, representing the least costly way to connect all vertices in a graph. 
In distributed environments, many works are based on the GHS algorithm~\cite{DBLP:journals/toplas/GallagerHS83}, which involves three steps: (1) each vertex starts as a single-vertex tree (fragment); (2) each fragment selects the lightest edge connecting it to another fragment, merging them; (3) reducing the number of fragments by at least half each iteration, achieving a single fragment representing the MST after at most $O(\log{n})$ rounds.

Distributed MST algorithms, if not optimized, may incur significant communication overhead, particularly when a fragment's diameter greatly exceeds the graph's diameter. in a linear graph with $n$ vertices where the first vertex connects to all others (resulting in a diameter of $1$), choosing a line as a fragment increases its diameter to $n-1$. If a line is chosen as a fragment, its diameter becomes $n-1$. This scenario requires the GHS algorithm to perform $O(n)$ rounds for merging. The Controlled-GHS algorithm~\cite{DBLP:journals/siamcomp/GarayKP98} addresses this by controlling fragment diameter growth. It merges fragments using specific methods when a fragment is small or few fragments remain, or by merging close fragments to minimize communication. Besides, some approaches focus on creating a low-diameter spanning tree first, then \mbox{applying Controlled-GHS to maintain low-diameter growth~\cite{DBLP:journals/dc/MashreghiK21}.}


\stitle{\cycledetection} entails identifying cycles of a specified length in a graph. For smaller cycles, there are numerous effective algorithms. The task of finding a 3-cycle, commonly known as a triangle, typically involves checking for common neighbors between pairs of vertices on each edge. Similarly, detecting a 4-cycle, or square, usually requires examining common neighbors between pairs of vertices on all wedges (a wedge being a path of two edges). However, efficiently detecting larger cycles, namely $k$-cycle for arbitrary $k$, aiming to reduce communication costs and iteration rounds, continues to be a challenging area of research.

Cycle detection, often based on a BFS approach where two traversals run from a vertex until they meet, hinges on minimizing communication overhead. A notable strategy, as described in~\cite{DBLP:journals/dc/Censor-HillelFS19}, involves sparsification—running BFS on a subgraph with edges randomly removed to reduce communication and computation costs. However, this can cause false negatives if edges in a cycle are removed. To counter this, the algorithm is executed multiple times on varied sparsified subgraphs, with the results aggregated to identify cycles. An enhanced method in~\cite{DBLP:conf/spaa/FraigniaudO17} improves accuracy by selectively removing edges less likely to form cycles, balancing efficiency with the likelihood of detecting cycles.

\comment{
\noindent\underline{\textbf{\cycledetection}} task requires to detect a cycle with specific length in a graph.
There are many effective algorithms to find all small cycles. For example, finding connected neighbors for $3$-cycle (i.e., triangle) detection and sending entire neighborhood to all neighbors for $4$-cycle (i.e., square) detection. However, detecting a large cycle within lower communication costs and iteration rounds is still an open problem with gradually tighter bound~\cite{DBLP:conf/isaac/GallM21}.

Intuitively, BFS algorithms can search a cycle if two different paths meet at the same vertex. Since running multiple BFS simultaneously is expensive in communications and bandwidth, a practical method is to execute BFS on a sparser subgraph by deleting some edges. This algorithm may cause a false negative, i.e., breaking a cycle from the original graph. But if the error is bounded, one can execute this process for multiple times, called a cycle tester\update{~\cite{DBLP:journals/dc/Censor-HillelFS19}}. Another kind of tester chooses a candidate edge susceptible to belong to some cycles, and then checks whether a cycle passing through this edge~\cite{DBLP:conf/spaa/FraigniaudO17}.
}

\stitle{\maximumflow} in graph theory focuses on computing the maximum flow achievable from a source vertex $s$ to a sink vertex $t$ in a graph. Each edge in this graph is assigned a certain capacity that limits the flow through it. The core methodology for determining maximum flow involves repeatedly identifying paths through the graph—known as augmenting paths—that can still carry more flow. 
This augmentation process is conducted iteratively: after each traversal, the flow is augmented along the identified path, and the residual capacities are updated. This process continues until no further augmenting paths can be identified, \mbox{indicating that the maximum flow has been achieved.}

In tackling parallelism challenges within maximum flow algorithms, optimizing the efficiency of graph traversals is crucial, particularly since each augmentation process involves a traversal that can be computationally intensive. The primary optimization strategy in this context aims to maximize the number of paths augmented in a single iteration, thereby enhancing overall efficiency.
In line with this strategy, it has been proven more practical and efficient for identifying augmenting paths on a sparse subgraph. For example, the algorithm in \cite{DBLP:journals/jpdc/PeretzF22} employs a method that extracts a spanning tree from the graph. It then concurrently finds multiple augmenting paths within this sub-structure. This approach not only boosts parallelism but also significantly reduces the number of iterative rounds required. Furthermore, as discussed in the context of SSSP tasks, flow problems can be reformulated as linear optimization problems and tackled using methods like gradient descent. This perspective allows for alternative solutions, as proposed in \cite{DBLP:journals/siamcomp/GhaffariKKLP18}, which computes an approximate solution with a capacity penalty. Such an approach, focusing on approximation rather than exact solutions, enables the algorithm to achieve satisfactory results in fewer rounds.

%% file: section/sec_Pattern_Matching.tex


In graph analysis, the task of pattern matching involves finding all subgraphs within a large data graph $G$ that are isomorphic to a given small pattern graph $p$. 
Triangle counting helps understand the graph's fundamental structure, while $k$-Clique identifies larger, more complex specific patterns. Subgraph matching and mining offer broader applications by matching general patterns. These tasks are computationally intensive but only require local structural information, making them well-suited for distributed computing with large datasets.
A survey presented in \cite{DBLP:journals/csur/BouhenniYNK21} offers an in-depth overview of distributed subgraph matching algorithms, focusing on programming models. Building on this, our survey highlights the primary challenges addressed by these studies and includes the latest research not covered in the previous survey.

Numerous studies have transformed essential operations in pattern matching into binary join operations~\cite{DBLP:journals/pvldb/LaiQYJLWHLQZZQZ19}, which are also inherently amenable to parallel processing. As illustrated in~\reffig{distribution}, this has resulted in only a small portion of research ($2.22\%$) focusing on the challenges of parallelizing pattern matching. In contrast, the issues of load balancing ($35.56\%$) as well as communication overhead ($48.89\%$) have garnered significantly more attention. Interestingly, fewer papers focus on the issue of bandwidth constraints ($13.33\%$) in pattern matching. This oversight could be critical, considering the vast amounts of intermediate data that pattern matching processes can generate. 

\comment{
\patternmatching~ refers to finding all subgraphs within a given graph that match the given pattern.
The focus of \patternmatching~ is usually on the graph's local information rather than global that, which is easy for parallelizing computational tasks. Thus little research focus on the challenges of parallelization (2.13\%).
More significant are the challenges of communication overhead (51.06\%) and load balancing (34.04\%).
The main reason is that such compute-intensive tasks typically have high computational complexity, especially the \subgraphmatching, which is an NP-hard task. This leads to extensive inter-vertex information exchange in distributed systems.
We notice that there's a lack of research on bandwidth constraints in current papers. Compared to bandwidth constraints, the issue of communication overhead is more significant, and it remains a key challenge that most research aims to address.
}


\comment{
\begin{itemize}
\item \trianglecounting: Here, the pattern $p$ is a specific structure, namely a triangle. It is important to note that while \cycledetection and \kclique could also be considered specific forms of pattern matching, they have been previously discussed in sections \refsec{traversal} and \refsec{cohesive_subgraph}, respectively.

\item \subgraphmatching: In this variant, the pattern graph $p$ can be any arbitrary graph, though it is usually small in size. This is the most general form of \patternmatching, which is the core operation while querying modern graph databases~\cite{DBLP:journals/csur/AnglesABHRV17}.

\item \subgraphmining: This approach extends beyond the identification of a single pattern, focusing instead on computing a sequence of patterns. This method is particularly useful in the fields of bioinformatics~\cite{alon2008-ic} and social network analysis~\cite{yan_icdm04}.

\item \simulation: This variant differs from the conventional isomorphic approach to pattern matching. Instead, it utilizes simulation~\cite{milner_simulation} as the core matching semantic, offering an alternative perspective on how patterns can be identified and matched in complex graphs.

\end{itemize}

These variants collectively encompass a broad spectrum of \patternmatching~ applications, each with its unique challenges and methodologies.
}

\stitle{\trianglecounting}, as implied by its name, is a specialized variant of pattern matching where the specific pattern
$p$ under consideration is a triangle. It is important to note that while cycle detection and $k$-clique could also be considered specific forms of pattern matching, they have been previously discussed in sections \refsec{traversal} and \refsec{cohesive_subgraph}, respectively.
Algorithms for computing triangles generally fall into two main categories:
(1) List-based algorithms: These algorithms \cite{DBLP:conf/ipps/StrauszVGBH22, DBLP:conf/hpec/HoangJCADGP19, DBLP:conf/ipps/SandersU23, DBLP:journals/jacm/ChangPSZ21, DBLP:conf/podc/ChangS19, DBLP:conf/hpec/GhoshH20, DBLP:conf/sc/SteilRIPSP21} primarily focus on traversing adjacency lists of two vertices to identify their common neighbors. 
(2) Map-based algorithms: In contrast, these algorithms \cite{DBLP:conf/hpec/Ghosh22, DBLP:journals/tpds/PandeyWZTZLLHDL21, DBLP:journals/corr/ArifuzzamanKM17, DBLP:journals/tpds/YasarRBC22} employ an auxiliary data structure, such as a hash map, to maintain each vertex's adjacency list, which involves a vertex checking whether its neighbors \mbox{appear in the auxiliary structures of other neighbors.} 

Load balancing poses a significant challenge in both list-based and map-based algorithms, especially when dealing with power-law graphs characterized by uneven vertex degree distribution.
The key of achieving workload balance lies in effective graph partitioning strategies. In list-based algorithms, 1D partitioning \cite{DBLP:conf/ipps/StrauszVGBH22, DBLP:conf/hpec/Ghosh22, DBLP:conf/hpec/GhoshH20} is commonly employed, focusing on partitioning either vertices or edges. Conversely, 2D partitioning \cite{DBLP:journals/corr/ArifuzzamanKM17, DBLP:journals/tpds/YasarRBC22, DBLP:journals/tpds/PandeyWZTZLLHDL21} is more prevalent in map-based algorithms, wherein the data is divided across the graph's two-dimensional adjacency matrix.
Optimization strategies vary based on the partitioning method chosen. The study in \cite{DBLP:journals/corr/ArifuzzamanKM17} begins with generating an adjacency matrix from sorted vertices to distribute tasks evenly across machines. It utilizes cyclic distribution to balance light and heavy tasks.
Similarly, \cite{DBLP:journals/tpds/YasarRBC22}adapts this approach in heterogeneous distributed environments. TRUST \cite{DBLP:journals/tpds/PandeyWZTZLLHDL21} adopts a distinct 2D distribution strategy, partitioning vertices first based on their one-hop and then using these to form two-hop neighbor partitions, to ensure balanced load distribution.
In \cite{DBLP:conf/ipps/StrauszVGBH22}, a cyclic distribution method akin to \cite{DBLP:journals/corr/ArifuzzamanKM17} is implemented for distributing sorted vertices in a 1D partitioning scheme. Meanwhile, other 1D partitioning algorithms like \cite{DBLP:conf/hpec/GhoshH20, DBLP:conf/hpec/Ghosh22} strive for a more equitable distribution of \mbox{edges to improve overall load balance.}

Optimizing communication overhead is critical in triangle counting, given its communication-intensive nature \cite{DBLP:conf/sc/SteilRIPSP21}. The efficiency in distributed environments is significantly impacted by the frequency and volume of communications.
One effective strategy is message aggregation, where small messages are consolidated into larger ones for collective transmission, reducing communication frequency and potentially optimizing bandwidth~\cite{DBLP:conf/ipps/SandersU23, DBLP:conf/hpec/Ghosh22, DBLP:conf/hpec/GhoshH20, DBLP:conf/sc/SteilRIPSP21}. 
Expander decomposition is another approach used in several studies \cite{DBLP:conf/podc/ChangS19, DBLP:conf/hpec/Ghosh22, DBLP:journals/jacm/ChangPSZ21}.  It segments graphs into partitions of varying densities, assigning computational tasks based on these densities. High-density partitions are processed within single machines to minimize communication overhead.
In \cite{DBLP:conf/ipps/StrauszVGBH22},caching RMA accesses are studied to reduce remote data requests and communication demands.
DistTC \cite{DBLP:conf/hpec/HoangJCADGP19} introduces a fully asynchronous triangle counting algorithm, using boundary vertices and their connecting edges as proxies replicated across machines to enable independent triangle counting without inter-machine communication.
Finally, Ancy Sarah Tom et al. \cite{DBLP:journals/corr/ArifuzzamanKM17} implement a pruning technique. Specifically, during the intersection of two adjacency lists of vertices $u_k$ and $u_j$, only the triangle-closing vertices, $u_k$, satisfying $k > j$ are considered, \mbox{reducing unnecessary computations and communications.}

\stitle{\kclique}
is defined as a subgraph that contains $k$ vertices, with each pair of vertices being interconnected by an edge.
As a local task, the computation of $k$-clique is well-suited for parallel processing. However, the high computational complexity of $k$-clique computations, typically requiring $O(k^2n^k)$ time in single-machine environments as noted by Shahrivari and Jalili \cite{DBLP:journals/tkde/ShahrivariJ21}, poses a significant challenge. During the computation, each vertex is involved in sending/receiving substantial amounts of information to/from its neighbors. Consequently, key focuses in optimizing $k$-clique computations include achieving load balance, reducing communication, and managing bandwidth constraints. 


For workload balancing, the study by Censor-Hillel et al. \cite{DBLP:conf/podc/Censor-HillelLV22} introduces the generalized partition trees, which are constructed by a partial-pass streaming algorithm: it adds vertices one by one to the current partition, and either continues if the number of edges is within a threshold, or creates a new partition if otherwise.
This method is designed to ensure equitable workload distribution, with each machine processing a similar number of edges.


Many research efforts in the field of $k$-clique computation are directed toward reducing communication overhead. A notable example presented in~\cite{DBLP:conf/podc/Censor-HillelLV22} introduces a two-tiered communication granularity, comprising auxiliary tokens and main tokens. 
The core strategy of this approach is to prioritize the transmission of main tokens between vertices, resorting to the transmission of auxiliary tokens only when absolutely necessary. By adopting this method, the algorithm effectively reduces the overall volume of communication.
Additionally, the algorithm in \cite{DBLP:conf/podc/Censor-HillelGL20} contributes to reducing communication overhead by eliminating vertices with lower degrees and their connected edges in each iteration. 
A similar pruning strategy is employed in KCminer \cite{DBLP:journals/tkde/ShahrivariJ21}. In this approach, vertices with fewer than $k$ neighbors are discarded before computation begins.

Bandwidth constraints pose significant challenges in computing $k$-clique. Matthias Bonne et al. \cite{DBLP:conf/icalp/BonneC19} address this by sending only digests of neighbors to avoid network congestion.
Another approach \cite{DBLP:conf/podc/Censor-HillelGL20} defers processing edges likely to require substantial computations early on, reducing initial information transmission that can lead to network bottlenecks.
Later in computation, removing vertices and edges gradually decreases communicated data, easing bandwidth limitations.

\stitle{\subgraphmatching} is another variant of pattern matching.
In this variant, the pattern graph $p$ can be any arbitrary graph, though it is usually small in size. This is the most general form of pattern matching, which is the core operation while querying modern graph databases~\cite{DBLP:journals/csur/AnglesABHRV17}.

\comment{
The super-linear nature of \subgraphmatching~ often leads to significant load skew, making load balancing a crucial aspect of the task. Several strategies have been developed to address this challenge effectively.
One such strategy is dynamic scheduling, which allows for workload balancing by enabling workers to ``steal''
unprocessed tasks from others once they have completed their tasks. This approach is utilized in studies like \cite{DBLP:conf/eurosys/ChenLZYYC18} and \cite{DBLP:conf/sigmod/YangL0H021}, enhancing efficiency by redistributing workloads dynamically across machines.
Additionally, the study in \cite{DBLP:conf/eurosys/ChenLZYYC18} employs a static partitioning method known as ``BDG Partitioning''. This method maintains the locality of the graph within a subgraph by coloring vertices through a multi-source distributed Breadth-First Search (BFS) with limited steps. This process results in the formation of subgraphs that share the same color, helping to balance the load.
In \cite{DBLP:conf/icde/WangGHYH19}, task splitting is implemented for vertices whose degree exceeds a predefined threshold. This approach generates local search tasks and divides larger tasks into smaller subtasks, effectively distributing the workload more evenly.
Lizhi Xiang et al. \cite{DBLP:conf/sc/XiangKSHS21} address load imbalance caused by processing vertices in the order of their IDs. They introduce a strategy that randomizes task execution paths, thereby preventing skew due to sequential processing.
}

The super-linear nature of subgraph matching often leads to significant load skew, making load balancing a crucial aspect of the task. 
Static heuristics are typically leveraged for this issue. In \cite{DBLP:conf/icde/WangGHYH19}, task splitting is implemented for vertices whose degree exceeds a predefined threshold. This method divides large tasks into smaller subtasks, promoting more balanced workload distribution.
Lizhi Xiang et al. \cite{DBLP:conf/sc/XiangKSHS21} address load imbalance caused by processing vertices in the order of their degrees. They introduce a strategy that randomizes task execution paths, thereby preventing skew due to sequential processing.
Besides static heuristics, dynamic scheduling presents an effective solution for balancing workloads during runtime. One such approach permits workers to ``steal'' unprocessed tasks from their peers after completing their assigned tasks. Implemented in \cite{DBLP:conf/sigmod/YangL0H021}, this method significantly enhances efficiency by \mbox{dynamically redistributing workloads among machines.}

The task of subgraph matching tends to become communication-intensive~\cite{DBLP:journals/pvldb/LaiQYJLWHLQZZQZ19} in the distributed context, and thus reducing communication overhead is a critical challenge. One effective approach is to utilize an optimizer to devise an execution plan that minimizes this estimated communication cost~\cite{DBLP:journals/pvldb/LaiQYJLWHLQZZQZ19}.
Adopting this strategy, the study in \cite{DBLP:journals/pvldb/LaiQLC15} proposes estimating the cost on a random graph. Based on this estimation, they demonstrate that decomposing the pattern into wedges (intersections of two edges) and subsequently joining the matches of these wedges is an instance-optimal method in terms of communication cost.
The approach in \cite{DBLP:journals/pvldb/AmmarMSJ18} employs a subgraph matching approach based on the worst-case optimal join algorithm \cite{DBLP:journals/jacm/NgoPRR18}.
The GLogS system \cite{DBLP:conf/usenix/LaiYWLMSLZYQ00C23} refines cost estimation by computing smaller pattern sizes on a sparsified graph, improving the accuracy of communication cost estimates.
Caching accessed vertex data locally is another effective solution, as it reduces redundant communication that occurs when the same vertex data is fetched repeatedly \cite{DBLP:conf/icde/WangGHYH19,DBLP:conf/sigmod/YangL0H021}. This method ensures that frequently accessed data is readily available, thereby minimizing unnecessary data transfers.
Yang et al. \cite{DBLP:conf/sigmod/YangL0H021} dynamically select between \push and \pull communication strategies to further reduce communication costs based on the context.
Furthermore, the study in~\cite{DBLP:conf/sc/XiangKSHS21} introduces a new data structure that facilitates the reuse of prefix paths. This structure significantly reduces data redundancy and lessens the volume of data transmitted between machines, contributing to more efficient communication in distributed subgraph matching.

\stitle{\subgraphmining}
goes a step further than merely identifying a single pattern; it involves the computation of a sequence of interconnected patterns. 
Essentially, the task of subgraph mining can be viewed as a series of subgraph matching~ processes. Consequently, the challenges encountered in subgraph mining closely mirror those in subgraph matching.

Achieving workload balance is critical in subgraph mining, prompting the development of various static and dynamic strategies.
Statically, the study in \cite{DBLP:conf/eurosys/ChenLZYYC18} introduces ``BDG Partitioning'' as a method for maintaining graph locality within subgraphs. This technique use multi-source distributed BFS with limited depth to color vertices and form subgraphs with balanced computational loads.
Dynamically, Arabesque, as presented in \cite{DBLP:conf/sosp/TeixeiraFSSZA15}, adopts a round-robin strategy for workload distribution. This approach sequentially and cyclically allocates tasks to machines, inherently balancing the load due to the randomness in task assignment.
Another dynamic approach is detailed in \cite{DBLP:conf/asplos/ChenQ23a}, where workloads are dynamically dispatched. This method involves maintaining a workload queue, into which intermediate results from completed tasks are subdivided into smaller tasks. These smaller tasks are then dynamically distributed to available working threads as needed.
G-Miner, also proposed in \cite{DBLP:conf/eurosys/ChenLZYYC18}, supports dynamic load balancing as well. It utilizes a strategy that allows workers to ``steal'' unprocessed tasks from others. This method ensures that all workers are actively engaged and effectively redistributes tasks to balance the overall workload across the system.

Subgraph mining necessitates strategies to minimize communication overhead. A widely adopted approach, similar to that of subgraph matching, involves caching frequently accessed vertex data~\cite{DBLP:conf/isca/ChenHXBCA21,DBLP:conf/eurosys/ChenLZYYC18,DBLP:conf/asplos/ChenQ23a}. This caching mechanism ensures that critical data is readily available, thereby reducing repetitive data requests across the network.
In addition to caching, the research in \cite{DBLP:conf/eurosys/ChenLZYYC18} introduces a task priority queue to organize tasks with common remote candidates. This grouping allows related data to be transferred in a single operation, reducing the frequency and volume of remote data fetches.
Furthermore, the study in \cite{DBLP:journals/datamine/TalukderZ16} explores the use of a personalized broadcast operation. This method involves selectively assigning data to relevant machines based on their specific needs or the tasks they are performing, aiming at minimizing unnecessary communication. 

%% file: section/sec8_Covering.tex
Covering tasks comprise a category of optimization problems centered around identifying sets of vertices or edges that adhere to specific constraints.
Minimum vertex covering seeks the smallest set of vertices covering all edges, maximum matching aims to find the largest set of non-overlapping edges, and graph coloring assigns colors to vertices such that no adjacent vertices share the same color, with each problem focusing on different aspects of graph structure and optimization.
These problems are NP-hard, making the algorithms computationally intensive and difficult to parallelize.

Consequently, over half of the reviewed papers focus on enhancing parallelism ($64.29\%$).
Additionally, the complexities of computation pose issues with load balancing, and $7.14\%$ of the surveyed papers target it.
Moreover, these algorithms often necessitate decisions based on neighboring vertices' data, making the reduction of communication overhead ($28.57\%$) another critical concern.

\stitle{\mvc} 
is challenging due to its NP-hard nature, which is further compounded in distributed environments.
Recent works~\cite{DBLP:conf/iscas/ChenL21} try to find a strict Nash Equilibrium state as an approximate, yet near-optimal, solution.
In this context, vertices act as strategic agents in a snowdrift game with adjacent vertices, representing a local optimization process. The game's strict Nash Equilibrium corresponds to a vertex cover, aligning the strategy with the overall objective.
This game-based algorithm is well-suited to the message-passing model, allowing each vertex to select a beneficial neighbor based on both past and present decisions through rational adjustments.

The level of parallelism achieved can significantly influence the rate of convergence.
With limited local storage, where a vertex can only retain information on one prior selection, the process may fail to reach a strict Nash Equilibrium and get stuck in a deadlock.
To overcome these limitations and expedite convergence, some algorithms employ randomization strategies—permitting vertices to select neighbors randomly instead of rationally, thereby introducing a chance to \textit{jump out} from the dead loop~\cite{DBLP:conf/iscas/ChenL21, DBLP:journals/chinaf/ChenL23}.
Further enhancements in parallelism can be realized by expanding the local memory capacity, allowing for the storage of additional previous choices~\cite{DBLP:journals/tsmc/SunQSCSWZ22}.

\stitle{\maximummatch} seeks the largest collection of edges in a graph where no two edges share a common vertex.
A traditional algorithm to find a maximum matching is the Hungary algorithm \cite{kuhn1955hungarian}, which iteratively searches for an augmenting path from each unmatched vertex.
An augmenting path, consisting of alternating matched and unmatched edges, is leveraged to increment the size of the matching.
The Hopcroft-Karp algorithm~\cite{DBLP:journals/fuin/Gabow17}, optimized for parallelism, can generate multiple augmenting paths within a single iteration.
While these algorithms are fundamentally designed for bipartite graphs, they can be adapted for use with general graphs.

The augmenting process is the bottleneck of the parallelism. An efficient augmenting path algorithm can reduce iteration rounds~\cite{DBLP:conf/wdag/AhmadiKO18}. Another algorithm is to locally simulate multiple augmenting paths, since these paths are not too long~\cite{DBLP:conf/icdcn/EvenMR15}. Recently, Sparse Matrix-Vector Multiplication (SpMV) has been well studied and the augmenting process can be rewritten by matrix operations~\cite{DBLP:conf/cluster/AzadB15}. The fast matrix calculation can significantly increase \mbox{parallelism with humongous computing resources.}

Another challenge is the high communication cost, stemming from the need for each augmenting path search to traverse the entire graph. However, since augmenting paths are typically short, operating on a strategically chosen subgraph with fewer vertices can be more efficient without compromising the quality of the result~\cite{DBLP:journals/dc/DaniGHP23}.

\stitle{\coloring} 
algorithms iteratively select an independent set (IS) for coloring. An IS is a set of vertices in which no two are adjacent, allowing for simultaneous conflict-free coloring. Each vertex in the IS greedily selects the lowest available color not used by its neighbors.
The IS can be obtained by some sampling methods~(e.g., Maximal Independent Set) or \mbox{priority methods.}

To enhance parallelism, one strategy is to identify a larger IS, coloring more vertices concurrently.
A vertex-cut-based algorithm partitions the graph into components by removing vertex-cuts, allowing IS algorithms to be efficiently applied to these smaller, more manageable components~\cite{DBLP:conf/icde/PengCHZXY16}.
Another approach optimizes the coloring priority scheme so that lower-priority vertices get colored earlier, reducing wait times for higher-priority decisions~\cite{DBLP:conf/ppopp/AlabandiPB20}.


Graph coloring algorithms often incur excessive communication overhead towards the end of the process due to increased dependency on a dwindling number of uncolored vertices.
A straightforward solution is to precompute frequently occurring structures, thereby coloring them in a single instance~\cite{DBLP:conf/icde/PengCHZXY16}, or to defer their coloring to a later stage~\cite{DBLP:conf/soda/ChechikM19}.
Building on this concept, a subgraph-based algorithm decomposes the main graph into several subgraphs, prioritizes the coloring of internal vertices, and subsequently resolves color conflicts that \mbox{arise among the boundary vertices~\cite{DBLP:conf/sc/BogleBDRS20}.}


%% file: section/applications.tex
\begin{figure}[t]
    \centering
    \includegraphics[width=0.93\textwidth]{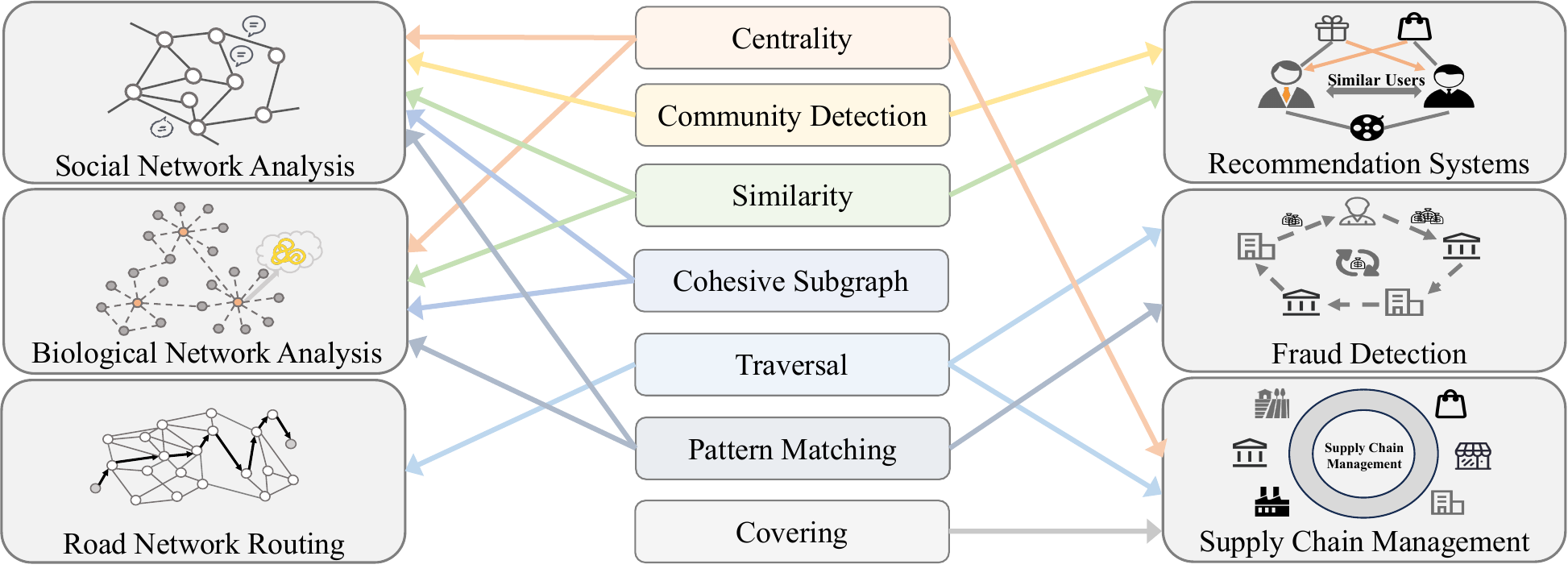}
    \caption{Application scenarios of graph tasks. Each arrow indicates that academic papers mention the application of graph tasks within the corresponding topic to the target scenarios.}
    \label{fig:app}
\end{figure}

Distributed graph tasks have numerous real-world applications, thanks to the expressive power of graphs and their ability to handle large-scale data efficiently in distributed environments. This section provides examples (Figure \ref{fig:app}), highlighting how distributed graph tasks are applied across various scenarios, as derived from applications mentioned in academic papers. Additionally, we provide descriptions of the performance differences between various graph tasks on certain topics and explain how these differences result in different application scenarios. 


\stitle{Centrality.} In social network analysis, PageRank and personalized PageRank identify influential individuals \cite{DBLP:conf/icml/Luo20,DBLP:conf/icassp/HeW21}. Closeness and betweenness centrality, calculated based on the shortest path, have lower efficiency but are more suitable for identifying key vertices in information transmission. For example, 
in biological network analysis, betweenness centrality identifies crucial genes or proteins that act as key intermediaries \cite{DBLP:journals/tcs/DurantW17}. In supply chain management, betweenness centrality finds important intermediaries in logistics pathways, improving efficiency and reducing costs \cite{DBLP:conf/ism2/WallmannG20}. Closeness centrality helps optimize logistics routes by identifying key vertices.

\stitle{Community Detection.} In {social network analysis}, these algorithms are used to identify densely connected communities \cite{DBLP:conf/cluster/ZengY18,azaouzi2017evidential,ma2018psplpa}. Users within the same community often share common interests, making community detection algorithms highly useful for {recommendation systems}. Louvain is an iterative method that optimizes modularity by progressively merging vertices and communities, ultimately forming a stable community structure. Due to its efficiency and scalability, Louvain is well-suited for handling large networks \cite{DBLP:conf/cluster/ZengY18,DBLP:journals/tjs/SattarA22}. LPA is fast but may produce unstable results in large networks, ideal for real-time data stream analysis \cite{DBLP:conf/hpec/0001FZHBLG19,DBLP:journals/concurrency/AttalMZ19,ma2018psplpa}. The connected components algorithm is low-cost and suitable for basic network analysis \cite{DBLP:journals/tpds/LulliCDLR17,DBLP:journals/tpds/JainFPGA17}.

\stitle{Similarity.} In {social network analysis}, similarity algorithms identify users with similar interests, helpful for making friends. In {biological networks}, it compares genes or proteins to find those with similar functions, aiding in research and drug development. In recommendation systems, it can be used to recommend other products similar to those a user has browsed or purchased, enhancing the user experience and increasing sales~\cite{DBLP:journals/corr/abs-2304-04015,DBLP:conf/icde/LuoGZY17}.

\stitle{Cohesive Subgraph.}
Cohesive subgraphs have significant applications in various networks. In social networks, they help identify tightly connected user groups, like interest-based communities~\cite{DBLP:conf/bigdataconf/MandalH17,DBLP:journals/jpdc/ChanSS21,DBLP:journals/pvldb/LiaoLJHXC22,DBLP:conf/sigmod/ShaoCC14}. In biological networks, they find gene groups with similar expressions, indicating potential functions \cite{DBLP:journals/tpds/MontresorPM13,DBLP:journals/jpdc/SvendsenMT15}. In e-commerce networks, they identify user groups with similar behavior, improving recommendation accuracy. Besides, different algorithms are also suited to some specific scenarios based on their trade-off between efficiency and subgraph density. For example, the maximum clique algorithm suits scenarios needing very tight connections but less time sensitivity, while $k$-Core works well for large networks due to its efficiency and scalability~\cite{DBLP:journals/pvldb/LiaoLJHXC22,DBLP:journals/tsc/XuCF16}.

\stitle{Traversal.} Traversal algorithms have diverse applications due to their different definitions. In road networks, BFS is used for simple tasks like finding the shortest path in an unweighted map~\cite{DBLP:journals/jpdc/DevismesJ16}, while SSSP handles complex networks, considering direction and traffic \cite{DBLP:journals/siamcomp/HenzingerKN21}. In supply chain management, the minimum spanning tree reduces total transportation costs, and maximum flow optimizes goods flow, increasing logistics throughput \cite{DBLP:journals/ton/JooLRS16}. Cycle detection identifies money laundering in financial networks and is used in operating systems to detect deadlocks~\cite{DBLP:conf/spaa/FraigniaudO17}.

\stitle{Pattern Matching.} In social network analysis, triangle counting evaluates community cohesion and aids clustering \cite{DBLP:journals/tpds/YasarRBC22,DBLP:conf/hpec/HoangJCADGP19,DBLP:journals/jacm/ChangPSZ21}, while $k$-Clique detection identifies tightly connected user groups~\cite{DBLP:journals/tkde/ShahrivariJ21}. In biological networks, $k$-Clique identifies protein interaction clusters, and subgraph matching locates specific specific gene clusters \cite{DBLP:journals/datamine/TalukderZ16,DBLP:conf/sosp/TeixeiraFSSZA15,DBLP:conf/eurosys/ChenLZYYC18}. In financial networks, subgraph matching and subgraph mining detect abnormal transaction patterns and potential fraudulent activities \cite{DBLP:conf/eurosys/ChenLZYYC18}.

\stitle{Covering.} The algorithms are commonly used for {optimization of resource allocation} and {supply chain management} \cite{DBLP:conf/wdag/AhmadiKO18}. Minimum vertex cover identifies key vertices to monitor, optimizing resource allocation \cite{DBLP:journals/chinaf/ChenL23}. For example, in sensor networks, minimum vertex cover helps in placing the least number of sensors to monitor all critical points. Maximum matching pairs vertices to maximize the number of edges, useful in optimizing resource distribution \cite{DBLP:conf/cluster/AzadB15}. Graph coloring assigns colors to vertices to minimize conflicts, essential in scheduling and resource allocation tasks \cite{DBLP:conf/sc/BogleBDRS20}.

%% file: section/sec10_Conclusion_and_Future_Directions.tex




This section aims to discuss prevalent research trends addressing various challenges in this domain, and offer perspectives on potential future research opportunities. 


\subsection{Computation Resource Efficiency} 
In the realm of distributed graph processing, the complete utilization of computation resources is identified as a primary objective. Numerous studies spanning a variety of topics have been conducted with the aim of optimizing resource utilization. Nevertheless, the effects of parallelism and load balancing have been found to vary significantly across these diverse topics.




\begin{itemize}
    \item \emph{Parallelism}: Paralleling graph tasks is the first step in distributed computing. It has been observed that not all topics focus on the challenges posed by parallelism. 
    Tasks that possess sequential dependencies, such as centrality, cohesive subgraph (like $k$-core and $k$-truss), and traversal, as well as those requiring global information like graph coloring, tend to allocate special attention to these challenges. 
    In such tasks, the state of each vertex in each step depends on the results of previous steps, or there is a necessity to consider the structure of the entire graph. 
    This makes it particularly challenging to circumvent conflicts and ensure the accuracy of algorithms under parallel computation. 
    Conversely, for tasks that are local in nature, such as similarity and pattern matching, parallelism is not the primary concern.

    \stitle{Opportunity.}  Current solutions aim to enhance the parallelism of algorithms but also introduce new challenges. In tasks such as $k$-core/$k$-truss, parallel computation of coreness/trussness might increase the number of iterations required for convergence. 
    In traversal and graph coloring tasks, additional computational costs are incurred to ensure algorithmic correctness when applied in distributed environments. 
    Furthermore, tasks adapting random walk-based algorithms can readily lead to local network congestion.
    It continues to be a significant challenge to parallelize algorithms while minimizing the negative impacts caused by such parallelization.


    \item \emph{Load Balance}: Load balancing is a prevalent challenge, with substantial research across all topics focusing on strategies to effectively balance workloads. Computation-intensive tasks, such as pattern matching and cohesive subgraph (like maximal clique), especially emphasize this challenge. Real-world graphs often exhibit a power-law degree distribution where the workload is typically proportional to the vertex degree. This results in a skewed workload distribution that is particularly problematic in computation-intensive tasks.

    \stitle{Opportunity.} The majority of existing work employs static partitioning strategies, which involve initially allocating a similar number of edges or vertices to each machine. However, a single static partitioning strategy may not be suitable for every type of graph data and algorithm, and it is challenging to ensure balanced workloads throughout the computation process. Developing dynamic load balance strategies is a valuable future direction.

\end{itemize}

\subsection{Network Resource Efficiency}

Our survey indicates that network resource efficiency, which encompasses communication overhead and bandwidth, follows a comparable pattern across different topics. 
There is a significant emphasis on minimizing communication overhead, yet bandwidth concerns receive comparatively less focus.

\begin{itemize}

    \item \emph{Communication Overhead}:  Communication overhead is a challenge that nearly all topics address, with more than half of the studies primarily dedicated to reducing it. Graph algorithms, particularly iterative ones, tend to produce substantial intermediate data throughout their execution. When vertices are distributed across separate machines, they require network communication to exchange data, making communication overhead a widespread issue in distributed computing.

    \stitle{Opportunity.} Minimizing communication overhead remains a critical challenge for many tasks within distributed settings. 
    Existing work typically employs one or two strategies to optimize communication overhead, such as the \subgraph~ model or pruning strategies. These methods are broadly applicable and generally suited for almost all tasks. Using multiple optimization techniques could further optimize communication overhead. 
    Moreover, the integration of deep learning techniques to predict and dynamically adjust communication patterns based on the current network state and workload distribution is also gaining increased attention.
    
    
    \item \emph{Bandwidth}: Relatively few studies concentrate on bandwidth issues; however, some computation-intensive tasks such as pattern matching, as well as algorithms utilizing random walks, have proposed methods to mitigate this challenge. 
    Bandwidth commonly is not a limiting factor for the performance of tasks that are not communication-intensive or when processing sparse graph data. Furthermore, strategies aimed at minimizing communication overhead and achieving load balance often help to alleviate bandwidth constraints.
    
    \stitle{Opportunity.} Nevertheless, as graph data scales continue to grow, the importance of bandwidth management is expected to escalate. Addressing this issue is inherently more complex than other challenges because it involves preventing network congestion while also maximizing bandwidth efficiency to avoid resource wastage. Future research will likely place increased emphasis on effectively tackling bandwidth challenges, potentially through the development of advanced network management techniques, like adaptive bandwidth allocation based on live traffic analysis.

\end{itemize}

%% file: section/conclusion.tex
Graphs can well represent relationships among entities. Analyzing and processing large-scale graph data has been applied in many applications, such as social network analysis, recommendation systems, and road network routing. Distributed graph processing provides a solution for efficiently handling large-scale graph data in the real world. 
To understand the state-of-the-art research of graph tasks in distributed environments and facilitate its development, in this paper, we conduct an extensive survey on distributed graph tasks. 

We first overview the existing distributed graph processing infrastructure. These tools facilitate the design of distributed algorithms, but it is still difficult to overcome the challenges arising from the inherent characteristics of distributed systems and graphs. We then analyze and summarize the main challenges and solutions for graph tasks in distributed environments.
Next, we provide a taxonomy of primary graph tasks and conduct a detailed analysis of their existing efforts on distributed environments, including challenges they focus on and unique insights for solving those challenges.
Finally, we discuss the research focus and the existing research gaps in the field of distributed graph processing and identify potential future research opportunities.


%% file: survey.bbl

\begin{thebibliography}{182}


\ifx \showCODEN    \undefined \def \showCODEN     #1{\unskip}     \fi
\ifx \showDOI      \undefined \def \showDOI       #1{#1}\fi
\ifx \showISBNx    \undefined \def \showISBNx     #1{\unskip}     \fi
\ifx \showISBNxiii \undefined \def \showISBNxiii  #1{\unskip}     \fi
\ifx \showISSN     \undefined \def \showISSN      #1{\unskip}     \fi
\ifx \showLCCN     \undefined \def \showLCCN      #1{\unskip}     \fi
\ifx \shownote     \undefined \def \shownote      #1{#1}          \fi
\ifx \showarticletitle \undefined \def \showarticletitle #1{#1}   \fi
\ifx \showURL      \undefined \def \showURL       {\relax}        \fi
\providecommand\bibfield[2]{#2}
\providecommand\bibinfo[2]{#2}
\providecommand\natexlab[1]{#1}
\providecommand\showeprint[2][]{arXiv:#2}

\bibitem[Abello et~al\mbox{.}(2006)]%
        {DBLP:journals/tvcg/AbelloHK06}
\bibfield{author}{\bibinfo{person}{James Abello}, \bibinfo{person}{Frank van Ham}, {and} \bibinfo{person}{Neeraj Krishnan}.} \bibinfo{year}{2006}\natexlab{}.
\newblock \showarticletitle{ASK-GraphView: {A} Large Scale Graph Visualization System}.
\newblock \bibinfo{journal}{\emph{TVCG}} \bibinfo{volume}{12}, \bibinfo{number}{5} (\bibinfo{year}{2006}), \bibinfo{pages}{669--676}.
\newblock


\bibitem[Ahmadi et~al\mbox{.}(2018)]%
        {DBLP:conf/wdag/AhmadiKO18}
\bibfield{author}{\bibinfo{person}{Mohamad Ahmadi}, \bibinfo{person}{Fabian Kuhn}, {and} \bibinfo{person}{Rotem Oshman}.} \bibinfo{year}{2018}\natexlab{}.
\newblock \showarticletitle{Distributed Approximate Maximum Matching in the {CONGEST} Model}. In \bibinfo{booktitle}{\emph{{DISC} 2018}} \emph{(\bibinfo{series}{LIPIcs}, Vol.~\bibinfo{volume}{121})}. \bibinfo{pages}{6:1--6:17}.
\newblock


\bibitem[Alabandi et~al\mbox{.}(2020)]%
        {DBLP:conf/ppopp/AlabandiPB20}
\bibfield{author}{\bibinfo{person}{Ghadeer Alabandi}, \bibinfo{person}{Evan Powers}, {and} \bibinfo{person}{Martin Burtscher}.} \bibinfo{year}{2020}\natexlab{}.
\newblock \showarticletitle{Increasing the parallelism of graph coloring via shortcutting}. In \bibinfo{booktitle}{\emph{PPoPP 2020}}. \bibinfo{pages}{262--275}.
\newblock


\bibitem[Alemi and Haghighi(2019)]%
        {DBLP:journals/is/AlemiH19}
\bibfield{author}{\bibinfo{person}{Mehdi Alemi} {and} \bibinfo{person}{Hassan Haghighi}.} \bibinfo{year}{2019}\natexlab{}.
\newblock \showarticletitle{KTMiner: distributed k-truss detection in big graphs}.
\newblock \bibinfo{journal}{\emph{Inf. Syst.}}  \bibinfo{volume}{83} (\bibinfo{year}{2019}), \bibinfo{pages}{195--216}.
\newblock


\bibitem[Ammar et~al\mbox{.}(2018)]%
        {DBLP:journals/pvldb/AmmarMSJ18}
\bibfield{author}{\bibinfo{person}{Khaled Ammar}, \bibinfo{person}{Frank McSherry}, \bibinfo{person}{Semih Salihoglu}, {and} \bibinfo{person}{Manas Joglekar}.} \bibinfo{year}{2018}\natexlab{}.
\newblock \showarticletitle{Distributed Evaluation of Subgraph Queries Using Worst-case Optimal and Low-Memory Dataflows}.
\newblock \bibinfo{journal}{\emph{Proc. {VLDB} Endow.}} \bibinfo{volume}{11}, \bibinfo{number}{6} (\bibinfo{year}{2018}), \bibinfo{pages}{691--704}.
\newblock


\bibitem[Angles et~al\mbox{.}(2017)]%
        {DBLP:journals/csur/AnglesABHRV17}
\bibfield{author}{\bibinfo{person}{Renzo Angles}, \bibinfo{person}{Marcelo Arenas}, \bibinfo{person}{Pablo Barcel{\'{o}}}, \bibinfo{person}{Aidan Hogan}, \bibinfo{person}{Juan~L. Reutter}, {and} \bibinfo{person}{Domagoj Vrgoc}.} \bibinfo{year}{2017}\natexlab{}.
\newblock \showarticletitle{Foundations of Modern Query Languages for Graph Databases}.
\newblock \bibinfo{journal}{\emph{{ACM} Comput. Surv.}} \bibinfo{volume}{50}, \bibinfo{number}{5} (\bibinfo{year}{2017}), \bibinfo{pages}{68:1--68:40}.
\newblock


\bibitem[Arifuzzaman et~al\mbox{.}(2017)]%
        {DBLP:journals/corr/ArifuzzamanKM17}
\bibfield{author}{\bibinfo{person}{Shaikh Arifuzzaman}, \bibinfo{person}{Maleq Khan}, {and} \bibinfo{person}{Madhav~V. Marathe}.} \bibinfo{year}{2017}\natexlab{}.
\newblock \showarticletitle{Distributed-Memory Parallel Algorithms for Counting and Listing Triangles in Big Graphs}.
\newblock \bibinfo{journal}{\emph{CoRR}}  \bibinfo{volume}{abs/1706.05151} (\bibinfo{year}{2017}).
\newblock
\showeprint[arXiv]{1706.05151}


\bibitem[Armstrong(2010)]%
        {armstrong2010erlang}
\bibfield{author}{\bibinfo{person}{Joe Armstrong}.} \bibinfo{year}{2010}\natexlab{}.
\newblock \showarticletitle{erlang}.
\newblock \bibinfo{journal}{\emph{Commun. ACM}} \bibinfo{volume}{53}, \bibinfo{number}{9} (\bibinfo{year}{2010}), \bibinfo{pages}{68--75}.
\newblock


\bibitem[Asuncion et~al\mbox{.}(2008)]%
        {DBLP:conf/nips/AsuncionSW08}
\bibfield{author}{\bibinfo{person}{Arthur~U. Asuncion}, \bibinfo{person}{Padhraic Smyth}, {and} \bibinfo{person}{Max Welling}.} \bibinfo{year}{2008}\natexlab{}.
\newblock \showarticletitle{Asynchronous Distributed Learning of Topic Models}. In \bibinfo{booktitle}{\emph{{NIPS} 2008}}. \bibinfo{publisher}{Curran Associates, Inc.}, \bibinfo{pages}{81--88}.
\newblock


\bibitem[Attal et~al\mbox{.}(2019)]%
        {DBLP:journals/concurrency/AttalMZ19}
\bibfield{author}{\bibinfo{person}{Jean{-}Philippe Attal}, \bibinfo{person}{Maria Malek}, {and} \bibinfo{person}{Marc Zolghadri}.} \bibinfo{year}{2019}\natexlab{}.
\newblock \showarticletitle{Parallel and distributed core label propagation with graph coloring}.
\newblock \bibinfo{journal}{\emph{Concurr. Comput. Pract. Exp.}} \bibinfo{volume}{31}, \bibinfo{number}{2} (\bibinfo{year}{2019}), \bibinfo{pages}{e4355}.
\newblock


\bibitem[Avery(2011)]%
        {avery2011giraph}
\bibfield{author}{\bibinfo{person}{Ching Avery}.} \bibinfo{year}{2011}\natexlab{}.
\newblock \showarticletitle{Giraph: Large-scale graph processing infrastructure on hadoop}.
\newblock \bibinfo{journal}{\emph{Proceedings of the Hadoop Summit. Santa Clara}} \bibinfo{volume}{11}, \bibinfo{number}{3} (\bibinfo{year}{2011}), \bibinfo{pages}{5--9}.
\newblock


\bibitem[Azad and Bulu{\c{c}}(2015)]%
        {DBLP:conf/cluster/AzadB15}
\bibfield{author}{\bibinfo{person}{Ariful Azad} {and} \bibinfo{person}{Aydin Bulu{\c{c}}}.} \bibinfo{year}{2015}\natexlab{}.
\newblock \showarticletitle{Distributed-Memory Algorithms for Maximal Cardinality Matching Using Matrix Algebra}. In \bibinfo{booktitle}{\emph{{CLUSTER} 2015}}. \bibinfo{pages}{398--407}.
\newblock


\bibitem[Azaouzi and Romdhane(2017)]%
        {azaouzi2017evidential}
\bibfield{author}{\bibinfo{person}{Mehdi Azaouzi} {and} \bibinfo{person}{Lotfi~Ben Romdhane}.} \bibinfo{year}{2017}\natexlab{}.
\newblock \showarticletitle{An evidential influence-based label propagation algorithm for distributed community detection in social networks}.
\newblock \bibinfo{journal}{\emph{Procedia computer science}}  \bibinfo{volume}{112} (\bibinfo{year}{2017}), \bibinfo{pages}{407--416}.
\newblock


\bibitem[Badam and Simmhan(2014)]%
        {DBLP:conf/comad/BadamS14}
\bibfield{author}{\bibinfo{person}{Nitin~Chandra Badam} {and} \bibinfo{person}{Yogesh Simmhan}.} \bibinfo{year}{2014}\natexlab{}.
\newblock \showarticletitle{Subgraph Rank: PageRank for Subgraph-Centric Distributed Graph Processing}. In \bibinfo{booktitle}{\emph{{COMAD} 2014}}. \bibinfo{pages}{38--49}.
\newblock


\bibitem[Badia et~al\mbox{.}(2022)]%
        {DBLP:journals/cse/BadiaCELL22}
\bibfield{author}{\bibinfo{person}{Rosa~M. Badia}, \bibinfo{person}{Javier Conejero}, \bibinfo{person}{Jorge Ejarque}, \bibinfo{person}{Daniele Lezzi}, {and} \bibinfo{person}{Francesc Lordan}.} \bibinfo{year}{2022}\natexlab{}.
\newblock \showarticletitle{PyCOMPSs as an Instrument for Translational Computer Science}.
\newblock \bibinfo{journal}{\emph{Comput. Sci. Eng.}} \bibinfo{volume}{24}, \bibinfo{number}{2} (\bibinfo{year}{2022}), \bibinfo{pages}{79--84}.
\newblock


\bibitem[Bal et~al\mbox{.}(1989)]%
        {DBLP:journals/csur/BalST89}
\bibfield{author}{\bibinfo{person}{Henri~E. Bal}, \bibinfo{person}{Jennifer~G. Steiner}, {and} \bibinfo{person}{Andrew~S. Tanenbaum}.} \bibinfo{year}{1989}\natexlab{}.
\newblock \showarticletitle{Programming Languages for Distributed Computing Systems}.
\newblock \bibinfo{journal}{\emph{{ACM} Comput. Surv.}} \bibinfo{volume}{21}, \bibinfo{number}{3} (\bibinfo{year}{1989}), \bibinfo{pages}{261--322}.
\newblock


\bibitem[Becker et~al\mbox{.}(2021)]%
        {DBLP:journals/siamcomp/BeckerFKL21}
\bibfield{author}{\bibinfo{person}{Ruben Becker}, \bibinfo{person}{Sebastian Forster}, \bibinfo{person}{Andreas Karrenbauer}, {and} \bibinfo{person}{Christoph Lenzen}.} \bibinfo{year}{2021}\natexlab{}.
\newblock \showarticletitle{Near-Optimal Approximate Shortest Paths and Transshipment in Distributed and Streaming Models}.
\newblock \bibinfo{journal}{\emph{{SIAM} J. Comput.}} \bibinfo{volume}{50}, \bibinfo{number}{3} (\bibinfo{year}{2021}), \bibinfo{pages}{815--856}.
\newblock


\bibitem[Bernaschi et~al\mbox{.}(2015)]%
        {DBLP:journals/jpdc/BernaschiCMV15}
\bibfield{author}{\bibinfo{person}{Massimo Bernaschi}, \bibinfo{person}{Giancarlo Carbone}, \bibinfo{person}{Enrico Mastrostefano}, {and} \bibinfo{person}{Flavio Vella}.} \bibinfo{year}{2015}\natexlab{}.
\newblock \showarticletitle{Solutions to the st-connectivity problem using a GPU-based distributed BFS}.
\newblock \bibinfo{journal}{\emph{J. Parallel and Distrib. Comput.}}  \bibinfo{volume}{76} (\bibinfo{year}{2015}), \bibinfo{pages}{145--153}.
\newblock


\bibitem[Besta et~al\mbox{.}(2020)]%
        {DBLP:conf/ipps/BestaKMKRHS20}
\bibfield{author}{\bibinfo{person}{Maciej Besta}, \bibinfo{person}{Raghavendra Kanakagiri}, \bibinfo{person}{Harun Mustafa}, \bibinfo{person}{Mikhail Karasikov}, \bibinfo{person}{Gunnar R{\"{a}}tsch}, \bibinfo{person}{Torsten Hoefler}, {and} \bibinfo{person}{Edgar Solomonik}.} \bibinfo{year}{2020}\natexlab{}.
\newblock \showarticletitle{Communication-Efficient Jaccard similarity for High-Performance Distributed Genome Comparisons}. In \bibinfo{booktitle}{\emph{{IPDPS} 2020}}. \bibinfo{pages}{1122--1132}.
\newblock


\bibitem[Blondel et~al\mbox{.}(2008)]%
        {blondel2008fast}
\bibfield{author}{\bibinfo{person}{Vincent~D Blondel}, \bibinfo{person}{Jean-Loup Guillaume}, \bibinfo{person}{Renaud Lambiotte}, {and} \bibinfo{person}{Etienne Lefebvre}.} \bibinfo{year}{2008}\natexlab{}.
\newblock \showarticletitle{Fast unfolding of communities in large networks}.
\newblock \bibinfo{journal}{\emph{Journal of statistical mechanics: theory and experiment}} \bibinfo{volume}{2008}, \bibinfo{number}{10} (\bibinfo{year}{2008}), \bibinfo{pages}{P10008}.
\newblock


\bibitem[Bogle et~al\mbox{.}(2020)]%
        {DBLP:conf/sc/BogleBDRS20}
\bibfield{author}{\bibinfo{person}{Ian Bogle}, \bibinfo{person}{Erik~G. Boman}, \bibinfo{person}{Karen~D. Devine}, \bibinfo{person}{Sivasankaran Rajamanickam}, {and} \bibinfo{person}{George~M. Slota}.} \bibinfo{year}{2020}\natexlab{}.
\newblock \showarticletitle{Distributed Memory Graph Coloring Algorithms for Multiple GPUs}. In \bibinfo{booktitle}{\emph{{IA3} 2020}}. \bibinfo{pages}{54--62}.
\newblock


\bibitem[Bonne and Censor{-}Hillel(2019)]%
        {DBLP:conf/icalp/BonneC19}
\bibfield{author}{\bibinfo{person}{Matthias Bonne} {and} \bibinfo{person}{Keren Censor{-}Hillel}.} \bibinfo{year}{2019}\natexlab{}.
\newblock \showarticletitle{Distributed Detection of Cliques in Dynamic Networks}. In \bibinfo{booktitle}{\emph{{ICALP} 2019}} \emph{(\bibinfo{series}{LIPIcs}, Vol.~\bibinfo{volume}{132})}. \bibinfo{pages}{132:1--132:15}.
\newblock


\bibitem[Borgwardt et~al\mbox{.}(2005)]%
        {DBLP:conf/ismb/BorgwardtOSVSK05}
\bibfield{author}{\bibinfo{person}{Karsten~M. Borgwardt}, \bibinfo{person}{Cheng~Soon Ong}, \bibinfo{person}{Stefan Sch{\"{o}}nauer}, \bibinfo{person}{S.~V.~N. Vishwanathan}, \bibinfo{person}{Alexander~J. Smola}, {and} \bibinfo{person}{Hans{-}Peter Kriegel}.} \bibinfo{year}{2005}\natexlab{}.
\newblock \showarticletitle{Protein function prediction via graph kernels}. In \bibinfo{booktitle}{\emph{ISMB 2005}}. \bibinfo{pages}{47--56}.
\newblock


\bibitem[Bouhenni et~al\mbox{.}(2022)]%
        {DBLP:journals/csur/BouhenniYNK21}
\bibfield{author}{\bibinfo{person}{Sarra Bouhenni}, \bibinfo{person}{Sa{\"{\i}}d Yahiaoui}, \bibinfo{person}{Nadia Nouali{-}Taboudjemat}, {and} \bibinfo{person}{Hamamache Kheddouci}.} \bibinfo{year}{2022}\natexlab{}.
\newblock \showarticletitle{A Survey on Distributed Graph Pattern Matching in Massive Graphs}.
\newblock \bibinfo{journal}{\emph{{ACM} Comput. Surv.}} \bibinfo{volume}{54}, \bibinfo{number}{2} (\bibinfo{year}{2022}), \bibinfo{pages}{36:1--36:35}.
\newblock


\bibitem[Brandes(2001)]%
        {Brandes2001Faster}
\bibfield{author}{\bibinfo{person}{Ulrik Brandes}.} \bibinfo{year}{2001}\natexlab{}.
\newblock \showarticletitle{A faster algorithm for betweenness centrality}.
\newblock \bibinfo{journal}{\emph{Journal of mathematical sociology}} \bibinfo{volume}{25}, \bibinfo{number}{2}, \bibinfo{pages}{163--177}.
\newblock


\bibitem[Brin and Page(1998)]%
        {pagerank}
\bibfield{author}{\bibinfo{person}{Sergey Brin} {and} \bibinfo{person}{Lawrence Page}.} \bibinfo{year}{1998}\natexlab{}.
\newblock \showarticletitle{The anatomy of a large-scale hypertextual web search engine}.
\newblock \bibinfo{journal}{\emph{Computer networks and ISDN systems}} \bibinfo{volume}{30}, \bibinfo{number}{1-7} (\bibinfo{year}{1998}), \bibinfo{pages}{107--117}.
\newblock


\bibitem[Broder et~al\mbox{.}(2000)]%
        {DBLP:journals/cn/BroderKMRRSTW00}
\bibfield{author}{\bibinfo{person}{Andrei~Z. Broder}, \bibinfo{person}{Ravi Kumar}, \bibinfo{person}{Farzin Maghoul}, \bibinfo{person}{Prabhakar Raghavan}, \bibinfo{person}{Sridhar Rajagopalan}, \bibinfo{person}{Raymie Stata}, \bibinfo{person}{Andrew Tomkins}, {and} \bibinfo{person}{Janet~L. Wiener}.} \bibinfo{year}{2000}\natexlab{}.
\newblock \showarticletitle{Graph structure in the Web}.
\newblock \bibinfo{journal}{\emph{Comput. Networks}} \bibinfo{volume}{33}, \bibinfo{number}{1-6} (\bibinfo{year}{2000}), \bibinfo{pages}{309--320}.
\newblock


\bibitem[Busato and Bombieri(2015)]%
        {DBLP:journals/tpds/BusatoB15}
\bibfield{author}{\bibinfo{person}{Federico Busato} {and} \bibinfo{person}{Nicola Bombieri}.} \bibinfo{year}{2015}\natexlab{}.
\newblock \showarticletitle{{BFS-4K:} An Efficient Implementation of {BFS} for Kepler {GPU} Architectures}.
\newblock \bibinfo{journal}{\emph{{IEEE} Trans. Parallel Distributed Syst.}} \bibinfo{volume}{26}, \bibinfo{number}{7} (\bibinfo{year}{2015}), \bibinfo{pages}{1826--1838}.
\newblock


\bibitem[Busato and Bombieri(2016)]%
        {DBLP:journals/tpds/BusatoB16}
\bibfield{author}{\bibinfo{person}{Federico Busato} {and} \bibinfo{person}{Nicola Bombieri}.} \bibinfo{year}{2016}\natexlab{}.
\newblock \showarticletitle{An Efficient Implementation of the Bellman-Ford Algorithm for Kepler {GPU} Architectures}.
\newblock \bibinfo{journal}{\emph{{IEEE} Trans. Parallel Distributed Syst.}} \bibinfo{volume}{27}, \bibinfo{number}{8} (\bibinfo{year}{2016}), \bibinfo{pages}{2222--2233}.
\newblock


\bibitem[Cao et~al\mbox{.}(2022)]%
        {DBLP:conf/ppopp/CaoWWLMYC22}
\bibfield{author}{\bibinfo{person}{Huanqi Cao}, \bibinfo{person}{Yuanwei Wang}, \bibinfo{person}{Haojie Wang}, \bibinfo{person}{Heng Lin}, \bibinfo{person}{Zixuan Ma}, \bibinfo{person}{Wanwang Yin}, {and} \bibinfo{person}{Wenguang Chen}.} \bibinfo{year}{2022}\natexlab{}.
\newblock \showarticletitle{Scaling graph traversal to 281 trillion edges with 40 million cores}. In \bibinfo{booktitle}{\emph{PPoPP 2022}}. \bibinfo{pages}{234--245}.
\newblock


\bibitem[Cao et~al\mbox{.}(2021)]%
        {DBLP:conf/podc/CaoFR21}
\bibfield{author}{\bibinfo{person}{Nairen Cao}, \bibinfo{person}{Jeremy~T. Fineman}, {and} \bibinfo{person}{Katina Russell}.} \bibinfo{year}{2021}\natexlab{}.
\newblock \showarticletitle{Brief Announcement: An Improved Distributed Approximate Single Source Shortest Paths Algorithm}. In \bibinfo{booktitle}{\emph{{PODC} 2021}}. \bibinfo{pages}{493--496}.
\newblock


\bibitem[Carbone et~al\mbox{.}(2015)]%
        {DBLP:journals/debu/CarboneKEMHT15}
\bibfield{author}{\bibinfo{person}{Paris Carbone}, \bibinfo{person}{Asterios Katsifodimos}, \bibinfo{person}{Stephan Ewen}, \bibinfo{person}{Volker Markl}, \bibinfo{person}{Seif Haridi}, {and} \bibinfo{person}{Kostas Tzoumas}.} \bibinfo{year}{2015}\natexlab{}.
\newblock \showarticletitle{Apache Flink{\texttrademark}: Stream and Batch Processing in a Single Engine}.
\newblock \bibinfo{journal}{\emph{{IEEE} Data Eng. Bull.}} \bibinfo{volume}{38}, \bibinfo{number}{4} (\bibinfo{year}{2015}), \bibinfo{pages}{28--38}.
\newblock


\bibitem[Censor{-}Hillel et~al\mbox{.}(2019)]%
        {DBLP:journals/dc/Censor-HillelFS19}
\bibfield{author}{\bibinfo{person}{Keren Censor{-}Hillel}, \bibinfo{person}{Eldar Fischer}, \bibinfo{person}{Gregory Schwartzman}, {and} \bibinfo{person}{Yadu Vasudev}.} \bibinfo{year}{2019}\natexlab{}.
\newblock \showarticletitle{Fast distributed algorithms for testing graph properties}.
\newblock \bibinfo{journal}{\emph{Distributed Comput.}} \bibinfo{volume}{32}, \bibinfo{number}{1} (\bibinfo{year}{2019}), \bibinfo{pages}{41--57}.
\newblock


\bibitem[Censor{-}Hillel et~al\mbox{.}(2020)]%
        {DBLP:conf/podc/Censor-HillelGL20}
\bibfield{author}{\bibinfo{person}{Keren Censor{-}Hillel}, \bibinfo{person}{Fran{\c{c}}ois~Le Gall}, {and} \bibinfo{person}{Dean Leitersdorf}.} \bibinfo{year}{2020}\natexlab{}.
\newblock \showarticletitle{On Distributed Listing of Cliques}. In \bibinfo{booktitle}{\emph{{PODC} 2020}}. \bibinfo{pages}{474--482}.
\newblock


\bibitem[Censor{-}Hillel et~al\mbox{.}(2022)]%
        {DBLP:conf/podc/Censor-HillelLV22}
\bibfield{author}{\bibinfo{person}{Keren Censor{-}Hillel}, \bibinfo{person}{Dean Leitersdorf}, {and} \bibinfo{person}{David Vulakh}.} \bibinfo{year}{2022}\natexlab{}.
\newblock \showarticletitle{Deterministic Near-Optimal Distributed Listing of Cliques}. In \bibinfo{booktitle}{\emph{{PODC} 2022}}. \bibinfo{pages}{271--280}.
\newblock


\bibitem[Chakaravarthy et~al\mbox{.}(2017)]%
        {DBLP:journals/tpds/ChakaravarthyCM17}
\bibfield{author}{\bibinfo{person}{Venkatesan~T. Chakaravarthy}, \bibinfo{person}{Fabio Checconi}, \bibinfo{person}{Prakash Murali}, \bibinfo{person}{Fabrizio Petrini}, {and} \bibinfo{person}{Yogish Sabharwal}.} \bibinfo{year}{2017}\natexlab{}.
\newblock \showarticletitle{Scalable Single Source Shortest Path Algorithms for Massively Parallel Systems}.
\newblock \bibinfo{journal}{\emph{{IEEE} Trans. Parallel Distributed Syst.}} \bibinfo{volume}{28}, \bibinfo{number}{7} (\bibinfo{year}{2017}), \bibinfo{pages}{2031--2045}.
\newblock


\bibitem[Chan et~al\mbox{.}(2021)]%
        {DBLP:journals/jpdc/ChanSS21}
\bibfield{author}{\bibinfo{person}{T-H~Hubert Chan}, \bibinfo{person}{Mauro Sozio}, {and} \bibinfo{person}{Bintao Sun}.} \bibinfo{year}{2021}\natexlab{}.
\newblock \showarticletitle{Distributed approximate k-core decomposition and min--max edge orientation: Breaking the diameter barrier}.
\newblock \bibinfo{journal}{\emph{J. Parallel and Distrib. Comput.}}  \bibinfo{volume}{147} (\bibinfo{year}{2021}), \bibinfo{pages}{87--99}.
\newblock


\bibitem[Chang et~al\mbox{.}(2021)]%
        {DBLP:journals/jacm/ChangPSZ21}
\bibfield{author}{\bibinfo{person}{Yi{-}Jun Chang}, \bibinfo{person}{Seth Pettie}, \bibinfo{person}{Thatchaphol Saranurak}, {and} \bibinfo{person}{Hengjie Zhang}.} \bibinfo{year}{2021}\natexlab{}.
\newblock \showarticletitle{Near-optimal Distributed Triangle Enumeration via Expander Decompositions}.
\newblock \bibinfo{journal}{\emph{J. {ACM}}} \bibinfo{volume}{68}, \bibinfo{number}{3} (\bibinfo{year}{2021}), \bibinfo{pages}{21:1--21:36}.
\newblock


\bibitem[Chang and Saranurak(2019)]%
        {DBLP:conf/podc/ChangS19}
\bibfield{author}{\bibinfo{person}{Yi{-}Jun Chang} {and} \bibinfo{person}{Thatchaphol Saranurak}.} \bibinfo{year}{2019}\natexlab{}.
\newblock \showarticletitle{Improved Distributed Expander Decomposition and Nearly Optimal Triangle Enumeration}. In \bibinfo{booktitle}{\emph{{PODC} 2019}}. \bibinfo{pages}{66--73}.
\newblock


\bibitem[Chapman et~al\mbox{.}(2010)]%
        {DBLP:conf/pgas/ChapmanCPPKKS10}
\bibfield{author}{\bibinfo{person}{Barbara~M. Chapman}, \bibinfo{person}{Tony Curtis}, \bibinfo{person}{Swaroop Pophale}, \bibinfo{person}{Stephen~W. Poole}, \bibinfo{person}{Jeffery~A. Kuehn}, \bibinfo{person}{Chuck Koelbel}, {and} \bibinfo{person}{Lauren Smith}.} \bibinfo{year}{2010}\natexlab{}.
\newblock \showarticletitle{Introducing OpenSHMEM: {SHMEM} for the {PGAS} community}. In \bibinfo{booktitle}{\emph{{PGAS} 2010}}. \bibinfo{publisher}{{ACM}}, \bibinfo{pages}{2}.
\newblock


\bibitem[Charles et~al\mbox{.}(2005)]%
        {DBLP:conf/oopsla/CharlesGSDKEPS05}
\bibfield{author}{\bibinfo{person}{Philippe Charles}, \bibinfo{person}{Christian Grothoff}, \bibinfo{person}{Vijay~A. Saraswat}, \bibinfo{person}{Christopher Donawa}, \bibinfo{person}{Allan Kielstra}, \bibinfo{person}{Kemal Ebcioglu}, \bibinfo{person}{Christoph von Praun}, {and} \bibinfo{person}{Vivek Sarkar}.} \bibinfo{year}{2005}\natexlab{}.
\newblock \showarticletitle{{X10:} an object-oriented approach to non-uniform cluster computing}. In \bibinfo{booktitle}{\emph{{OOPSLA} 2005}}. \bibinfo{publisher}{{ACM}}, \bibinfo{pages}{519--538}.
\newblock


\bibitem[Chechik and Mukhtar(2019)]%
        {DBLP:conf/soda/ChechikM19}
\bibfield{author}{\bibinfo{person}{Shiri Chechik} {and} \bibinfo{person}{Doron Mukhtar}.} \bibinfo{year}{2019}\natexlab{}.
\newblock \showarticletitle{Optimal Distributed Coloring Algorithms for Planar Graphs in the {LOCAL} model}. In \bibinfo{booktitle}{\emph{{SODA} 2019}}. \bibinfo{pages}{787--804}.
\newblock


\bibitem[Chechik and Mukhtar(2022)]%
        {DBLP:journals/dc/ChechikM22}
\bibfield{author}{\bibinfo{person}{Shiri Chechik} {and} \bibinfo{person}{Doron Mukhtar}.} \bibinfo{year}{2022}\natexlab{}.
\newblock \showarticletitle{Single-source shortest paths in the {CONGEST} model with improved bounds}.
\newblock \bibinfo{journal}{\emph{Distributed Comput.}} \bibinfo{volume}{35}, \bibinfo{number}{4} (\bibinfo{year}{2022}), \bibinfo{pages}{357--374}.
\newblock


\bibitem[Chen et~al\mbox{.}(2018)]%
        {DBLP:conf/eurosys/ChenLZYYC18}
\bibfield{author}{\bibinfo{person}{Hongzhi Chen}, \bibinfo{person}{Miao Liu}, \bibinfo{person}{Yunjian Zhao}, \bibinfo{person}{Xiao Yan}, \bibinfo{person}{Da Yan}, {and} \bibinfo{person}{James Cheng}.} \bibinfo{year}{2018}\natexlab{}.
\newblock \showarticletitle{G-Miner: an efficient task-oriented graph mining system}. In \bibinfo{booktitle}{\emph{EuroSys 2018}}. \bibinfo{pages}{32:1--32:12}.
\newblock


\bibitem[Chen and Li(2021)]%
        {DBLP:conf/iscas/ChenL21}
\bibfield{author}{\bibinfo{person}{Jie Chen} {and} \bibinfo{person}{Xiang Li}.} \bibinfo{year}{2021}\natexlab{}.
\newblock \showarticletitle{A Minimal Memory Game-Based Distributed Algorithm to Vertex Cover of Networks}. In \bibinfo{booktitle}{\emph{{ISCAS} 2021}}. \bibinfo{pages}{1--5}.
\newblock


\bibitem[Chen and Li(2023)]%
        {DBLP:journals/chinaf/ChenL23}
\bibfield{author}{\bibinfo{person}{Jie Chen} {and} \bibinfo{person}{Xiang Li}.} \bibinfo{year}{2023}\natexlab{}.
\newblock \showarticletitle{Toward the minimum vertex cover of complex networks using distributed potential games}.
\newblock \bibinfo{journal}{\emph{Sci. China Inf. Sci.}} \bibinfo{volume}{66}, \bibinfo{number}{1} (\bibinfo{year}{2023}), \bibinfo{pages}{112205}.
\newblock


\bibitem[Chen and Qian(2023)]%
        {DBLP:conf/asplos/ChenQ23a}
\bibfield{author}{\bibinfo{person}{Jingji Chen} {and} \bibinfo{person}{Xuehai Qian}.} \bibinfo{year}{2023}\natexlab{}.
\newblock \showarticletitle{Khuzdul: Efficient and Scalable Distributed Graph Pattern Mining Engine}. In \bibinfo{booktitle}{\emph{{ASPLOS} 2023}}. \bibinfo{pages}{413--426}.
\newblock


\bibitem[Chen et~al\mbox{.}(2014)]%
        {DBLP:conf/bigdataconf/ChenCC14}
\bibfield{author}{\bibinfo{person}{Pei{-}Ling Chen}, \bibinfo{person}{Chung{-}Kuang Chou}, {and} \bibinfo{person}{Ming{-}Syan Chen}.} \bibinfo{year}{2014}\natexlab{}.
\newblock \showarticletitle{Distributed algorithms for k-truss decomposition}. In \bibinfo{booktitle}{\emph{{IEEE} BigData 2014}}. \bibinfo{pages}{471--480}.
\newblock


\bibitem[Chen et~al\mbox{.}(2021)]%
        {DBLP:conf/isca/ChenHXBCA21}
\bibfield{author}{\bibinfo{person}{Xuhao Chen}, \bibinfo{person}{Tianhao Huang}, \bibinfo{person}{Shuotao Xu}, \bibinfo{person}{Thomas Bourgeat}, \bibinfo{person}{Chanwoo Chung}, {and} \bibinfo{person}{Arvind}.} \bibinfo{year}{2021}\natexlab{}.
\newblock \showarticletitle{FlexMiner: {A} Pattern-Aware Accelerator for Graph Pattern Mining}. In \bibinfo{booktitle}{\emph{{ISCA} 2021}}. \bibinfo{pages}{581--594}.
\newblock


\bibitem[Clauset et~al\mbox{.}(2009)]%
        {DBLP:journals/siamrev/ClausetSN09}
\bibfield{author}{\bibinfo{person}{Aaron Clauset}, \bibinfo{person}{Cosma~Rohilla Shalizi}, {and} \bibinfo{person}{Mark E.~J. Newman}.} \bibinfo{year}{2009}\natexlab{}.
\newblock \showarticletitle{Power-Law Distributions in Empirical Data}.
\newblock \bibinfo{journal}{\emph{{SIAM} Rev.}} \bibinfo{volume}{51}, \bibinfo{number}{4} (\bibinfo{year}{2009}), \bibinfo{pages}{661--703}.
\newblock


\bibitem[Cosulschi et~al\mbox{.}(2015)]%
        {DBLP:journals/ijcsa/CosulschiGSS15}
\bibfield{author}{\bibinfo{person}{Mirel Cosulschi}, \bibinfo{person}{Mihai Gabroveanu}, \bibinfo{person}{Florin Slabu}, {and} \bibinfo{person}{Adriana Sbircea}.} \bibinfo{year}{2015}\natexlab{}.
\newblock \showarticletitle{Scaling Up a Distributed Computing Of Similarity Coefficient with Mapreduce}.
\newblock \bibinfo{journal}{\emph{Int. J. Comput. Sci. Appl.}} \bibinfo{volume}{12}, \bibinfo{number}{2} (\bibinfo{year}{2015}), \bibinfo{pages}{81--98}.
\newblock


\bibitem[Crescenzi et~al\mbox{.}(2020)]%
        {DBLP:conf/infocom/CrescenziFP20}
\bibfield{author}{\bibinfo{person}{Pierluigi Crescenzi}, \bibinfo{person}{Pierre Fraigniaud}, {and} \bibinfo{person}{Ami Paz}.} \bibinfo{year}{2020}\natexlab{}.
\newblock \showarticletitle{Simple and Fast Distributed Computation of Betweenness Centrality}. In \bibinfo{booktitle}{\emph{{INFOCOM} 2020}}. \bibinfo{pages}{337--346}.
\newblock


\bibitem[Czarnul et~al\mbox{.}(2020)]%
        {DBLP:journals/sp/CzarnulPD20}
\bibfield{author}{\bibinfo{person}{Pawel Czarnul}, \bibinfo{person}{Jerzy Proficz}, {and} \bibinfo{person}{Krzysztof Drypczewski}.} \bibinfo{year}{2020}\natexlab{}.
\newblock \showarticletitle{Survey of Methodologies, Approaches, and Challenges in Parallel Programming Using High-Performance Computing Systems}.
\newblock \bibinfo{journal}{\emph{Sci. Program.}}  \bibinfo{volume}{2020} (\bibinfo{year}{2020}), \bibinfo{pages}{4176794:1--4176794:19}.
\newblock


\bibitem[Dai et~al\mbox{.}(2019)]%
        {DBLP:journals/tcad/DaiHCZSL00Y19}
\bibfield{author}{\bibinfo{person}{Guohao Dai}, \bibinfo{person}{Tianhao Huang}, \bibinfo{person}{Yuze Chi}, \bibinfo{person}{Jishen Zhao}, \bibinfo{person}{Guangyu Sun}, \bibinfo{person}{Yongpan Liu}, \bibinfo{person}{Yu Wang}, \bibinfo{person}{Yuan Xie}, {and} \bibinfo{person}{Huazhong Yang}.} \bibinfo{year}{2019}\natexlab{}.
\newblock \showarticletitle{GraphH: {A} Processing-in-Memory Architecture for Large-Scale Graph Processing}.
\newblock \bibinfo{journal}{\emph{{IEEE} Trans. Comput. Aided Des. Integr. Circuits Syst.}} \bibinfo{volume}{38}, \bibinfo{number}{4} (\bibinfo{year}{2019}), \bibinfo{pages}{640--653}.
\newblock


\bibitem[Dani et~al\mbox{.}(2023)]%
        {DBLP:journals/dc/DaniGHP23}
\bibfield{author}{\bibinfo{person}{Varsha Dani}, \bibinfo{person}{Aayush Gupta}, \bibinfo{person}{Thomas~P. Hayes}, {and} \bibinfo{person}{Seth Pettie}.} \bibinfo{year}{2023}\natexlab{}.
\newblock \showarticletitle{Wake up and join me! An energy-efficient algorithm for maximal matching in radio networks}.
\newblock \bibinfo{journal}{\emph{Distributed Comput.}} \bibinfo{volume}{36}, \bibinfo{number}{3} (\bibinfo{year}{2023}), \bibinfo{pages}{373--384}.
\newblock


\bibitem[Dean and Ghemawat(2004)]%
        {DBLP:conf/osdi/DeanG04}
\bibfield{author}{\bibinfo{person}{Jeffrey Dean} {and} \bibinfo{person}{Sanjay Ghemawat}.} \bibinfo{year}{2004}\natexlab{}.
\newblock \showarticletitle{MapReduce: Simplified Data Processing on Large Clusters}. In \bibinfo{booktitle}{\emph{{OSDI} 2004}}. \bibinfo{pages}{137--150}.
\newblock


\bibitem[Devismes and Johnen(2016)]%
        {DBLP:journals/jpdc/DevismesJ16}
\bibfield{author}{\bibinfo{person}{St{\'e}phane Devismes} {and} \bibinfo{person}{Colette Johnen}.} \bibinfo{year}{2016}\natexlab{}.
\newblock \showarticletitle{Silent self-stabilizing BFS tree algorithms revisited}.
\newblock \bibinfo{journal}{\emph{J. Parallel and Distrib. Comput.}}  \bibinfo{volume}{97} (\bibinfo{year}{2016}), \bibinfo{pages}{11--23}.
\newblock


\bibitem[Dominguez{-}Sal et~al\mbox{.}(2010)]%
        {DBLP:conf/tpctc/Dominguez-SalMMBL10}
\bibfield{author}{\bibinfo{person}{David Dominguez{-}Sal}, \bibinfo{person}{Norbert Mart{\'{\i}}nez{-}Bazan}, \bibinfo{person}{Victor Munt{\'{e}}s{-}Mulero}, \bibinfo{person}{Pere Baleta}, {and} \bibinfo{person}{Josep~Llu{\'{\i}}s Larriba{-}Pey}.} \bibinfo{year}{2010}\natexlab{}.
\newblock \showarticletitle{A Discussion on the Design of Graph Database Benchmarks}. In \bibinfo{booktitle}{\emph{{TPCTC} 2010}} \emph{(\bibinfo{series}{Lecture Notes in Computer Science}, Vol.~\bibinfo{volume}{6417})}. \bibinfo{publisher}{Springer}, \bibinfo{pages}{25--40}.
\newblock


\bibitem[Durant and Wagner(2017)]%
        {DBLP:journals/tcs/DurantW17}
\bibfield{author}{\bibinfo{person}{Kevin Durant} {and} \bibinfo{person}{Stephan Wagner}.} \bibinfo{year}{2017}\natexlab{}.
\newblock \showarticletitle{On the distribution of betweenness centrality in random trees}.
\newblock \bibinfo{journal}{\emph{Theoretical Computer Science}}  \bibinfo{volume}{699} (\bibinfo{year}{2017}), \bibinfo{pages}{33--52}.
\newblock


\bibitem[Elkin(2020)]%
        {DBLP:journals/jacm/Elkin20a}
\bibfield{author}{\bibinfo{person}{Michael Elkin}.} \bibinfo{year}{2020}\natexlab{}.
\newblock \showarticletitle{Distributed Exact Shortest Paths in Sublinear Time}.
\newblock \bibinfo{journal}{\emph{J. {ACM}}} \bibinfo{volume}{67}, \bibinfo{number}{3} (\bibinfo{year}{2020}), \bibinfo{pages}{15:1--15:36}.
\newblock


\bibitem[Even et~al\mbox{.}(2015)]%
        {DBLP:conf/icdcn/EvenMR15}
\bibfield{author}{\bibinfo{person}{Guy Even}, \bibinfo{person}{Moti Medina}, {and} \bibinfo{person}{Dana Ron}.} \bibinfo{year}{2015}\natexlab{}.
\newblock \showarticletitle{Distributed Maximum Matching in Bounded Degree Graphs}. In \bibinfo{booktitle}{\emph{Proceedings of the 2015 International Conference on Distributed Computing and Networking, {ICDCN} 2015}}. \bibinfo{pages}{18:1--18:10}.
\newblock


\bibitem[Fan et~al\mbox{.}(2021)]%
        {DBLP:journals/pvldb/FanHLLLLQ0WXYYY21}
\bibfield{author}{\bibinfo{person}{Wenfei Fan}, \bibinfo{person}{Tao He}, \bibinfo{person}{Longbin Lai}, \bibinfo{person}{Xue Li}, \bibinfo{person}{Yong Li}, \bibinfo{person}{Zhao Li}, \bibinfo{person}{Zhengping Qian}, \bibinfo{person}{Chao Tian}, \bibinfo{person}{Lei Wang}, \bibinfo{person}{Jingbo Xu}, \bibinfo{person}{Youyang Yao}, \bibinfo{person}{Qiang Yin}, \bibinfo{person}{Wenyuan Yu}, \bibinfo{person}{Kai Zeng}, \bibinfo{person}{Kun Zhao}, \bibinfo{person}{Jingren Zhou}, \bibinfo{person}{Diwen Zhu}, {and} \bibinfo{person}{Rong Zhu}.} \bibinfo{year}{2021}\natexlab{}.
\newblock \showarticletitle{GraphScope: {A} Unified Engine For Big Graph Processing}.
\newblock \bibinfo{journal}{\emph{Proc. {VLDB} Endow.}} \bibinfo{volume}{14}, \bibinfo{number}{12} (\bibinfo{year}{2021}), \bibinfo{pages}{2879--2892}.
\newblock


\bibitem[Feng et~al\mbox{.}(2018)]%
        {DBLP:journals/dpd/FengCLQZY18}
\bibfield{author}{\bibinfo{person}{Xing Feng}, \bibinfo{person}{Lijun Chang}, \bibinfo{person}{Xuemin Lin}, \bibinfo{person}{Lu Qin}, \bibinfo{person}{Wenjie Zhang}, {and} \bibinfo{person}{Long Yuan}.} \bibinfo{year}{2018}\natexlab{}.
\newblock \showarticletitle{Distributed computing connected components with linear communication cost}.
\newblock \bibinfo{journal}{\emph{Distributed Parallel Databases}} \bibinfo{volume}{36}, \bibinfo{number}{3} (\bibinfo{year}{2018}), \bibinfo{pages}{555--592}.
\newblock


\bibitem[Forster and Nanongkai(2018)]%
        {DBLP:conf/focs/ForsterN18}
\bibfield{author}{\bibinfo{person}{Sebastian Forster} {and} \bibinfo{person}{Danupon Nanongkai}.} \bibinfo{year}{2018}\natexlab{}.
\newblock \showarticletitle{A Faster Distributed Single-Source Shortest Paths Algorithm}. In \bibinfo{booktitle}{\emph{{FOCS} 2018}}. \bibinfo{pages}{686--697}.
\newblock


\bibitem[Forum(1993)]%
        {forum1994mpi}
\bibfield{author}{\bibinfo{person}{Message~P Forum}.} \bibinfo{year}{1993}\natexlab{}.
\newblock \bibinfo{title}{MPI: A message passing interface}.
\newblock , \bibinfo{numpages}{878-883}~pages.
\newblock


\bibitem[Fournet et~al\mbox{.}(2002)]%
        {fournet2002jocaml}
\bibfield{author}{\bibinfo{person}{C{\'e}dric Fournet}, \bibinfo{person}{Fabrice Le~Fessant}, \bibinfo{person}{Luc Maranget}, {and} \bibinfo{person}{Alan Schmitt}.} \bibinfo{year}{2002}\natexlab{}.
\newblock \showarticletitle{JoCaml: A language for concurrent distributed and mobile programming}.
\newblock In \bibinfo{booktitle}{\emph{International School on Advanced Functional Programming}}. \bibinfo{pages}{129--158}.
\newblock


\bibitem[Fraigniaud and Olivetti(2017)]%
        {DBLP:conf/spaa/FraigniaudO17}
\bibfield{author}{\bibinfo{person}{Pierre Fraigniaud} {and} \bibinfo{person}{Dennis Olivetti}.} \bibinfo{year}{2017}\natexlab{}.
\newblock \showarticletitle{Distributed Detection of Cycles}. In \bibinfo{booktitle}{\emph{Proceedings of the 29th {ACM} Symposium on Parallelism in Algorithms and Architectures, {SPAA} 2017}}. \bibinfo{pages}{153--162}.
\newblock


\bibitem[Gabow(2017)]%
        {DBLP:journals/fuin/Gabow17}
\bibfield{author}{\bibinfo{person}{Harold~N. Gabow}.} \bibinfo{year}{2017}\natexlab{}.
\newblock \showarticletitle{The Weighted Matching Approach to Maximum Cardinality Matching}.
\newblock \bibinfo{journal}{\emph{Fundam. Informaticae}} \bibinfo{volume}{154}, \bibinfo{number}{1-4} (\bibinfo{year}{2017}), \bibinfo{pages}{109--130}.
\newblock


\bibitem[Gallager et~al\mbox{.}(1983)]%
        {DBLP:journals/toplas/GallagerHS83}
\bibfield{author}{\bibinfo{person}{Robert~G. Gallager}, \bibinfo{person}{Pierre~A. Humblet}, {and} \bibinfo{person}{Philip~M. Spira}.} \bibinfo{year}{1983}\natexlab{}.
\newblock \showarticletitle{A Distributed Algorithm for Minimum-Weight Spanning Trees}.
\newblock \bibinfo{journal}{\emph{{ACM} Trans. Program. Lang. Syst.}} \bibinfo{volume}{5}, \bibinfo{number}{1} (\bibinfo{year}{1983}), \bibinfo{pages}{66--77}.
\newblock


\bibitem[Garay et~al\mbox{.}(1998)]%
        {DBLP:journals/siamcomp/GarayKP98}
\bibfield{author}{\bibinfo{person}{Juan~A. Garay}, \bibinfo{person}{Shay Kutten}, {and} \bibinfo{person}{David Peleg}.} \bibinfo{year}{1998}\natexlab{}.
\newblock \showarticletitle{A Sublinear Time Distributed Algorithm for Minimum-Weight Spanning Trees}.
\newblock \bibinfo{journal}{\emph{{SIAM} J. Comput.}} \bibinfo{volume}{27}, \bibinfo{number}{1} (\bibinfo{year}{1998}), \bibinfo{pages}{302--316}.
\newblock


\bibitem[Ghaffari et~al\mbox{.}(2018)]%
        {DBLP:journals/siamcomp/GhaffariKKLP18}
\bibfield{author}{\bibinfo{person}{Mohsen Ghaffari}, \bibinfo{person}{Andreas Karrenbauer}, \bibinfo{person}{Fabian Kuhn}, \bibinfo{person}{Christoph Lenzen}, {and} \bibinfo{person}{Boaz Patt{-}Shamir}.} \bibinfo{year}{2018}\natexlab{}.
\newblock \showarticletitle{Near-Optimal Distributed Maximum Flow}.
\newblock \bibinfo{journal}{\emph{{SIAM} J. Comput.}} \bibinfo{volume}{47}, \bibinfo{number}{6} (\bibinfo{year}{2018}), \bibinfo{pages}{2078--2117}.
\newblock


\bibitem[Ghosh(2022)]%
        {DBLP:conf/hpec/Ghosh22}
\bibfield{author}{\bibinfo{person}{Sayan Ghosh}.} \bibinfo{year}{2022}\natexlab{}.
\newblock \showarticletitle{Improved Distributed-memory Triangle Counting by Exploiting the Graph Structure}. In \bibinfo{booktitle}{\emph{{HPEC} 2022}}. \bibinfo{pages}{1--6}.
\newblock


\bibitem[Ghosh and Halappanavar(2020)]%
        {DBLP:conf/hpec/GhoshH20}
\bibfield{author}{\bibinfo{person}{Sayan Ghosh} {and} \bibinfo{person}{Mahantesh Halappanavar}.} \bibinfo{year}{2020}\natexlab{}.
\newblock \showarticletitle{TriC: Distributed-memory Triangle Counting by Exploiting the Graph Structure}. In \bibinfo{booktitle}{\emph{{HPEC} 2020}}. \bibinfo{pages}{1--6}.
\newblock


\bibitem[Ghosh et~al\mbox{.}(2018)]%
        {DBLP:conf/ipps/GhoshHTKLCKG18}
\bibfield{author}{\bibinfo{person}{Sayan Ghosh}, \bibinfo{person}{Mahantesh Halappanavar}, \bibinfo{person}{Antonino Tumeo}, \bibinfo{person}{Ananth Kalyanaraman}, \bibinfo{person}{Hao Lu}, \bibinfo{person}{Daniel~G. Chavarr{\'{\i}}a{-}Miranda}, \bibinfo{person}{Arif Khan}, {and} \bibinfo{person}{Assefaw~Hadish Gebremedhin}.} \bibinfo{year}{2018}\natexlab{}.
\newblock \showarticletitle{Distributed Louvain Algorithm for Graph Community Detection}. In \bibinfo{booktitle}{\emph{2018 {IEEE} International Parallel and Distributed Processing Symposium, {IPDPS} 2018}}. \bibinfo{pages}{885--895}.
\newblock


\bibitem[Goldberg and Harrelson(2005)]%
        {DBLP:conf/soda/GoldbergH05}
\bibfield{author}{\bibinfo{person}{Andrew~V. Goldberg} {and} \bibinfo{person}{Chris Harrelson}.} \bibinfo{year}{2005}\natexlab{}.
\newblock \showarticletitle{Computing the shortest path: \emph{A} search meets graph theory}. In \bibinfo{booktitle}{\emph{{SODA} 2005}}. \bibinfo{pages}{156--165}.
\newblock


\bibitem[Gonzalez et~al\mbox{.}(2012)]%
        {DBLP:conf/osdi/GonzalezLGBG12}
\bibfield{author}{\bibinfo{person}{Joseph~E. Gonzalez}, \bibinfo{person}{Yucheng Low}, \bibinfo{person}{Haijie Gu}, \bibinfo{person}{Danny Bickson}, {and} \bibinfo{person}{Carlos Guestrin}.} \bibinfo{year}{2012}\natexlab{}.
\newblock \showarticletitle{PowerGraph: Distributed Graph-Parallel Computation on Natural Graphs}. In \bibinfo{booktitle}{\emph{{OSDI} 2012}}. \bibinfo{pages}{17--30}.
\newblock


\bibitem[Gonzalez et~al\mbox{.}(2014)]%
        {DBLP:conf/osdi/GonzalezXDCFS14}
\bibfield{author}{\bibinfo{person}{Joseph~E. Gonzalez}, \bibinfo{person}{Reynold~S. Xin}, \bibinfo{person}{Ankur Dave}, \bibinfo{person}{Daniel Crankshaw}, \bibinfo{person}{Michael~J. Franklin}, {and} \bibinfo{person}{Ion Stoica}.} \bibinfo{year}{2014}\natexlab{}.
\newblock \showarticletitle{GraphX: Graph Processing in a Distributed Dataflow Framework}. In \bibinfo{booktitle}{\emph{{OSDI} 2014}}. \bibinfo{pages}{599--613}.
\newblock


\bibitem[Gropp(2001)]%
        {DBLP:conf/hipc/Gropp01}
\bibfield{author}{\bibinfo{person}{William Gropp}.} \bibinfo{year}{2001}\natexlab{}.
\newblock \showarticletitle{Learning from the Success of {MPI}}. In \bibinfo{booktitle}{\emph{HiPC 2001}} \emph{(\bibinfo{series}{Lecture Notes in Computer Science}, Vol.~\bibinfo{volume}{2228})}. \bibinfo{publisher}{Springer}, \bibinfo{pages}{81--94}.
\newblock


\bibitem[Guo et~al\mbox{.}(2017)]%
        {DBLP:conf/sigmod/GuoCCLL17}
\bibfield{author}{\bibinfo{person}{Tao Guo}, \bibinfo{person}{Xin Cao}, \bibinfo{person}{Gao Cong}, \bibinfo{person}{Jiaheng Lu}, {and} \bibinfo{person}{Xuemin Lin}.} \bibinfo{year}{2017}\natexlab{}.
\newblock \showarticletitle{Distributed Algorithms on Exact Personalized PageRank}. In \bibinfo{booktitle}{\emph{{SIGMOD} 2017}}. \bibinfo{pages}{479--494}.
\newblock


\bibitem[Gupta and Sundaram(2023)]%
        {DBLP:conf/aaai/GuptaS23}
\bibfield{author}{\bibinfo{person}{Shubhankar Gupta} {and} \bibinfo{person}{Suresh Sundaram}.} \bibinfo{year}{2023}\natexlab{}.
\newblock \showarticletitle{Moving-Landmark Assisted Distributed Learning Based Decentralized Cooperative Localization {(DL-DCL)} with Fault Tolerance}. In \bibinfo{booktitle}{\emph{{AAAI} 2023}}. \bibinfo{pages}{6175--6182}.
\newblock


\bibitem[Han and Daudjee(2015)]%
        {DBLP:journals/pvldb/HanD15}
\bibfield{author}{\bibinfo{person}{Minyang Han} {and} \bibinfo{person}{Khuzaima Daudjee}.} \bibinfo{year}{2015}\natexlab{}.
\newblock \showarticletitle{Giraph Unchained: Barrierless Asynchronous Parallel Execution in Pregel-like Graph Processing Systems}.
\newblock \bibinfo{journal}{\emph{Proc. {VLDB} Endow.}} \bibinfo{volume}{8}, \bibinfo{number}{9} (\bibinfo{year}{2015}), \bibinfo{pages}{950--961}.
\newblock


\bibitem[He and Wai(2021)]%
        {DBLP:conf/icassp/HeW21}
\bibfield{author}{\bibinfo{person}{Yiran He} {and} \bibinfo{person}{Hoi{-}To Wai}.} \bibinfo{year}{2021}\natexlab{}.
\newblock \showarticletitle{Provably Fast Asynchronous And Distributed Algorithms For Pagerank Centrality Computation}. In \bibinfo{booktitle}{\emph{{ICASSP} 2021}}. \bibinfo{pages}{5050--5054}.
\newblock


\bibitem[Heidari et~al\mbox{.}(2018)]%
        {DBLP:journals/csur/HeidariSCB18}
\bibfield{author}{\bibinfo{person}{Safiollah Heidari}, \bibinfo{person}{Yogesh Simmhan}, \bibinfo{person}{Rodrigo~N. Calheiros}, {and} \bibinfo{person}{Rajkumar Buyya}.} \bibinfo{year}{2018}\natexlab{}.
\newblock \showarticletitle{Scalable Graph Processing Frameworks: {A} Taxonomy and Open Challenges}.
\newblock \bibinfo{journal}{\emph{{ACM} Comput. Surv.}} \bibinfo{volume}{51}, \bibinfo{number}{3} (\bibinfo{year}{2018}), \bibinfo{pages}{60:1--60:53}.
\newblock


\bibitem[Henzinger et~al\mbox{.}(2016)]%
        {DBLP:journals/siamcomp/HenzingerKN21}
\bibfield{author}{\bibinfo{person}{Monika Henzinger}, \bibinfo{person}{Sebastian Krinninger}, {and} \bibinfo{person}{Danupon Nanongkai}.} \bibinfo{year}{2016}\natexlab{}.
\newblock \showarticletitle{A deterministic almost-tight distributed algorithm for approximating single-source shortest paths}.
\newblock  (\bibinfo{year}{2016}), \bibinfo{pages}{489--498}.
\newblock


\bibitem[Hoang et~al\mbox{.}(2019a)]%
        {DBLP:conf/hpec/HoangJCADGP19}
\bibfield{author}{\bibinfo{person}{Loc Hoang}, \bibinfo{person}{Vishwesh Jatala}, \bibinfo{person}{Xuhao Chen}, \bibinfo{person}{Udit Agarwal}, \bibinfo{person}{Roshan Dathathri}, \bibinfo{person}{Gurbinder Gill}, {and} \bibinfo{person}{Keshav Pingali}.} \bibinfo{year}{2019}\natexlab{a}.
\newblock \showarticletitle{DistTC: High Performance Distributed Triangle Counting}. In \bibinfo{booktitle}{\emph{{HPEC} 2019}}. \bibinfo{pages}{1--7}.
\newblock


\bibitem[Hoang et~al\mbox{.}(2019b)]%
        {DBLP:conf/ppopp/HoangPDGYPR19}
\bibfield{author}{\bibinfo{person}{Loc Hoang}, \bibinfo{person}{Matteo Pontecorvi}, \bibinfo{person}{Roshan Dathathri}, \bibinfo{person}{Gurbinder Gill}, \bibinfo{person}{Bozhi You}, \bibinfo{person}{Keshav Pingali}, {and} \bibinfo{person}{Vijaya Ramachandran}.} \bibinfo{year}{2019}\natexlab{b}.
\newblock \showarticletitle{A round-efficient distributed betweenness centrality algorithm}. In \bibinfo{booktitle}{\emph{PPoPP 2019}}. \bibinfo{pages}{272--286}.
\newblock


\bibitem[Holzer and Wattenhofer(2012)]%
        {DBLP:conf/podc/HolzerW12}
\bibfield{author}{\bibinfo{person}{Stephan Holzer} {and} \bibinfo{person}{Roger Wattenhofer}.} \bibinfo{year}{2012}\natexlab{}.
\newblock \showarticletitle{Optimal distributed all pairs shortest paths and applications}. In \bibinfo{booktitle}{\emph{{PODC} 2012}}. \bibinfo{pages}{355--364}.
\newblock


\bibitem[Hua et~al\mbox{.}(2016)]%
        {DBLP:conf/icdcs/HuaFAQLSJ16}
\bibfield{author}{\bibinfo{person}{Qiang{-}Sheng Hua}, \bibinfo{person}{Haoqiang Fan}, \bibinfo{person}{Ming Ai}, \bibinfo{person}{Lixiang Qian}, \bibinfo{person}{Yangyang Li}, \bibinfo{person}{Xuanhua Shi}, {and} \bibinfo{person}{Hai Jin}.} \bibinfo{year}{2016}\natexlab{}.
\newblock \showarticletitle{Nearly Optimal Distributed Algorithm for Computing Betweenness Centrality}. In \bibinfo{booktitle}{\emph{36th {IEEE} International Conference on Distributed Computing Systems, {ICDCS} 2016}}. \bibinfo{pages}{271--280}.
\newblock


\bibitem[Jain et~al\mbox{.}(2017)]%
        {DBLP:journals/tpds/JainFPGA17}
\bibfield{author}{\bibinfo{person}{Chirag Jain}, \bibinfo{person}{Patrick Flick}, \bibinfo{person}{Tony Pan}, \bibinfo{person}{Oded Green}, {and} \bibinfo{person}{Srinivas Aluru}.} \bibinfo{year}{2017}\natexlab{}.
\newblock \showarticletitle{An Adaptive Parallel Algorithm for Computing Connected Components}.
\newblock \bibinfo{journal}{\emph{{IEEE} Trans. Parallel Distributed Syst.}} \bibinfo{volume}{28}, \bibinfo{number}{9} (\bibinfo{year}{2017}), \bibinfo{pages}{2428--2439}.
\newblock


\bibitem[Jamour et~al\mbox{.}(2018)]%
        {DBLP:journals/tpds/JamourSK18}
\bibfield{author}{\bibinfo{person}{Fuad~T. Jamour}, \bibinfo{person}{Spiros Skiadopoulos}, {and} \bibinfo{person}{Panos Kalnis}.} \bibinfo{year}{2018}\natexlab{}.
\newblock \showarticletitle{Parallel Algorithm for Incremental Betweenness Centrality on Large Graphs}.
\newblock \bibinfo{journal}{\emph{{IEEE} Trans. Parallel Distributed Syst.}} \bibinfo{volume}{29}, \bibinfo{number}{3} (\bibinfo{year}{2018}), \bibinfo{pages}{659--672}.
\newblock


\bibitem[Jeh and Widom(2002)]%
        {DBLP:conf/kdd/JehW02}
\bibfield{author}{\bibinfo{person}{Glen Jeh} {and} \bibinfo{person}{Jennifer Widom}.} \bibinfo{year}{2002}\natexlab{}.
\newblock \showarticletitle{SimRank: a measure of structural-context similarity}. In \bibinfo{booktitle}{\emph{KDD 2002}}. \bibinfo{pages}{538--543}.
\newblock


\bibitem[Jiang(2016)]%
        {DBLP:journals/tpds/Jiang16}
\bibfield{author}{\bibinfo{person}{Yichuan Jiang}.} \bibinfo{year}{2016}\natexlab{}.
\newblock \showarticletitle{A Survey of Task Allocation and Load Balancing in Distributed Systems}.
\newblock \bibinfo{journal}{\emph{{IEEE} Trans. Parallel Distributed Syst.}} \bibinfo{volume}{27}, \bibinfo{number}{2} (\bibinfo{year}{2016}), \bibinfo{pages}{585--599}.
\newblock


\bibitem[Joo et~al\mbox{.}(2016)]%
        {DBLP:journals/ton/JooLRS16}
\bibfield{author}{\bibinfo{person}{Changhee Joo}, \bibinfo{person}{Xiaojun Lin}, \bibinfo{person}{Jiho Ryu}, {and} \bibinfo{person}{Ness~B. Shroff}.} \bibinfo{year}{2016}\natexlab{}.
\newblock \showarticletitle{Distributed Greedy Approximation to Maximum Weighted Independent Set for Scheduling With Fading Channels}.
\newblock \bibinfo{journal}{\emph{{IEEE/ACM} Trans. Netw.}} \bibinfo{volume}{24}, \bibinfo{number}{3} (\bibinfo{year}{2016}), \bibinfo{pages}{1476--1488}.
\newblock


\bibitem[Kalavri et~al\mbox{.}(2018)]%
        {DBLP:journals/tkde/KalavriVH18}
\bibfield{author}{\bibinfo{person}{Vasiliki Kalavri}, \bibinfo{person}{Vladimir Vlassov}, {and} \bibinfo{person}{Seif Haridi}.} \bibinfo{year}{2018}\natexlab{}.
\newblock \showarticletitle{High-Level Programming Abstractions for Distributed Graph Processing}.
\newblock \bibinfo{journal}{\emph{{IEEE} Trans. Knowl. Data Eng.}} \bibinfo{volume}{30}, \bibinfo{number}{2} (\bibinfo{year}{2018}), \bibinfo{pages}{305--324}.
\newblock


\bibitem[Karypis and Kumar(1998)]%
        {karypis1997metis}
\bibfield{author}{\bibinfo{person}{George Karypis} {and} \bibinfo{person}{Vipin Kumar}.} \bibinfo{year}{1998}\natexlab{}.
\newblock \showarticletitle{A software package for partitioning unstructured graphs, partitioning meshes, and computing fill-reducing orderings of sparse matrices}.
\newblock \bibinfo{journal}{\emph{University of Minnesota, Department of Computer Science and Engineering, Army HPC Research Center, Minneapolis, MN}}  \bibinfo{volume}{38} (\bibinfo{year}{1998}), \bibinfo{pages}{7--1}.
\newblock


\bibitem[Kuhn(1955)]%
        {kuhn1955hungarian}
\bibfield{author}{\bibinfo{person}{Harold~W Kuhn}.} \bibinfo{year}{1955}\natexlab{}.
\newblock \showarticletitle{The Hungarian method for the assignment problem}.
\newblock \bibinfo{journal}{\emph{Nav. Res. Logist. Q.}} \bibinfo{volume}{2}, \bibinfo{number}{1-2} (\bibinfo{year}{1955}), \bibinfo{pages}{83--97}.
\newblock


\bibitem[Lai et~al\mbox{.}(2015)]%
        {DBLP:journals/pvldb/LaiQLC15}
\bibfield{author}{\bibinfo{person}{Longbin Lai}, \bibinfo{person}{Lu Qin}, \bibinfo{person}{Xuemin Lin}, {and} \bibinfo{person}{Lijun Chang}.} \bibinfo{year}{2015}\natexlab{}.
\newblock \showarticletitle{Scalable Subgraph Enumeration in MapReduce}.
\newblock \bibinfo{journal}{\emph{Proc. {VLDB} Endow.}} \bibinfo{volume}{8}, \bibinfo{number}{10} (\bibinfo{year}{2015}), \bibinfo{pages}{974--985}.
\newblock


\bibitem[Lai et~al\mbox{.}(2019)]%
        {DBLP:journals/pvldb/LaiQYJLWHLQZZQZ19}
\bibfield{author}{\bibinfo{person}{Longbin Lai}, \bibinfo{person}{Zhu Qing}, \bibinfo{person}{Zhengyi Yang}, \bibinfo{person}{Xin Jin}, \bibinfo{person}{Zhengmin Lai}, \bibinfo{person}{Ran Wang}, \bibinfo{person}{Kongzhang Hao}, \bibinfo{person}{Xuemin Lin}, \bibinfo{person}{Lu Qin}, \bibinfo{person}{Wenjie Zhang}, \bibinfo{person}{Ying Zhang}, \bibinfo{person}{Zhengping Qian}, {and} \bibinfo{person}{Jingren Zhou}.} \bibinfo{year}{2019}\natexlab{}.
\newblock \showarticletitle{Distributed Subgraph Matching on Timely Dataflow}.
\newblock \bibinfo{journal}{\emph{Proc. {VLDB} Endow.}} \bibinfo{volume}{12}, \bibinfo{number}{10} (\bibinfo{year}{2019}), \bibinfo{pages}{1099--1112}.
\newblock


\bibitem[Lai et~al\mbox{.}(2023)]%
        {DBLP:conf/usenix/LaiYWLMSLZYQ00C23}
\bibfield{author}{\bibinfo{person}{Longbin Lai}, \bibinfo{person}{Yufan Yang}, \bibinfo{person}{Zhibin Wang}, \bibinfo{person}{Yuxuan Liu}, \bibinfo{person}{Haotian Ma}, \bibinfo{person}{Sijie Shen}, \bibinfo{person}{Bingqing Lyu}, \bibinfo{person}{Xiaoli Zhou}, \bibinfo{person}{Wenyuan Yu}, \bibinfo{person}{Zhengping Qian}, \bibinfo{person}{Chen Tian}, \bibinfo{person}{Sheng Zhong}, \bibinfo{person}{Yeh{-}Ching Chung}, {and} \bibinfo{person}{Jingren Zhou}.} \bibinfo{year}{2023}\natexlab{}.
\newblock \showarticletitle{GLogS: Interactive Graph Pattern Matching Query At Large Scale}. In \bibinfo{booktitle}{\emph{{USENIX} {ATC} 2023}}. \bibinfo{publisher}{{USENIX} Association}, \bibinfo{pages}{53--69}.
\newblock


\bibitem[Lakhotia et~al\mbox{.}(2018)]%
        {DBLP:conf/usenix/LakhotiaKP18}
\bibfield{author}{\bibinfo{person}{Kartik Lakhotia}, \bibinfo{person}{Rajgopal Kannan}, {and} \bibinfo{person}{Viktor~K. Prasanna}.} \bibinfo{year}{2018}\natexlab{}.
\newblock \showarticletitle{Accelerating PageRank using Partition-Centric Processing}. In \bibinfo{booktitle}{\emph{{USENIX} {ATC} 2018}}. \bibinfo{pages}{427--440}.
\newblock


\bibitem[Lamm and Sanders(2022)]%
        {DBLP:conf/ipps/Lamm022}
\bibfield{author}{\bibinfo{person}{Sebastian Lamm} {and} \bibinfo{person}{Peter Sanders}.} \bibinfo{year}{2022}\natexlab{}.
\newblock \showarticletitle{Communication-efficient Massively Distributed Connected Components}. In \bibinfo{booktitle}{\emph{{IPDPS} 2022}}. \bibinfo{pages}{302--312}.
\newblock


\bibitem[Lenzen et~al\mbox{.}(2019)]%
        {DBLP:journals/dc/LenzenPP19}
\bibfield{author}{\bibinfo{person}{Christoph Lenzen}, \bibinfo{person}{Boaz Patt{-}Shamir}, {and} \bibinfo{person}{David Peleg}.} \bibinfo{year}{2019}\natexlab{}.
\newblock \showarticletitle{Distributed distance computation and routing with small messages}.
\newblock \bibinfo{journal}{\emph{Distributed Comput.}} \bibinfo{volume}{32}, \bibinfo{number}{2} (\bibinfo{year}{2019}), \bibinfo{pages}{133--157}.
\newblock


\bibitem[Li et~al\mbox{.}(2023)]%
        {DBLP:conf/icde/LiM0LYQ0Z23}
\bibfield{author}{\bibinfo{person}{Xue Li}, \bibinfo{person}{Ke Meng}, \bibinfo{person}{Lu Qin}, \bibinfo{person}{Longbin Lai}, \bibinfo{person}{Wenyuan Yu}, \bibinfo{person}{Zhengping Qian}, \bibinfo{person}{Xuemin Lin}, {and} \bibinfo{person}{Jingren Zhou}.} \bibinfo{year}{2023}\natexlab{}.
\newblock \showarticletitle{Flash: {A} Framework for Programming Distributed Graph Processing Algorithms}. In \bibinfo{booktitle}{\emph{{ICDE} 2023}}. \bibinfo{publisher}{{IEEE}}, \bibinfo{pages}{232--244}.
\newblock


\bibitem[Li et~al\mbox{.}(2015)]%
        {DBLP:journals/pvldb/LiFLCCL15}
\bibfield{author}{\bibinfo{person}{Zhenguo Li}, \bibinfo{person}{Yixiang Fang}, \bibinfo{person}{Qin Liu}, \bibinfo{person}{Jiefeng Cheng}, \bibinfo{person}{Reynold Cheng}, {and} \bibinfo{person}{John C.~S. Lui}.} \bibinfo{year}{2015}\natexlab{}.
\newblock \showarticletitle{Walking in the Cloud: Parallel SimRank at Scale}.
\newblock \bibinfo{journal}{\emph{Proc. {VLDB} Endow.}} \bibinfo{volume}{9}, \bibinfo{number}{1} (\bibinfo{year}{2015}), \bibinfo{pages}{24--35}.
\newblock


\bibitem[Liao et~al\mbox{.}(2022)]%
        {DBLP:journals/pvldb/LiaoLJHXC22}
\bibfield{author}{\bibinfo{person}{Xuankun Liao}, \bibinfo{person}{Qing Liu}, \bibinfo{person}{Jiaxin Jiang}, \bibinfo{person}{Xin Huang}, \bibinfo{person}{Jianliang Xu}, {and} \bibinfo{person}{Byron Choi}.} \bibinfo{year}{2022}\natexlab{}.
\newblock \showarticletitle{Distributed D-core Decomposition over Large Directed Graphs}.
\newblock \bibinfo{journal}{\emph{Proc. {VLDB} Endow.}} \bibinfo{volume}{15}, \bibinfo{number}{8} (\bibinfo{year}{2022}), \bibinfo{pages}{1546--1558}.
\newblock


\bibitem[Lin(2019)]%
        {DBLP:conf/www/Lin19}
\bibfield{author}{\bibinfo{person}{Wenqing Lin}.} \bibinfo{year}{2019}\natexlab{}.
\newblock \showarticletitle{Distributed Algorithms for Fully Personalized PageRank on Large Graphs}. In \bibinfo{booktitle}{\emph{{WWW} 2019}}. \bibinfo{pages}{1084--1094}.
\newblock


\bibitem[Liu et~al\mbox{.}(2023)]%
        {DBLP:conf/icde/LiuLHXG23}
\bibfield{author}{\bibinfo{person}{Qing Liu}, \bibinfo{person}{Xuankun Liao}, \bibinfo{person}{Xin Huang}, \bibinfo{person}{Jianliang Xu}, {and} \bibinfo{person}{Yunjun Gao}.} \bibinfo{year}{2023}\natexlab{}.
\newblock \showarticletitle{Distributed ({\(\alpha\)}, {\(\beta\)})-Core Decomposition over Bipartite Graphs}. In \bibinfo{booktitle}{\emph{{ICDE} 2023}}. \bibinfo{pages}{909--921}.
\newblock


\bibitem[Liu et~al\mbox{.}(2019)]%
        {DBLP:conf/hpec/0001FZHBLG19}
\bibfield{author}{\bibinfo{person}{Xu Liu}, \bibinfo{person}{Jesun~Sahariar Firoz}, \bibinfo{person}{Marcin Zalewski}, \bibinfo{person}{Mahantesh Halappanavar}, \bibinfo{person}{Kevin~J. Barker}, \bibinfo{person}{Andrew Lumsdaine}, {and} \bibinfo{person}{Assefaw~H. Gebremedhin}.} \bibinfo{year}{2019}\natexlab{}.
\newblock \showarticletitle{Distributed Direction-Optimizing Label Propagation for Community Detection}. In \bibinfo{booktitle}{\emph{{HPEC} 2019}}. \bibinfo{pages}{1--6}.
\newblock


\bibitem[Low et~al\mbox{.}(2012)]%
        {DBLP:journals/pvldb/LowGKBGH12}
\bibfield{author}{\bibinfo{person}{Yucheng Low}, \bibinfo{person}{Joseph Gonzalez}, \bibinfo{person}{Aapo Kyrola}, \bibinfo{person}{Danny Bickson}, \bibinfo{person}{Carlos Guestrin}, {and} \bibinfo{person}{Joseph~M. Hellerstein}.} \bibinfo{year}{2012}\natexlab{}.
\newblock \showarticletitle{Distributed GraphLab: {A} Framework for Machine Learning in the Cloud}.
\newblock \bibinfo{journal}{\emph{Proc. {VLDB} Endow.}} \bibinfo{volume}{5}, \bibinfo{number}{8} (\bibinfo{year}{2012}), \bibinfo{pages}{716--727}.
\newblock


\bibitem[Lulli et~al\mbox{.}(2017)]%
        {DBLP:journals/tpds/LulliCDLR17}
\bibfield{author}{\bibinfo{person}{Alessandro Lulli}, \bibinfo{person}{Emanuele Carlini}, \bibinfo{person}{Patrizio Dazzi}, \bibinfo{person}{Claudio Lucchese}, {and} \bibinfo{person}{Laura Ricci}.} \bibinfo{year}{2017}\natexlab{}.
\newblock \showarticletitle{Fast Connected Components Computation in Large Graphs by Vertex Pruning}.
\newblock \bibinfo{journal}{\emph{{IEEE} Trans. Parallel Distributed Syst.}} \bibinfo{volume}{28}, \bibinfo{number}{3} (\bibinfo{year}{2017}), \bibinfo{pages}{760--773}.
\newblock


\bibitem[Luo(2019)]%
        {DBLP:conf/aaai/Luo19}
\bibfield{author}{\bibinfo{person}{Siqiang Luo}.} \bibinfo{year}{2019}\natexlab{}.
\newblock \showarticletitle{Distributed PageRank Computation: An Improved Theoretical Study}. In \bibinfo{booktitle}{\emph{{AAAI} 2019}}. \bibinfo{pages}{4496--4503}.
\newblock


\bibitem[Luo(2020)]%
        {DBLP:conf/icml/Luo20}
\bibfield{author}{\bibinfo{person}{Siqiang Luo}.} \bibinfo{year}{2020}\natexlab{}.
\newblock \showarticletitle{Improved Communication Cost in Distributed PageRank Computation - {A} Theoretical Study}. In \bibinfo{booktitle}{\emph{{ICML} 2020}} \emph{(\bibinfo{series}{Proceedings of Machine Learning Research}, Vol.~\bibinfo{volume}{119})}. \bibinfo{pages}{6459--6467}.
\newblock


\bibitem[Luo et~al\mbox{.}(2022)]%
        {DBLP:journals/isci/LuoWK22}
\bibfield{author}{\bibinfo{person}{Siqiang Luo}, \bibinfo{person}{Xiaowei Wu}, {and} \bibinfo{person}{Ben Kao}.} \bibinfo{year}{2022}\natexlab{}.
\newblock \showarticletitle{Distributed PageRank computation with improved round complexities}.
\newblock \bibinfo{journal}{\emph{Inf. Sci.}}  \bibinfo{volume}{607} (\bibinfo{year}{2022}), \bibinfo{pages}{109--125}.
\newblock


\bibitem[Luo and Zhu(2023)]%
        {DBLP:journals/corr/abs-2304-04015}
\bibfield{author}{\bibinfo{person}{Siqiang Luo} {and} \bibinfo{person}{Zulun Zhu}.} \bibinfo{year}{2023}\natexlab{}.
\newblock \showarticletitle{Massively Parallel Single-Source SimRanks in o(log n) Rounds}.
\newblock \bibinfo{journal}{\emph{CoRR}}  \bibinfo{volume}{abs/2304.04015} (\bibinfo{year}{2023}).
\newblock
\showeprint[arXiv]{2304.04015}


\bibitem[Luo et~al\mbox{.}(2017)]%
        {DBLP:conf/icde/LuoGZY17}
\bibfield{author}{\bibinfo{person}{Xiongcai Luo}, \bibinfo{person}{Jun Gao}, \bibinfo{person}{Chang Zhou}, {and} \bibinfo{person}{Jeffrey~Xu Yu}.} \bibinfo{year}{2017}\natexlab{}.
\newblock \showarticletitle{UniWalk: Unidirectional Random Walk Based Scalable SimRank Computation over Large Graph}. In \bibinfo{booktitle}{\emph{{ICDE} 2017}}. \bibinfo{pages}{325--336}.
\newblock


\bibitem[Ma et~al\mbox{.}(2018)]%
        {ma2018psplpa}
\bibfield{author}{\bibinfo{person}{Tinghuai Ma}, \bibinfo{person}{Mingliang Yue}, \bibinfo{person}{Jingjing Qu}, \bibinfo{person}{Yuan Tian}, \bibinfo{person}{Abdullah Al-Dhelaan}, {and} \bibinfo{person}{Mznah Al-Rodhaan}.} \bibinfo{year}{2018}\natexlab{}.
\newblock \showarticletitle{PSPLPA: Probability and similarity based parallel label propagation algorithm on spark}.
\newblock \bibinfo{journal}{\emph{Physica A: Statistical Mechanics and its Applications}}  \bibinfo{volume}{503} (\bibinfo{year}{2018}), \bibinfo{pages}{366--378}.
\newblock


\bibitem[Maleki et~al\mbox{.}(2016)]%
        {DBLP:conf/ppopp/MalekiNLGPP16}
\bibfield{author}{\bibinfo{person}{Saeed Maleki}, \bibinfo{person}{Donald Nguyen}, \bibinfo{person}{Andrew Lenharth}, \bibinfo{person}{Mar{\'{\i}}a~Jes{\'{u}}s Garzar{\'{a}}n}, \bibinfo{person}{David~A. Padua}, {and} \bibinfo{person}{Keshav Pingali}.} \bibinfo{year}{2016}\natexlab{}.
\newblock \showarticletitle{{DSMR:} a shared and distributed memory algorithm for single-source shortest path problem}. In \bibinfo{booktitle}{\emph{PPoPP 2016}}. \bibinfo{pages}{39:1--39:2}.
\newblock


\bibitem[Malewicz et~al\mbox{.}(2010)]%
        {DBLP:conf/sigmod/MalewiczABDHLC10}
\bibfield{author}{\bibinfo{person}{Grzegorz Malewicz}, \bibinfo{person}{Matthew~H. Austern}, \bibinfo{person}{Aart J.~C. Bik}, \bibinfo{person}{James~C. Dehnert}, \bibinfo{person}{Ilan Horn}, \bibinfo{person}{Naty Leiser}, {and} \bibinfo{person}{Grzegorz Czajkowski}.} \bibinfo{year}{2010}\natexlab{}.
\newblock \showarticletitle{Pregel: a system for large-scale graph processing}. In \bibinfo{booktitle}{\emph{{SIGMOD} 2010}}. \bibinfo{pages}{135--146}.
\newblock


\bibitem[Malliaros et~al\mbox{.}(2020)]%
        {DBLP:journals/vldb/MalliarosGPV20}
\bibfield{author}{\bibinfo{person}{Fragkiskos~D. Malliaros}, \bibinfo{person}{Christos Giatsidis}, \bibinfo{person}{Apostolos~N. Papadopoulos}, {and} \bibinfo{person}{Michalis Vazirgiannis}.} \bibinfo{year}{2020}\natexlab{}.
\newblock \showarticletitle{The core decomposition of networks: theory, algorithms and applications}.
\newblock \bibinfo{journal}{\emph{{VLDB} J.}} \bibinfo{volume}{29}, \bibinfo{number}{1} (\bibinfo{year}{2020}), \bibinfo{pages}{61--92}.
\newblock


\bibitem[Mandal and Hasan(2017)]%
        {DBLP:conf/bigdataconf/MandalH17}
\bibfield{author}{\bibinfo{person}{Aritra Mandal} {and} \bibinfo{person}{Mohammad~Al Hasan}.} \bibinfo{year}{2017}\natexlab{}.
\newblock \showarticletitle{A distributed k-core decomposition algorithm on spark}. In \bibinfo{booktitle}{\emph{{IEEE} BigData 2017}}. \bibinfo{pages}{976--981}.
\newblock


\bibitem[Mashreghi and King(2021)]%
        {DBLP:journals/dc/MashreghiK21}
\bibfield{author}{\bibinfo{person}{Ali Mashreghi} {and} \bibinfo{person}{Valerie King}.} \bibinfo{year}{2021}\natexlab{}.
\newblock \showarticletitle{Broadcast and minimum spanning tree with o(m) messages in the asynchronous {CONGEST} model}.
\newblock \bibinfo{journal}{\emph{Distributed Comput.}} \bibinfo{volume}{34}, \bibinfo{number}{4} (\bibinfo{year}{2021}), \bibinfo{pages}{283--299}.
\newblock


\bibitem[Modi et~al\mbox{.}(2005)]%
        {DBLP:journals/ai/ModiSTY05}
\bibfield{author}{\bibinfo{person}{Pragnesh~Jay Modi}, \bibinfo{person}{Wei{-}Min Shen}, \bibinfo{person}{Milind Tambe}, {and} \bibinfo{person}{Makoto Yokoo}.} \bibinfo{year}{2005}\natexlab{}.
\newblock \showarticletitle{Adopt: asynchronous distributed constraint optimization with quality guarantees}.
\newblock \bibinfo{journal}{\emph{Artif. Intell.}} \bibinfo{volume}{161}, \bibinfo{number}{1-2} (\bibinfo{year}{2005}), \bibinfo{pages}{149--180}.
\newblock


\bibitem[Montresor et~al\mbox{.}(2013)]%
        {DBLP:journals/tpds/MontresorPM13}
\bibfield{author}{\bibinfo{person}{Alberto Montresor}, \bibinfo{person}{Francesco~De Pellegrini}, {and} \bibinfo{person}{Daniele Miorandi}.} \bibinfo{year}{2013}\natexlab{}.
\newblock \showarticletitle{Distributed k-Core Decomposition}.
\newblock \bibinfo{journal}{\emph{{IEEE} Trans. Parallel Distributed Syst.}} \bibinfo{volume}{24}, \bibinfo{number}{2} (\bibinfo{year}{2013}), \bibinfo{pages}{288--300}.
\newblock


\bibitem[Moritz et~al\mbox{.}(2018)]%
        {DBLP:conf/osdi/MoritzNWTLLEYPJ18}
\bibfield{author}{\bibinfo{person}{Philipp Moritz}, \bibinfo{person}{Robert Nishihara}, \bibinfo{person}{Stephanie Wang}, \bibinfo{person}{Alexey Tumanov}, \bibinfo{person}{Richard Liaw}, \bibinfo{person}{Eric Liang}, \bibinfo{person}{Melih Elibol}, \bibinfo{person}{Zongheng Yang}, \bibinfo{person}{William Paul}, \bibinfo{person}{Michael~I. Jordan}, {and} \bibinfo{person}{Ion Stoica}.} \bibinfo{year}{2018}\natexlab{}.
\newblock \showarticletitle{Ray: {A} Distributed Framework for Emerging {AI} Applications}. In \bibinfo{booktitle}{\emph{{OSDI} 2018}}. \bibinfo{publisher}{{USENIX} Association}, \bibinfo{pages}{561--577}.
\newblock


\bibitem[Ngo et~al\mbox{.}(2018)]%
        {DBLP:journals/jacm/NgoPRR18}
\bibfield{author}{\bibinfo{person}{Hung~Q. Ngo}, \bibinfo{person}{Ely Porat}, \bibinfo{person}{Christopher R{\'{e}}}, {and} \bibinfo{person}{Atri Rudra}.} \bibinfo{year}{2018}\natexlab{}.
\newblock \showarticletitle{Worst-case Optimal Join Algorithms}.
\newblock \bibinfo{journal}{\emph{J. {ACM}}} \bibinfo{volume}{65}, \bibinfo{number}{3} (\bibinfo{year}{2018}), \bibinfo{pages}{16:1--16:40}.
\newblock


\bibitem[Oracle(2014)]%
        {RN16}
\bibfield{author}{\bibinfo{person}{Oracle}.} \bibinfo{year}{2014}\natexlab{}.
\newblock \bibinfo{title}{Remote Method Invocation Home}.
\newblock
\newblock
\urldef\tempurl%
\url{https://www.oracle.com}
\showURL{%
\tempurl}
\newblock
\shownote{Accessed: January 15, 2024}.


\bibitem[Pandey et~al\mbox{.}(2021)]%
        {DBLP:journals/tpds/PandeyWZTZLLHDL21}
\bibfield{author}{\bibinfo{person}{Santosh Pandey}, \bibinfo{person}{Zhibin Wang}, \bibinfo{person}{Sheng Zhong}, \bibinfo{person}{Chen Tian}, \bibinfo{person}{Bolong Zheng}, \bibinfo{person}{Xiaoye~S. Li}, \bibinfo{person}{Lingda Li}, \bibinfo{person}{Adolfy Hoisie}, \bibinfo{person}{Caiwen Ding}, \bibinfo{person}{Dong Li}, {and} \bibinfo{person}{Hang Liu}.} \bibinfo{year}{2021}\natexlab{}.
\newblock \showarticletitle{Trust: Triangle Counting Reloaded on GPUs}.
\newblock \bibinfo{journal}{\emph{{IEEE} Trans. Parallel Distributed Syst.}} \bibinfo{volume}{32}, \bibinfo{number}{11} (\bibinfo{year}{2021}), \bibinfo{pages}{2646--2660}.
\newblock


\bibitem[Pearce et~al\mbox{.}(2014)]%
        {DBLP:conf/sc/PearceGA14}
\bibfield{author}{\bibinfo{person}{Roger~A. Pearce}, \bibinfo{person}{Maya~B. Gokhale}, {and} \bibinfo{person}{Nancy~M. Amato}.} \bibinfo{year}{2014}\natexlab{}.
\newblock \showarticletitle{Faster Parallel Traversal of Scale Free Graphs at Extreme Scale with Vertex Delegates}. In \bibinfo{booktitle}{\emph{{SC} 2014,}}. \bibinfo{publisher}{{IEEE} Computer Society}, \bibinfo{pages}{549--559}.
\newblock


\bibitem[Peng et~al\mbox{.}(2016)]%
        {DBLP:conf/icde/PengCHZXY16}
\bibfield{author}{\bibinfo{person}{Yun Peng}, \bibinfo{person}{Byron Choi}, \bibinfo{person}{Bingsheng He}, \bibinfo{person}{Shuigeng Zhou}, \bibinfo{person}{Ruzhi Xu}, {and} \bibinfo{person}{Xiaohui Yu}.} \bibinfo{year}{2016}\natexlab{}.
\newblock \showarticletitle{VColor: {A} practical vertex-cut based approach for coloring large graphs}. In \bibinfo{booktitle}{\emph{{ICDE} 2016}}. \bibinfo{pages}{97--108}.
\newblock


\bibitem[Peretz and Fischler(2022)]%
        {DBLP:journals/jpdc/PeretzF22}
\bibfield{author}{\bibinfo{person}{Yossi Peretz} {and} \bibinfo{person}{Yigal Fischler}.} \bibinfo{year}{2022}\natexlab{}.
\newblock \showarticletitle{A fast parallel max-flow algorithm}.
\newblock \bibinfo{journal}{\emph{J. Parallel Distr. Com.}}  \bibinfo{volume}{169} (\bibinfo{year}{2022}), \bibinfo{pages}{226--241}.
\newblock


\bibitem[Raghavan et~al\mbox{.}(2007)]%
        {raghavan2007near}
\bibfield{author}{\bibinfo{person}{Usha~Nandini Raghavan}, \bibinfo{person}{R{\'e}ka Albert}, {and} \bibinfo{person}{Soundar Kumara}.} \bibinfo{year}{2007}\natexlab{}.
\newblock \showarticletitle{Near linear time algorithm to detect community structures in large-scale networks}.
\newblock \bibinfo{journal}{\emph{Physical review E}} \bibinfo{volume}{76}, \bibinfo{number}{3} (\bibinfo{year}{2007}), \bibinfo{pages}{036106}.
\newblock


\bibitem[Reed et~al\mbox{.}(2023)]%
        {DBLP:journals/tac/ReedRBP23}
\bibfield{author}{\bibinfo{person}{Emily~A. Reed}, \bibinfo{person}{Guilherme Ramos}, \bibinfo{person}{Paul Bogdan}, {and} \bibinfo{person}{S{\'{e}}rgio Pequito}.} \bibinfo{year}{2023}\natexlab{}.
\newblock \showarticletitle{A Scalable Distributed Dynamical Systems Approach to Learn the Strongly Connected Components and Diameter of Networks}.
\newblock \bibinfo{journal}{\emph{{IEEE} Trans. Autom. Control.}} \bibinfo{volume}{68}, \bibinfo{number}{5} (\bibinfo{year}{2023}), \bibinfo{pages}{3099--3106}.
\newblock


\bibitem[Remis et~al\mbox{.}(2018)]%
        {DBLP:journals/jpdc/RemisGAN18}
\bibfield{author}{\bibinfo{person}{Luis Remis}, \bibinfo{person}{Mar{\'{\i}}a~Jes{\'{u}}s Garzar{\'{a}}n}, \bibinfo{person}{Rafael Asenjo}, {and} \bibinfo{person}{Angeles~G. Navarro}.} \bibinfo{year}{2018}\natexlab{}.
\newblock \showarticletitle{Exploiting social network graph characteristics for efficient {BFS} on heterogeneous chips}.
\newblock \bibinfo{journal}{\emph{J. Parallel Distributed Comput.}}  \bibinfo{volume}{120} (\bibinfo{year}{2018}), \bibinfo{pages}{282--294}.
\newblock


\bibitem[Roy et~al\mbox{.}(2013)]%
        {DBLP:conf/sosp/RoyMZ13}
\bibfield{author}{\bibinfo{person}{Amitabha Roy}, \bibinfo{person}{Ivo Mihailovic}, {and} \bibinfo{person}{Willy Zwaenepoel}.} \bibinfo{year}{2013}\natexlab{}.
\newblock \showarticletitle{X-Stream: edge-centric graph processing using streaming partitions}. In \bibinfo{booktitle}{\emph{{SOSP} 2013}}. \bibinfo{pages}{472--488}.
\newblock


\bibitem[Rozhon et~al\mbox{.}(2022)]%
        {DBLP:conf/stoc/RozhonGHZL22}
\bibfield{author}{\bibinfo{person}{V{\'{a}}clav Rozhon}, \bibinfo{person}{Christoph Grunau}, \bibinfo{person}{Bernhard Haeupler}, \bibinfo{person}{Goran Zuzic}, {and} \bibinfo{person}{Jason Li}.} \bibinfo{year}{2022}\natexlab{}.
\newblock \showarticletitle{Undirected (1+\emph{{\(\epsilon\)}})-shortest paths via minor-aggregates: near-optimal deterministic parallel and distributed algorithms}. In \bibinfo{booktitle}{\emph{{STOC} 2022}}. \bibinfo{pages}{478--487}.
\newblock


\bibitem[Sakr et~al\mbox{.}(2013)]%
        {DBLP:journals/csur/SakrLF13}
\bibfield{author}{\bibinfo{person}{Sherif Sakr}, \bibinfo{person}{Anna Liu}, {and} \bibinfo{person}{Ayman~G. Fayoumi}.} \bibinfo{year}{2013}\natexlab{}.
\newblock \showarticletitle{The family of mapreduce and large-scale data processing systems}.
\newblock \bibinfo{journal}{\emph{{ACM} Comput. Surv.}} \bibinfo{volume}{46}, \bibinfo{number}{1} (\bibinfo{year}{2013}), \bibinfo{pages}{11:1--11:44}.
\newblock


\bibitem[Salihoglu and Widom(2013)]%
        {DBLP:conf/ssdbm/SalihogluW13}
\bibfield{author}{\bibinfo{person}{Semih Salihoglu} {and} \bibinfo{person}{Jennifer Widom}.} \bibinfo{year}{2013}\natexlab{}.
\newblock \showarticletitle{{GPS:} a graph processing system}. In \bibinfo{booktitle}{\emph{{SSDBM} 2013}}. \bibinfo{pages}{22:1--22:12}.
\newblock


\bibitem[Sanders and Uhl(2023)]%
        {DBLP:conf/ipps/SandersU23}
\bibfield{author}{\bibinfo{person}{Peter Sanders} {and} \bibinfo{person}{Tim~Niklas Uhl}.} \bibinfo{year}{2023}\natexlab{}.
\newblock \showarticletitle{Engineering a Distributed-Memory Triangle Counting Algorithm}. In \bibinfo{booktitle}{\emph{{IPDPS} 2023}}. \bibinfo{pages}{702--712}.
\newblock


\bibitem[Sariy{\"{u}}ce et~al\mbox{.}(2013)]%
        {DBLP:conf/cluster/SariyuceSKC13}
\bibfield{author}{\bibinfo{person}{Ahmet~Erdem Sariy{\"{u}}ce}, \bibinfo{person}{Erik Saule}, \bibinfo{person}{Kamer Kaya}, {and} \bibinfo{person}{{\"{U}}mit~V. {\c{C}}ataly{\"{u}}rek}.} \bibinfo{year}{2013}\natexlab{}.
\newblock \showarticletitle{{STREAMER:} {A} distributed framework for incremental closeness centrality computation}. In \bibinfo{booktitle}{\emph{{CLUSTER} 2013}}. \bibinfo{pages}{1--8}.
\newblock


\bibitem[Sarma et~al\mbox{.}(2015)]%
        {DBLP:journals/tcs/SarmaMPU15}
\bibfield{author}{\bibinfo{person}{Atish~Das Sarma}, \bibinfo{person}{Anisur~Rahaman Molla}, \bibinfo{person}{Gopal Pandurangan}, {and} \bibinfo{person}{Eli Upfal}.} \bibinfo{year}{2015}\natexlab{}.
\newblock \showarticletitle{Fast distributed PageRank computation}.
\newblock \bibinfo{journal}{\emph{Theor. Comput. Sci.}}  \bibinfo{volume}{561} (\bibinfo{year}{2015}), \bibinfo{pages}{113--121}.
\newblock


\bibitem[Sattar and Arifuzzaman(2018)]%
        {DBLP:conf/dasc/SattarA18}
\bibfield{author}{\bibinfo{person}{Naw~Safrin Sattar} {and} \bibinfo{person}{Shaikh Arifuzzaman}.} \bibinfo{year}{2018}\natexlab{}.
\newblock \showarticletitle{Parallelizing Louvain Algorithm: Distributed Memory Challenges}. In \bibinfo{booktitle}{\emph{DASC/PiCom/DataCom/CyberSciTech 2018}}. \bibinfo{pages}{695--701}.
\newblock


\bibitem[Sattar and Arifuzzaman(2022)]%
        {DBLP:journals/tjs/SattarA22}
\bibfield{author}{\bibinfo{person}{Naw~Safrin Sattar} {and} \bibinfo{person}{Shaikh Arifuzzaman}.} \bibinfo{year}{2022}\natexlab{}.
\newblock \showarticletitle{Scalable distributed Louvain algorithm for community detection in large graphs}.
\newblock \bibinfo{journal}{\emph{J. Supercomput.}} \bibinfo{volume}{78}, \bibinfo{number}{7} (\bibinfo{year}{2022}), \bibinfo{pages}{10275--10309}.
\newblock


\bibitem[Shahrivari and Jalili(2021)]%
        {DBLP:journals/tkde/ShahrivariJ21}
\bibfield{author}{\bibinfo{person}{Saeed Shahrivari} {and} \bibinfo{person}{Saeed Jalili}.} \bibinfo{year}{2021}\natexlab{}.
\newblock \showarticletitle{Efficient Distributed k-Clique Mining for Large Networks Using MapReduce}.
\newblock \bibinfo{journal}{\emph{{IEEE} Trans. Knowl. Data Eng.}} \bibinfo{volume}{33}, \bibinfo{number}{3} (\bibinfo{year}{2021}), \bibinfo{pages}{964--974}.
\newblock


\bibitem[Shao et~al\mbox{.}(2014)]%
        {DBLP:conf/sigmod/ShaoCC14}
\bibfield{author}{\bibinfo{person}{Yingxia Shao}, \bibinfo{person}{Lei Chen}, {and} \bibinfo{person}{Bin Cui}.} \bibinfo{year}{2014}\natexlab{}.
\newblock \showarticletitle{Efficient cohesive subgraphs detection in parallel}. In \bibinfo{booktitle}{\emph{{SIGMOD} 2014}}. \bibinfo{pages}{613--624}.
\newblock


\bibitem[Sharma et~al\mbox{.}(2017)]%
        {DBLP:conf/www/SharmaSG17}
\bibfield{author}{\bibinfo{person}{Aneesh Sharma}, \bibinfo{person}{C. Seshadhri}, {and} \bibinfo{person}{Ashish Goel}.} \bibinfo{year}{2017}\natexlab{}.
\newblock \showarticletitle{When Hashes Met Wedges: {A} Distributed Algorithm for Finding High Similarity Vectors}. In \bibinfo{booktitle}{\emph{{WWW} 2017}}. \bibinfo{publisher}{{ACM}}, \bibinfo{pages}{431--440}.
\newblock


\bibitem[Shiraz et~al\mbox{.}(2013)]%
        {DBLP:journals/comsur/ShirazGKB13}
\bibfield{author}{\bibinfo{person}{Muhammad Shiraz}, \bibinfo{person}{Abdullah Gani}, \bibinfo{person}{Rashid~Hafeez Khokhar}, {and} \bibinfo{person}{Rajkumar Buyya}.} \bibinfo{year}{2013}\natexlab{}.
\newblock \showarticletitle{A Review on Distributed Application Processing Frameworks in Smart Mobile Devices for Mobile Cloud Computing}.
\newblock \bibinfo{journal}{\emph{{IEEE} Commun. Surv. Tutorials}} \bibinfo{volume}{15}, \bibinfo{number}{3} (\bibinfo{year}{2013}), \bibinfo{pages}{1294--1313}.
\newblock


\bibitem[Simmhan et~al\mbox{.}(2014)]%
        {DBLP:conf/europar/SimmhanKWNRRP14}
\bibfield{author}{\bibinfo{person}{Yogesh Simmhan}, \bibinfo{person}{Alok~Gautam Kumbhare}, \bibinfo{person}{Charith Wickramaarachchi}, \bibinfo{person}{Soonil Nagarkar}, \bibinfo{person}{Santosh Ravi}, \bibinfo{person}{Cauligi~S. Raghavendra}, {and} \bibinfo{person}{Viktor~K. Prasanna}.} \bibinfo{year}{2014}\natexlab{}.
\newblock \showarticletitle{GoFFish: {A} Sub-graph Centric Framework for Large-Scale Graph Analytics}. In \bibinfo{booktitle}{\emph{Euro-Par 2014}} \emph{(\bibinfo{series}{Lecture Notes in Computer Science}, Vol.~\bibinfo{volume}{8632})}. \bibinfo{pages}{451--462}.
\newblock


\bibitem[Soudani et~al\mbox{.}(2019)]%
        {DBLP:journals/kais/SoudaniFN19}
\bibfield{author}{\bibinfo{person}{Nasrin~Mazaheri Soudani}, \bibinfo{person}{Afsaneh Fatemi}, {and} \bibinfo{person}{Mohammadali Nematbakhsh}.} \bibinfo{year}{2019}\natexlab{}.
\newblock \showarticletitle{PPR-partitioning: a distributed graph partitioning algorithm based on the personalized PageRank vectors in vertex-centric systems}.
\newblock \bibinfo{journal}{\emph{Knowl. Inf. Syst.}} \bibinfo{volume}{61}, \bibinfo{number}{2} (\bibinfo{year}{2019}), \bibinfo{pages}{847--871}.
\newblock


\bibitem[Steil et~al\mbox{.}(2021)]%
        {DBLP:conf/sc/SteilRIPSP21}
\bibfield{author}{\bibinfo{person}{Trevor Steil}, \bibinfo{person}{Tahsin Reza}, \bibinfo{person}{Keita Iwabuchi}, \bibinfo{person}{Benjamin~W. Priest}, \bibinfo{person}{Geoffrey Sanders}, {and} \bibinfo{person}{Roger Pearce}.} \bibinfo{year}{2021}\natexlab{}.
\newblock \showarticletitle{TriPoll: computing surveys of triangles in massive-scale temporal graphs with metadata}. In \bibinfo{booktitle}{\emph{{SC} 2021}}. \bibinfo{publisher}{{ACM}}, \bibinfo{pages}{67}.
\newblock


\bibitem[Strausz et~al\mbox{.}(2022)]%
        {DBLP:conf/ipps/StrauszVGBH22}
\bibfield{author}{\bibinfo{person}{Andr{\'{a}}s Strausz}, \bibinfo{person}{Flavio Vella}, \bibinfo{person}{Salvatore~Di Girolamo}, \bibinfo{person}{Maciej Besta}, {and} \bibinfo{person}{Torsten Hoefler}.} \bibinfo{year}{2022}\natexlab{}.
\newblock \showarticletitle{Asynchronous Distributed-Memory Triangle Counting and {LCC} with {RMA} Caching}. In \bibinfo{booktitle}{\emph{{IPDPS} 2022}}. \bibinfo{pages}{291--301}.
\newblock


\bibitem[Sun et~al\mbox{.}(2022)]%
        {DBLP:journals/tsmc/SunQSCSWZ22}
\bibfield{author}{\bibinfo{person}{Changhao Sun}, \bibinfo{person}{Huaxin Qiu}, \bibinfo{person}{Wei Sun}, \bibinfo{person}{Qian Chen}, \bibinfo{person}{Li Su}, \bibinfo{person}{Xiaochu Wang}, {and} \bibinfo{person}{Qingrui Zhou}.} \bibinfo{year}{2022}\natexlab{}.
\newblock \showarticletitle{Better Approximation for Distributed Weighted Vertex Cover via Game-Theoretic Learning}.
\newblock \bibinfo{journal}{\emph{{IEEE} Trans. Syst. Man Cybern. Syst.}} \bibinfo{volume}{52}, \bibinfo{number}{8} (\bibinfo{year}{2022}), \bibinfo{pages}{5308--5319}.
\newblock


\bibitem[Svendsen et~al\mbox{.}(2015)]%
        {DBLP:journals/jpdc/SvendsenMT15}
\bibfield{author}{\bibinfo{person}{Michael Svendsen}, \bibinfo{person}{Arko~Provo Mukherjee}, {and} \bibinfo{person}{Srikanta Tirthapura}.} \bibinfo{year}{2015}\natexlab{}.
\newblock \showarticletitle{Mining maximal cliques from a large graph using MapReduce: Tackling highly uneven subproblem sizes}.
\newblock \bibinfo{journal}{\emph{J. Parallel Distributed Comput.}}  \bibinfo{volume}{79-80} (\bibinfo{year}{2015}), \bibinfo{pages}{104--114}.
\newblock


\bibitem[Talukder and Zaki(2016)]%
        {DBLP:journals/datamine/TalukderZ16}
\bibfield{author}{\bibinfo{person}{Nilothpal Talukder} {and} \bibinfo{person}{Mohammed~J. Zaki}.} \bibinfo{year}{2016}\natexlab{}.
\newblock \showarticletitle{A distributed approach for graph mining in massive networks}.
\newblock \bibinfo{journal}{\emph{Data Min. Knowl. Discov.}} \bibinfo{volume}{30}, \bibinfo{number}{5} (\bibinfo{year}{2016}), \bibinfo{pages}{1024--1052}.
\newblock


\bibitem[Teixeira et~al\mbox{.}(2015)]%
        {DBLP:conf/sosp/TeixeiraFSSZA15}
\bibfield{author}{\bibinfo{person}{Carlos H.~C. Teixeira}, \bibinfo{person}{Alexandre~J. Fonseca}, \bibinfo{person}{Marco Serafini}, \bibinfo{person}{Georgos Siganos}, \bibinfo{person}{Mohammed~J. Zaki}, {and} \bibinfo{person}{Ashraf Aboulnaga}.} \bibinfo{year}{2015}\natexlab{}.
\newblock \showarticletitle{Arabesque: a system for distributed graph mining}. In \bibinfo{booktitle}{\emph{{SOSP} 2015}}. \bibinfo{pages}{425--440}.
\newblock


\bibitem[The MPI~Forum(1993)]%
        {the1993mpi}
\bibfield{author}{\bibinfo{person}{CORPORATE The MPI~Forum}.} \bibinfo{year}{1993}\natexlab{}.
\newblock \showarticletitle{MPI: a message passing interface}. In \bibinfo{booktitle}{\emph{Proceedings of the 1993 ACM/IEEE Conference on Supercomputing}}. \bibinfo{pages}{878--883}.
\newblock


\bibitem[Tian et~al\mbox{.}(2013)]%
        {DBLP:journals/pvldb/TianBCTM13}
\bibfield{author}{\bibinfo{person}{Yuanyuan Tian}, \bibinfo{person}{Andrey Balmin}, \bibinfo{person}{Severin~Andreas Corsten}, \bibinfo{person}{Shirish Tatikonda}, {and} \bibinfo{person}{John McPherson}.} \bibinfo{year}{2013}\natexlab{}.
\newblock \showarticletitle{From "Think Like a Vertex" to "Think Like a Graph"}.
\newblock \bibinfo{journal}{\emph{Proc. {VLDB} Endow.}} \bibinfo{volume}{7}, \bibinfo{number}{3} (\bibinfo{year}{2013}), \bibinfo{pages}{193--204}.
\newblock


\bibitem[Tomita et~al\mbox{.}(2004)]%
        {DBLP:conf/cocoon/TomitaTT04}
\bibfield{author}{\bibinfo{person}{Etsuji Tomita}, \bibinfo{person}{Akira Tanaka}, {and} \bibinfo{person}{Haruhisa Takahashi}.} \bibinfo{year}{2004}\natexlab{}.
\newblock \showarticletitle{The Worst-Case Time Complexity for Generating All Maximal Cliques}. In \bibinfo{booktitle}{\emph{{COCOON} 2004}} \emph{(\bibinfo{series}{Lecture Notes in Computer Science}, Vol.~\bibinfo{volume}{3106})}. \bibinfo{pages}{161--170}.
\newblock


\bibitem[Toshniwal et~al\mbox{.}(2014)]%
        {toshniwal2014storm}
\bibfield{author}{\bibinfo{person}{Ankit Toshniwal}, \bibinfo{person}{Siddarth Taneja}, \bibinfo{person}{Amit Shukla}, \bibinfo{person}{Karthik Ramasamy}, \bibinfo{person}{Jignesh~M Patel}, \bibinfo{person}{Sanjeev Kulkarni}, \bibinfo{person}{Jason Jackson}, \bibinfo{person}{Krishna Gade}, \bibinfo{person}{Maosong Fu}, \bibinfo{person}{Jake Donham}, {et~al\mbox{.}}} \bibinfo{year}{2014}\natexlab{}.
\newblock \showarticletitle{Storm@ twitter}. In \bibinfo{booktitle}{\emph{Proceedings of SIGMOD}}. \bibinfo{pages}{147--156}.
\newblock


\bibitem[Trinder et~al\mbox{.}(2002)]%
        {DBLP:journals/jfp/TrinderLP02}
\bibfield{author}{\bibinfo{person}{Philip~W. Trinder}, \bibinfo{person}{Hans{-}Wolfgang Loidl}, {and} \bibinfo{person}{Robert~F. Pointon}.} \bibinfo{year}{2002}\natexlab{}.
\newblock \showarticletitle{Parallel and Distributed Haskells}.
\newblock \bibinfo{journal}{\emph{J. Funct. Program.}} \bibinfo{volume}{12}, \bibinfo{number}{4{\&}5} (\bibinfo{year}{2002}), \bibinfo{pages}{469--510}.
\newblock


\bibitem[Wallmann and Gerschberger(2020)]%
        {DBLP:conf/ism2/WallmannG20}
\bibfield{author}{\bibinfo{person}{Christian Wallmann} {and} \bibinfo{person}{Markus Gerschberger}.} \bibinfo{year}{2020}\natexlab{}.
\newblock \showarticletitle{The association between network centrality measures and supply chain performance: The case of distribution networks}. In \bibinfo{booktitle}{\emph{{ISM} 2020}}, Vol.~\bibinfo{volume}{180}. \bibinfo{publisher}{Elsevier}, \bibinfo{pages}{172--179}.
\newblock


\bibitem[Wang et~al\mbox{.}(2022)]%
        {DBLP:conf/sc/WangCMYC22}
\bibfield{author}{\bibinfo{person}{Yuanwei Wang}, \bibinfo{person}{Huanqi Cao}, \bibinfo{person}{Zixuan Ma}, \bibinfo{person}{Wanwang Yin}, {and} \bibinfo{person}{Wenguang Chen}.} \bibinfo{year}{2022}\natexlab{}.
\newblock \showarticletitle{Scaling Graph 500 {SSSP} to 140 Trillion Edges with over 40 Million Cores}. In \bibinfo{booktitle}{\emph{{SC} 2022}}. \bibinfo{pages}{19:1--19:15}.
\newblock


\bibitem[Wang et~al\mbox{.}(2020)]%
        {DBLP:journals/pvldb/0012XFC00M20}
\bibfield{author}{\bibinfo{person}{Yue Wang}, \bibinfo{person}{Ruiqi Xu}, \bibinfo{person}{Zonghao Feng}, \bibinfo{person}{Yulin Che}, \bibinfo{person}{Lei Chen}, \bibinfo{person}{Qiong Luo}, {and} \bibinfo{person}{Rui Mao}.} \bibinfo{year}{2020}\natexlab{}.
\newblock \showarticletitle{{DISK:} {A} Distributed Framework for Single-Source SimRank with Accuracy Guarantee}.
\newblock \bibinfo{journal}{\emph{Proc. {VLDB} Endow.}} \bibinfo{volume}{14}, \bibinfo{number}{3} (\bibinfo{year}{2020}), \bibinfo{pages}{351--363}.
\newblock


\bibitem[Wang et~al\mbox{.}(2019)]%
        {DBLP:conf/icde/WangGHYH19}
\bibfield{author}{\bibinfo{person}{Zhaokang Wang}, \bibinfo{person}{Rong Gu}, \bibinfo{person}{Weiwei Hu}, \bibinfo{person}{Chunfeng Yuan}, {and} \bibinfo{person}{Yihua Huang}.} \bibinfo{year}{2019}\natexlab{}.
\newblock \showarticletitle{{BENU:} Distributed Subgraph Enumeration with Backtracking-Based Framework}. In \bibinfo{booktitle}{\emph{{ICDE} 2019}}. \bibinfo{pages}{136--147}.
\newblock


\bibitem[Wehmuth and Ziviani(2013)]%
        {DBLP:journals/cn/WehmuthZ13}
\bibfield{author}{\bibinfo{person}{Klaus Wehmuth} {and} \bibinfo{person}{Artur Ziviani}.} \bibinfo{year}{2013}\natexlab{}.
\newblock \showarticletitle{{DACCER:} Distributed Assessment of the Closeness Centrality Ranking in complex networks}.
\newblock \bibinfo{journal}{\emph{Comput. Networks}} \bibinfo{volume}{57}, \bibinfo{number}{13} (\bibinfo{year}{2013}), \bibinfo{pages}{2536--2548}.
\newblock


\bibitem[Weng et~al\mbox{.}(2022)]%
        {DBLP:journals/tpds/WengZLPL22}
\bibfield{author}{\bibinfo{person}{Tongfeng Weng}, \bibinfo{person}{Xu Zhou}, \bibinfo{person}{Kenli Li}, \bibinfo{person}{Peng Peng}, {and} \bibinfo{person}{Keqin Li}.} \bibinfo{year}{2022}\natexlab{}.
\newblock \showarticletitle{Efficient Distributed Approaches to Core Maintenance on Large Dynamic Graphs}.
\newblock \bibinfo{journal}{\emph{{IEEE} Trans. Parallel Distributed Syst.}} \bibinfo{volume}{33}, \bibinfo{number}{1} (\bibinfo{year}{2022}), \bibinfo{pages}{129--143}.
\newblock


\bibitem[White(2012)]%
        {white2012hadoop}
\bibfield{author}{\bibinfo{person}{Tom White}.} \bibinfo{year}{2012}\natexlab{}.
\newblock \bibinfo{booktitle}{\emph{Hadoop: The definitive guide}}.
\newblock \bibinfo{publisher}{" O'Reilly Media, Inc."}.
\newblock


\bibitem[Xiang et~al\mbox{.}(2021)]%
        {DBLP:conf/sc/XiangKSHS21}
\bibfield{author}{\bibinfo{person}{Lizhi Xiang}, \bibinfo{person}{Arif Khan}, \bibinfo{person}{Edoardo Serra}, \bibinfo{person}{Mahantesh Halappanavar}, {and} \bibinfo{person}{Aravind Sukumaran{-}Rajam}.} \bibinfo{year}{2021}\natexlab{}.
\newblock \showarticletitle{cuTS: scaling subgraph isomorphism on distributed multi-GPU systems using trie based data structure}. In \bibinfo{booktitle}{\emph{{SC} 2021}}. \bibinfo{pages}{69}.
\newblock


\bibitem[Xu et~al\mbox{.}(2016)]%
        {DBLP:journals/tsc/XuCF16}
\bibfield{author}{\bibinfo{person}{Yanyan Xu}, \bibinfo{person}{James Cheng}, {and} \bibinfo{person}{Ada~Wai{-}Chee Fu}.} \bibinfo{year}{2016}\natexlab{}.
\newblock \showarticletitle{Distributed Maximal Clique Computation and Management}.
\newblock \bibinfo{journal}{\emph{{IEEE} Trans. Serv. Comput.}} \bibinfo{volume}{9}, \bibinfo{number}{1} (\bibinfo{year}{2016}), \bibinfo{pages}{110--122}.
\newblock


\bibitem[Xu et~al\mbox{.}(2014)]%
        {DBLP:conf/bigdata/XuCFB14}
\bibfield{author}{\bibinfo{person}{Yanyan Xu}, \bibinfo{person}{James Cheng}, \bibinfo{person}{Ada~Wai{-}Chee Fu}, {and} \bibinfo{person}{Yingyi Bu}.} \bibinfo{year}{2014}\natexlab{}.
\newblock \showarticletitle{Distributed Maximal Clique Computation}. In \bibinfo{booktitle}{\emph{IEEE BigData 2014}}. \bibinfo{pages}{160--167}.
\newblock


\bibitem[Yan et~al\mbox{.}(2014)]%
        {DBLP:journals/pvldb/YanCLN14}
\bibfield{author}{\bibinfo{person}{Da Yan}, \bibinfo{person}{James Cheng}, \bibinfo{person}{Yi Lu}, {and} \bibinfo{person}{Wilfred Ng}.} \bibinfo{year}{2014}\natexlab{}.
\newblock \showarticletitle{Blogel: {A} Block-Centric Framework for Distributed Computation on Real-World Graphs}.
\newblock \bibinfo{journal}{\emph{Proc. {VLDB} Endow.}} \bibinfo{volume}{7}, \bibinfo{number}{14} (\bibinfo{year}{2014}), \bibinfo{pages}{1981--1992}.
\newblock


\bibitem[Yan et~al\mbox{.}(2015)]%
        {DBLP:conf/www/YanCLN15}
\bibfield{author}{\bibinfo{person}{Da Yan}, \bibinfo{person}{James Cheng}, \bibinfo{person}{Yi Lu}, {and} \bibinfo{person}{Wilfred Ng}.} \bibinfo{year}{2015}\natexlab{}.
\newblock \showarticletitle{Effective Techniques for Message Reduction and Load Balancing in Distributed Graph Computation}. In \bibinfo{booktitle}{\emph{{WWW} 2015}}. \bibinfo{pages}{1307--1317}.
\newblock


\bibitem[Yang et~al\mbox{.}(2021)]%
        {DBLP:conf/sigmod/YangL0H021}
\bibfield{author}{\bibinfo{person}{Zhengyi Yang}, \bibinfo{person}{Longbin Lai}, \bibinfo{person}{Xuemin Lin}, \bibinfo{person}{Kongzhang Hao}, {and} \bibinfo{person}{Wenjie Zhang}.} \bibinfo{year}{2021}\natexlab{}.
\newblock \showarticletitle{{HUGE:} An Efficient and Scalable Subgraph Enumeration System}. In \bibinfo{booktitle}{\emph{{SIGMOD} 2021}}. \bibinfo{pages}{2049--2062}.
\newblock


\bibitem[Yasar et~al\mbox{.}(2022)]%
        {DBLP:journals/tpds/YasarRBC22}
\bibfield{author}{\bibinfo{person}{Abdurrahman Yasar}, \bibinfo{person}{Sivasankaran Rajamanickam}, \bibinfo{person}{Jonathan~W. Berry}, {and} \bibinfo{person}{{\"{U}}mit~V. {\c{C}}ataly{\"{u}}rek}.} \bibinfo{year}{2022}\natexlab{}.
\newblock \showarticletitle{A Block-Based Triangle Counting Algorithm on Heterogeneous Environments}.
\newblock \bibinfo{journal}{\emph{{IEEE} Trans. Parallel Distributed Syst.}} \bibinfo{volume}{33}, \bibinfo{number}{2} (\bibinfo{year}{2022}), \bibinfo{pages}{444--458}.
\newblock


\bibitem[Yu et~al\mbox{.}(2020)]%
        {DBLP:conf/sigmod/Yu0KLLCY20}
\bibfield{author}{\bibinfo{person}{Ziqiang Yu}, \bibinfo{person}{Xiaohui Yu}, \bibinfo{person}{Nick Koudas}, \bibinfo{person}{Yang Liu}, \bibinfo{person}{Yifan Li}, \bibinfo{person}{Yueting Chen}, {and} \bibinfo{person}{Dingyu Yang}.} \bibinfo{year}{2020}\natexlab{}.
\newblock \showarticletitle{Distributed Processing of k Shortest Path Queries over Dynamic Road Networks}. In \bibinfo{booktitle}{\emph{{SIGMOD} 2020}}. \bibinfo{pages}{665--679}.
\newblock


\bibitem[Zaharia et~al\mbox{.}(2012)]%
        {DBLP:conf/nsdi/ZahariaCDDMMFSS12}
\bibfield{author}{\bibinfo{person}{Matei Zaharia}, \bibinfo{person}{Mosharaf Chowdhury}, \bibinfo{person}{Tathagata Das}, \bibinfo{person}{Ankur Dave}, \bibinfo{person}{Justin Ma}, \bibinfo{person}{Murphy McCauly}, \bibinfo{person}{Michael~J. Franklin}, \bibinfo{person}{Scott Shenker}, {and} \bibinfo{person}{Ion Stoica}.} \bibinfo{year}{2012}\natexlab{}.
\newblock \showarticletitle{Resilient Distributed Datasets: {A} Fault-Tolerant Abstraction for In-Memory Cluster Computing}. In \bibinfo{booktitle}{\emph{{NSDI} 2012}}. \bibinfo{pages}{15--28}.
\newblock


\bibitem[Zaharia et~al\mbox{.}(2010)]%
        {DBLP:conf/hotcloud/ZahariaCFSS10}
\bibfield{author}{\bibinfo{person}{Matei Zaharia}, \bibinfo{person}{Mosharaf Chowdhury}, \bibinfo{person}{Michael~J. Franklin}, \bibinfo{person}{Scott Shenker}, {and} \bibinfo{person}{Ion Stoica}.} \bibinfo{year}{2010}\natexlab{}.
\newblock \showarticletitle{Spark: Cluster Computing with Working Sets}. In \bibinfo{booktitle}{\emph{HotCloud 2010}}.
\newblock


\bibitem[Zaharia et~al\mbox{.}(2016)]%
        {DBLP:journals/cacm/ZahariaXWDADMRV16}
\bibfield{author}{\bibinfo{person}{Matei Zaharia}, \bibinfo{person}{Reynold~S. Xin}, \bibinfo{person}{Patrick Wendell}, \bibinfo{person}{Tathagata Das}, \bibinfo{person}{Michael Armbrust}, \bibinfo{person}{Ankur Dave}, \bibinfo{person}{Xiangrui Meng}, \bibinfo{person}{Josh Rosen}, \bibinfo{person}{Shivaram Venkataraman}, \bibinfo{person}{Michael~J. Franklin}, \bibinfo{person}{Ali Ghodsi}, \bibinfo{person}{Joseph Gonzalez}, \bibinfo{person}{Scott Shenker}, {and} \bibinfo{person}{Ion Stoica}.} \bibinfo{year}{2016}\natexlab{}.
\newblock \showarticletitle{Apache Spark: a unified engine for big data processing}.
\newblock \bibinfo{journal}{\emph{Commun. {ACM}}} \bibinfo{volume}{59}, \bibinfo{number}{11} (\bibinfo{year}{2016}), \bibinfo{pages}{56--65}.
\newblock


\bibitem[Zeng and Yu(2018)]%
        {DBLP:conf/cluster/ZengY18}
\bibfield{author}{\bibinfo{person}{Jianping Zeng} {and} \bibinfo{person}{Hongfeng Yu}.} \bibinfo{year}{2018}\natexlab{}.
\newblock \showarticletitle{A Scalable Distributed Louvain Algorithm for Large-Scale Graph Community Detection}. In \bibinfo{booktitle}{\emph{{CLUSTER} 2018}}. \bibinfo{pages}{268--278}.
\newblock


\bibitem[Zhang et~al\mbox{.}(2017)]%
        {DBLP:journals/vldb/ZhangYWTCS17}
\bibfield{author}{\bibinfo{person}{Dongxiang Zhang}, \bibinfo{person}{Dingyu Yang}, \bibinfo{person}{Yuan Wang}, \bibinfo{person}{Kian{-}Lee Tan}, \bibinfo{person}{Jian Cao}, {and} \bibinfo{person}{Heng~Tao Shen}.} \bibinfo{year}{2017}\natexlab{}.
\newblock \showarticletitle{Distributed shortest path query processing on dynamic road networks}.
\newblock \bibinfo{journal}{\emph{{VLDB} J.}} \bibinfo{volume}{26}, \bibinfo{number}{3} (\bibinfo{year}{2017}), \bibinfo{pages}{399--419}.
\newblock


\bibitem[Zhou et~al\mbox{.}(2021)]%
        {DBLP:journals/tpds/ZhouGG21}
\bibfield{author}{\bibinfo{person}{Tian Zhou}, \bibinfo{person}{Lixin Gao}, {and} \bibinfo{person}{Xiaohong Guan}.} \bibinfo{year}{2021}\natexlab{}.
\newblock \showarticletitle{A Fault-Tolerant Distributed Framework for Asynchronous Iterative Computations}.
\newblock \bibinfo{journal}{\emph{{IEEE} Trans. Parallel Distributed Syst.}} \bibinfo{volume}{32}, \bibinfo{number}{8} (\bibinfo{year}{2021}), \bibinfo{pages}{2062--2073}.
\newblock


\bibitem[Zhu et~al\mbox{.}(2020)]%
        {DBLP:journals/fgcs/ZhuHLM20}
\bibfield{author}{\bibinfo{person}{Huanzhou Zhu}, \bibinfo{person}{Ligang He}, \bibinfo{person}{Matthew Leeke}, {and} \bibinfo{person}{Rui Mao}.} \bibinfo{year}{2020}\natexlab{}.
\newblock \showarticletitle{WolfGraph: The edge-centric graph processing on {GPU}}.
\newblock \bibinfo{journal}{\emph{Future Gener. Comput. Syst.}}  \bibinfo{volume}{111} (\bibinfo{year}{2020}), \bibinfo{pages}{552--569}.
\newblock


\bibitem[Zuzic et~al\mbox{.}(2022)]%
        {DBLP:conf/soda/ZuzicGYHS22}
\bibfield{author}{\bibinfo{person}{Goran Zuzic}, \bibinfo{person}{Gramoz Goranci}, \bibinfo{person}{Mingquan Ye}, \bibinfo{person}{Bernhard Haeupler}, {and} \bibinfo{person}{Xiaorui Sun}.} \bibinfo{year}{2022}\natexlab{}.
\newblock \showarticletitle{Universally-Optimal Distributed Shortest Paths and Transshipment via Graph-Based {\(\mathscr{l}\)}\({}_{\mbox{1}}\)-Oblivious Routing}. In \bibinfo{booktitle}{\emph{{SODA} 2022}}. \bibinfo{pages}{2549--2579}.
\newblock


\end{thebibliography}
